\documentclass{aa}
\usepackage[english]{babel}

\usepackage{hyperref}
\usepackage{txfonts}
\usepackage{epsfig}
\usepackage{graphicx,natbib,subfigure}
\usepackage{wasysym}
\usepackage{ulem}
\usepackage[usenames,dvipsnames,svgnames,table]{xcolor}

\bibpunct{(}{)}{;}{a}{}{,}

\hypersetup{
    colorlinks=true,
    citecolor=blue,
    linkcolor=red,
    urlcolor=black
    }

\def\ga{\,\hbox{\hbox{$ > $}\kern -0.8em \lower 1.0ex\hbox{$\sim$}}\,}
\def\la{\,\hbox{\hbox{$ < $}\kern -0.8em \lower 1.0ex\hbox{$\sim$}}\,}
\def\aap{A\&A}
\def\mnras{MNRAS}

\def\msol{\,{\rm M}_\odot}              

\newcommand{\vect}[1]{\mathbf{#1}}

\renewcommand{\div}{\boldsymbol{\nabla}\cdot}

\newcommand{\rot}[1]{\boldsymbol{\nabla}\times\vect{#1}}

\newcommand{\nable}{\boldsymbol{\nabla}}

\newcommand{\trm}[1]{\textrm{#1}}
\newcommand{\refeq}[1]{Eq.~(\ref{#1})}
\newcommand{\refeqp}[1]{Eq.~\ref{#1}}
\newcommand{\refig}[1]{Fig.~\ref{#1}}
\newcommand{\soll}{\astrosun}
\newcommand{\tff}{\trm{t}_{\trm{free-fall}}}

\newcommand{\noprint}[1]{{}}

\begin{document}

\def\nat{Nature }
\def\apj{Astrophys. J. }
\def\apjs{Astrophys. J., Suppl. Ser. }
\def\apjl{Astrophys. J., Lett. }
\def\apss{Astrophys. and Space Science}

\title{Ambipolar diffusion in low-mass star formation. I. General comparison with the ideal MHD case.}

\author{J. Masson \inst{1} \and
        G. Chabrier \inst{1,2} \and 
        P. Hennebelle \inst{3} \and
        N. Vaytet \inst{2}  \and
        B. Commer\c{c}on \inst{2}  }

\institute{School of Physics, University of Exeter, Exeter, EX4 4QL, UK \and
\'Ecole Normale Sup\'erieure de Lyon, CRAL, UMR CNRS 5574, Universit\'e de Lyon, 46 All\'ee d'Italie, 69364 Lyon Cedex 07, France \and 
Laboratoire de radioastronomie, UMR CNRS 8112, \'Ecole Normale Sup\'erieure et Observatoire de Paris, 24 rue Lhomond, 75231 Paris Cedex 05, France} 

\authorrunning{J. Masson et al.}
\titlerunning{Ambipolar diffusion in low-mass star formation. I.}

\date{}

\abstract
{Angular momentum transport and the formation of rotationally supported structures are major issues in our understanding of protostellar core formation. 
    Whereas purely hydrodynamical simulations lead to large Keplerian disks, ideal magnetohydrodynamics (MHD) models yield the opposite result, with essentially no disk formation. 
This stems from the flux-freezing condition in ideal MHD, which leads to strong magnetic braking.

In this paper, we provide a more accurate description of the evolution of the magnetic flux redistribution by including resistive terms in the MHD equations. 
We focus more particularly on the effect of ambipolar diffusion on the properties of the first Larson core and its surrounding structure, 
exploring various initial magnetisations and magnetic field versus rotation axis orientations of a $1\,\msol$ collapsing prestellar dense core.

We used the non-ideal magnetohydrodynamics version of the adaptive mesh refinement code {\ttfamily RAMSES}  to carry out these calculations. 
The resistivities required to calculate the ambipolar diffusion terms were computed using a reduced chemical network of charged, neutral, and grain species.

Including ambipolar diffusion leads to the formation of a magnetic diffusion barrier (also known as the decoupling stage) in the vicinity of the core, 
which prevents accumulation of magnetic flux in and around the core and amplification of the field above $0.1$~G. 
The mass and radius of the first Larson core, however, remain similar between ideal and non-ideal MHD models. 
This diffusion plateau, preventing further amplification of the field and reorganising the field topology, has crucial consequences for magnetic braking processes, allowing the formation of disk structures. 
Magnetically supported outflows launched in ideal MHD models are weakened or even disappear when using non-ideal MHD.
In contrast to ideal MHD calculations, misalignment between the initial rotation axis and the magnetic field direction does not significantly affect the results for a given magnetisation, 
showing that the physical dissipation processes truly dominate numerical diffusion.

We demonstrate severe limits of the ideal MHD formalism; it yields unphysical behaviours in the long-term evolution of the system. 
This includes counter-rotation inside the outflow or magnetic tower, interchange instabilities, and flux redistribution triggered by numerical diffusion. 
These effects are not observed in non-ideal MHD.
Disks with Keplerian velocity profiles are found to form around the protostar in all our non-ideal MHD simulations, 
with a final mass and size that strongly depend on the initial magnetisation. 
This ranges from a few $10^{-2}\msol$ and $\sim 20$-30 au for the most magnetised case ($\mu=2$)
 to $\sim 2 \times 10^{-1}\msol$ and $\sim 40$-80 au for a lower magnetisation ($\mu=5$). 
In all cases, these disks remain significantly smaller than disks found in pure hydrodynamical simulations.

Ambipolar diffusion thus bears a crucial impact on the regulation of magnetic flux 
and angular momentum transport during the collapse of a prestellar core and the formation of the resulting protostellar core-disk system,
enabling the formation and growth of rotationally supported structures.}


\maketitle

\section{Introduction}
In the past few decades, star formation studies have been struggling to properly describe the mechanism of angular momentum transport, regulated by angular momentum conservation and magnetic braking.
The formation of rotationally supported structures, that is, the protoplanetary disks that are expected to give birth to planets and/or binaries, is
directly related to the amount of angular momentum in protostellar systems.
While there is plenty of observational evidence for large Keplerian disks\footnote{See Appendix~\ref{app:imhddiskprofile} for details on our definition of a Keplerian disk.} around Class-II and Class-I objects, their presence around younger Class-0 objects is still subject to debate (as discussed in \citealt{2010A&A...512A..40M} or \citealt{2012Natur.492...83T,2013ApJ...771...48T}, see the review by \citealt{2014prpl.conf..173L}).

The simplified framework of the ideal magnetohydrodynamics (MHD) used in the first studies led to
the disappearance of the large Keplerian disks
that are easily formed
in hydrodynamical simulations as a result of the very effective magnetic braking
created by a strong pile-up of magnetic flux towards the centre of the collapsing system
(\citealt{GalliShuLizano2006}, \citealt{2003ApJ...599..363A}, \citealt{2007Ap&SS.311...75P,HennebelleTeyssier2008,2004ApJ...616..266M,HennebelleFromang2008,commercon10}). In these simulations, disks were found to form only for unrealistically weak magnetic field intensities \citep[corresponding to a mass-to-flux ratio more than 10 times the critical value derived by][]{MouschoviasSpitzer1976}. 
Other consequences of the magnetic flux freezing assumption inherent to the ideal MHD approximation include the strong resulting magnetisation of protostars compared to the low observed values in stars \citep{crutcher_review}, the generation of violent interchange instabilities at the protostar-disk interface \citep{2014ApJ...793..130L}, the growth of the pseudo-disk \citep{HennebelleFromang2008}, and the distortion of the upper layer of the pseudo-disk due to reconnection and the split monopole \citep{GalliShu1993,2014ApJ...793..130L}.

In the framework of ideal MHD, there is no possibility to regulate the magnetic flux pile-up and its consequences, except for the intrinsic numerical resistivity due to both the numerical method used to solve the induction equation and the numerical grid resolution.
To circumvent this problem, the magnetic field redistribution needs
to be correctly addressed in the framework of complete non-ideal MHD. Following the pioneering work of \citet{MestelSpitzer56}, the study of non-ideal MHD effects has been the focus of intense research in the recent years,
as illustrated by the studies of \citet{DuffinPudritz} and \citet{MellonLi2009} for ambipolar diffusion, or \citet{Machida_etal06} and \citet{MachidaMatsumotoDisk} for a generic resistivity. While it is still unclear if non-ideal MHD can, by itself, solve the problems regarding the formation of disks \citep{Krasnopolsky}, a physical dissipative scale for the magnetic flux definitely improves the regulation of magnetic flux pile-up in star formation studies.

In this work, we study the effects of ambipolar diffusion in the context of star formation, more specifically, of low-mass star formation, and highlight the differences compared to ideal MHD simulations. The influence of turbulent initial conditions will be studied in a forthcoming paper (Paper II). The article is organised as follows. In Sect.~\ref{2} we discuss the framework and numerical setup of the study. In Sect.~\ref{sec:Results} we focus on the general description and properties of the collapsing core until formation and early evolution of the first Larson core, for various cases. In Sect.~\ref{sec:longtermevo} we discuss the long-term evolution of the structures and highlight the limits of the ideal MHD framework that are due to numerical issues. In Sect.~\ref{sec:disks}, we examine the formation of rotationally supported structures in non-ideal MHD. Last, we summarise our findings in Sect.~\ref{sec:Conclusions}.

\section{General context\label{2}}
\subsection{Physical framework}
In star formation, the ionisation fraction is low, and
quasi-equilibrium holds between the Lorentz force and the plasma-neutrals friction force as a result of collisions. Therefore, we can drop the pressure and gravitational forces for charged particles
when
writing the equations of motion for each fluid particle.
In the case of positively charged species of number density $n_{\trm{i}}$, velocity $\mathbf{v_{\trm{i}}}$, atomic number $Z_{\trm{i}}$ , and collision rate $\nu_{\trm{i}j}$ with the species $j$, and using the subscript $\trm{e}$ for the negatively charged species, the equations of motion read:

\begin{equation}
\left\{
    \begin{array}{rcl}
        Z_{\trm{i}} e n_{\trm{i}}(\mathbf{E}+\mathbf{v_{\trm{i}}} \times \mathbf{B}) -\rho_{\trm{i}} \sum_{j\trm{=\{e,n\}}} \nu_{\trm{i}j}(\mathbf{v_{\trm{i}}}-\mathbf{v}_{{j}}) &=&0 \\
        -e n_{\trm{e}}(\mathbf{E}+\mathbf{v_{\trm{e}}} \times \mathbf{B}) -\rho_{\trm{e}}\sum_{j\trm{=\{i,n\}}} \nu_{\trm{e}j}(\mathbf{v_{\trm{e}}}-\mathbf{v}_{j}) &=&0~, 
\end{array} \right.
\label{deuxeqs}
\end{equation}
while the generalized Ohm's law is written \citep[see e.g.][]{balbus2001} 
\begin{equation}
    \partial_t{\mathbf{B}} = \nable \times \Big[  \mathbf{v_{\trm{n}}} \times \mathbf{B} - \frac{\mathbf{J} \times \mathbf{B}}{e n_{\trm{e}}} + \frac{\big[(\nable \times \mathbf{B}) \times \mathbf{B} \big]  \times \mathbf{B}}{\gamma_{\trm{AD}} \rho \rho_{\trm{i}}} - \frac{\mathbf{J}}{\sigma_{\parallel}}  \Big]
\label{cavient}
\end{equation}
with
\begin{align}
    &\gamma_{\trm{AD}} = \frac{<\sigma_{\trm{in}} v_{\trm{i}}>}{(m_{\trm{i}}+m_{\trm{n}})} ~~\text{the drag coefficient and} \label{dragcoef}\\
    &\sigma_{\parallel} = \frac{n_{\trm{e}}e^2}{n_{\trm{n}}  m_{\trm{e}}<\sigma_{\trm{en}} v_{\trm{e}}>} ~~\text{the electrical conductivity.}
\end{align}
The symbols $\mathbf{v_{\trm{n}}}$, $\rho,$ and $\rho_{\trm{i}}$ denote the velocity of neutral particles, the total density, and the ion density, respectively. In this framework, the drag force per unit volume exerted by the neutrals on the ions compensates for the Lorentz force and reads $\vect{F_{\trm{drag}}}=\gamma_{\trm{AD}} \rho_i \rho (\vect{v_{\trm{i}}}-\vect{v_{\trm{n}}})$. When considering multiple species \citep[see][]{KunzMouschovias2009,nakano2002}, Ohm's law can be written as
\begin{align}
    \frac{\partial \mathbf{B}}{\partial t} =& \nable \times \Bigg[ \mathbf{v_{\trm{n}}} \times \mathbf{B} \nonumber \\
      -& \eta_{\Omega} (\nable \times \mathbf{B}) \nonumber \\
      -& \eta_{\trm{H}} \left\{ (\nable \times \mathbf{B})\times \frac{\mathbf{B}}{||\vect{B}||} \right\} \nonumber \\
      -& \eta_{\trm{AD}} \frac{\mathbf{B}}{||\vect{B}||}  \times \left\{ (\nable \times \mathbf{B})\times \frac{\mathbf{B}}{||\vect{B}||} \right\}  \Bigg] \label{induc_1} ~,
\end{align}
with $||~||$ standing for the L2 norm, and the Ohmic, Hall, and ambipolar diffusivities are defined as 
\begin{align}
\eta_{\Omega} &= \frac{1}{\sigma_{\parallel}} ~,\\
\eta_{\trm{H}} &= \frac{\sigma_{\trm{H}}}{\sigma_{\bot}^2+ \sigma_{\trm{H}}^2} ~, \\
\eta_{\trm{AD}} &= \frac{\sigma_{\bot}}{\sigma_{\bot}^2+ \sigma_{\trm{H}}^2} - \frac{1}{\sigma_{\parallel}}.
\end{align}
The parallel, perpendicular, and Hall conductivity components ($\sigma_{\parallel}$, $\sigma_{\bot}$, and $\sigma_{\mathrm{H}}$, respectively) compose the conductivity tensor 
\begin{align}
    \vect{\sigma} = \left(
    \begin{array}{ccc}
        \sigma_{\bot} & -\sigma_{\trm{H}}  & 0 \\
        \sigma_{\trm{H}} & \sigma_{\bot}  & 0 \\
        0 & 0 & \sigma_{\parallel}
    \end{array} \right) 
\end{align}
from Faraday's law $\vect{j}~=~\vect{\sigma} \left( \vect{E}+ \vect{v}\times \vect{B} \right)$. These resistivities reduce to the values displayed in 
\refeq{cavient} in a three-fluid description. In this case,
\begin{align}
\eta_{\Omega}   &=  \frac{1}{\sigma_{\parallel}}        ~,\\
\eta_{\trm{H}}  &=  \frac{||\vect{B}||}{e n_{\trm{e}}}        ~,\\
\eta_{\trm{AD}} &=   \frac{||\vect{B}||^2}{\gamma_{\trm{AD}} \rho \rho_{\trm{i}}}       ~.
\end{align}

The present work is essentially devoted to the first collapse and the formation of the first core. At this stage and on this
scale, only ambipolar diffusion plays a role. In the following, we therefore only focus on this term. Rewriting the above equations with $\eta_{\Omega}=\eta_{\trm{H}}=0$ and
\begin{equation}
    \mathbf{\bar{v}} = \mathbf{v_{\trm{n}}}+\frac{\eta_{\trm{AD}}}{||\vect{B}||^2}\big[(\nable \times \mathbf{B}) \times \mathbf{B} \big] \label{eq_ref3}
,\end{equation}
\refeq{induc_1} reduces to
\begin{align}
    \partial_t{\mathbf{B}} = \nable \times \Big[  \mathbf{\bar{v}} \times \mathbf{B} \Big] \label{rewritten_induction}.
\end{align}
In contrast to a Laplace operator, as in Ohmic dissipation, \refeq{rewritten_induction} does not allow magnetic reconnection, since it corresponds to another {\it \textup{flux-freezing}} condition at a different speed $\vect{\bar{v}}$. We stress that this is true only under the
one-fluid approximation, as pointed out by \citet{2012ARep...56..138T}.

\subsection{Magnetic braking \label{magbrak}}
The main effect of magnetic braking in the context of prestellar core collapse is to slow down
the rotationally supported structures that develop
during the gravitational collapse of matter with
non-zero net angular momentum; this occurs because of the conservation of angular momentum. One of the consequences is that the formation of large and rapidly rotating disks is hampered, as found in purely hydrodynamical simulations. The braking arises essentially from the magnetic tension that
prevents the distortion of the field lines,
which are coupled to the flow through the ionised species.
The angular momentum is transported along the field lines at a speed
close to
the Alfv\'en speed, depending on the topology of the magnetic field. The braking efficiency can be characterised by deriving an Alembert equation of propagation for the angular momentum \citep[e.g.][]{magneticbreakingZ,gillis2,mousse78,mousse792,mousse80}, but can be estimated by order-of-magnitude calculations \citep{joos}. In cylindrical coordinates, with the $z$-axis aligned with the mean angular momentum direction vector, the angular momentum flux due to the magnetic field reads \citep[]{joos}
\begin{align}
   \vect{F_l} = - \frac{B_{\theta} r}{4 \pi} \vect{B} ~.
   \label{eqfluxmb}
\end{align}
It is proportional to the radial and poloidal
components of the field $B_{r}$ and $B_{z}$, while scaling with the square
of the toroidal component $B_{\theta}$, making this latter
the main focus of magnetic braking studies.

\subsection{Numerical setup}

\subsubsection{Methods}

To carry out our study of collapsing magnetised molecular cloud cores,
we used the adaptive mesh refinement (AMR) code {\ttfamily RAMSES} \citep{teyssier,fromang} in its non-ideal MHD extension \citep{masson_nimhd}. {\ttfamily RAMSES}
solves the complete set of MHD equations (self-gravity, Euler's fluid flow equations, and the induction equation with non-ideal terms) using the constrained transport method, which preserves the divergence-free condition for the magnetic field to machine precision. The adaptive mesh is extremely well suited to protostellar collapse calculations, where many levels of refinement are needed to efficiently describe spatial scales spanning $10^{4}-10^{5}$ orders of magnitude in a single simulation. AMR is also a powerful tool for fragmentation studies and disk formation in a turbulent medium where nested grids are difficult to use.

\noindent The grid refinement criterion is based on the Jeans mass, ensuring the Jeans length is always sampled by at least eight cells. The coarse grid has a resolution of $32^{3}$, and 11 levels of AMR were used, resulting in a maximum resolution of 0.15 au at the finest level.
Our general set of equations includes the conservation of mass, the mean neutral gas dynamics (Euler equation), the induction equation with the ambipolar diffusion electromotive force, self-gravity through the Poisson equation, and the divergence-free constraint:
\begin{align}
\frac{\partial \rho}{\partial t} + {\nable} . \left[ \rho \mathbf{v} \right] &= 0 \label{eqcont}~, \\
\frac{\partial \rho \mathbf{v}}{\partial t} + \div{\left[ \rho \vect{v} \otimes  \vect{v} + P \mathbb{I} - \vect{B} \otimes \vect{B}    \right]} &= 0 \label{eqqtmvt}~, \\
\frac{\partial \mathbf{B}}{\partial t} - {\nable} \times ( \mathbf{v} \times \mathbf{B}) - {\nable} \times \mathbf{E}_{\trm{AD}} &= 0 \label{eqdbdt}~, \\
{\nable}\cdot {\mathbf{B}} &= 0 \label{eqdivb}.
\end{align}Here $\rho$ is the mean fluid density, $\vect{v}$ the mean fluid velocity, $P$ the thermal pressure of the gas, and $\mathbf{E}_{\trm{AD}}=-\eta_{\trm{AD}} \frac{\mathbf{B}}{||\vect{B}||}  \times \left\{ (\nable \times \mathbf{B})\times \frac{\mathbf{B}}{||\vect{B}||} \right\}$ is the electromagnetic force due to the ambipolar diffusion, as derived in \refeq{induc_1}. The energy equation is approximated by a barotropic equation of state (see below).

Magnetic resistivities are calculated using a reduced chemical network including neutral and charged species, as well as dust grains. We followed \citet{KunzMouschovias2009} to compute the relevant charged species abundances including grains sizes in a classical MRN distribution \citep{mathis}, which were sampled using 50 bins. For an exhaustive description of the chemical model used and its application in the context of star formation, we refer to \citet{chimie}. We computed a three-dimensional table of density, temperature, and magnetic field dependent resistivities covering the ranges $10^{-24} < \rho < 10^{-10}~\text{g~cm}^{-3}$, $5 < T < 2000$ K, and $10^{-6} < B < 10^{2}$ G, respectively. During the simulations, the resistivities in each grid cell are interpolated on-the-fly according to the local state variables, which greatly reduces computational cost but implies thermodynamical equilibrium.

\subsubsection{Initial conditions}

We adopted initial conditions similar to those in \citet{commercon10}, who followed \citet{1979ApJ...234..289B}. A magnetised uniform-density sphere of molecular gas, rotating about the $z$-axis with solid 
body rotation, is placed in a surrounding medium a hundred times less dense with equal pressure. The prestellar core mass has a mass of 1 $\text{M}_{\odot}$, a radius $R_{0} = 2500$ au\footnote{The initial condition corresponds to a ratio of the thermal over gravitational energies
    $\alpha=0.25$ and a density of $9.4 \times 10^{-18}$~g~cm$^{-3}$.} and a ratio of rotational over gravitational energy of $\beta_{\trm{rot}} = 0.02$. The magnetic field is initially parallel to, and invariant along, the rotation $z$-axis. The field strength is stronger in a cylinder of radius $R_{0}$ (with the dense core at its centre) than in the surrounding medium, with $B_z(r>R_0) = B_z(R_0)*100^{2/3}$, where the factor of 100 comes from the difference in density between the core and the surroundings\footnote{This is chosen to try to reproduce the dragging-in of field lines that would occur in the formation of the dense core \citep[see][for example]{magneticbreakingZ}, 
while also retaining in the simplest manner the divergence-free condition for the MHD.}.
We define a mass-to-flux ratio parameter similar to the one defined by \citet{MouschoviasSpitzer1976} to measure the importance of the magnetisation in the core:
\begin{align}
    \mu_{\textrm{}}(r) &= \frac{\frac{\int_0^r dM}{\int_0^r d\phi_B}}{\left(\frac{M}{\phi}\right)_{\trm{crit}}}\label{mumumu} ~,
\end{align}
with the critical value $\left(\frac{M}{\phi}\right)_{\trm{crit}}=\frac{0.53}{3 \pi}\left(   \frac{5}{G}   \right)^{0.5} $ \citep{MouschoviasSpitzer1976}. We note that $\mu(r=R_0)$ is strictly equal to the theoretical value for a homogeneous cloud permeated by vertical field lines.

Even though a Bonnor-Ebert (BE) density profile may better fit observations of dense cores \citep[see][]{2000prpl.conf...59A,2002A&A...393..927B} and has an analytical foundation \citep{ShuLi,1977ApJ...218..834H}, we assumed a magnetised medium permeated by straight parallel field lines, and it is hardly possible to end up with a BE density profile without bending the field lines
during
the formation of the density enhancement (the dense core).
Some authors \citep[][e.g.,]{vanloofalle} found non-linear density enhancements in simulated turbulent molecular sheets via slow-mode magnetic waves, which left the magnetic field unchanged in the cores, but it remains unclear whether this process does occur in molecular clouds. Our own tests tend to show that the initial density profile is not a critical matter because the infalling
material adopts a BE-like density profile very shortly after the beginning of the dense core collapse.

As the aim of the present paper is to focus on the effect of non-ideal MHD effects, notably ambipolar diffusion, on prestellar core collapse and disk formation,
we used  a barotropic equation of state (EOS) to mimic the effect of radiative transfer instead of solving the full set of radiation magnetohydrodynamics equations. This was done for the sake of simplicity and to save computational cost. The gas pressure is thus related to the density as
\begin{align}
    \frac{P}{\rho} = c_{\trm{s}}^2 \sqrt{ 1 + \left(\frac{n_{\trm{H}}}{10^{-13} \, \textrm{g~cm}^{-3}}\right)^{\frac{4}{3}}} \label{barotrop_eq} ~,
\end{align}
where $c_{\trm{s}}$ is the gas sound speed and $n_{\trm{H}}$ the number density of atoms. After an initial phase of isothermal collapse up to a density $n_{\trm{H}} \simeq 10^{-13} \,~\textrm{g~cm}^{-3}$, the first hydrostatic Larson core \citep{Larson1969} forms when the gas becomes optically thick enough 
 to stop radiative cooling, which was until then counter-balancing the compressional heating. This adiabatic phase is reproduced with the barotropic EOS by using a 
polytropic index $n$ for a gas (which is equal to the ratio of specific heats $\gamma$) $n=\frac{5}{3}$, higher than the critical value for stability against gravitational collapse $n_{\textrm{critic}}=\frac{4}{3}$.
The temperature and density of the gas begin to rise inside the core as it accretes material from the surrounding envelope. The energy sink provided by the dissociation of $\text{H}_{2}$ molecules, which occurs around $T\sim$2000 K, triggers a second phase of collapse \citep[]{Larson1969}. We
here focus on the properties of the first Larson core. 
Our general approach allows us to accurately describe the first Larson core and its surroundings and the magnetic flux transport without needing to introduce a sink particle that would limit the resolution.  It is beyond the scope of the
present paper to include long-term evolution, which leads to the formation of the second core. This would imply the accurate treatment of jets, high-energy radiation, 
 and remaining non-ideal MHD terms  and will be addressed in a forthcoming study.{}

We have performed eight simulations for which we varied the mass-to-flux ratio $\mu=2$ and 5, the angle between the rotation axis and the magnetic field initial direction ($0$ and $40$ degree) and used ideal (iMHD) or non-ideal (niMHD) magnetohydrodynamics (accounting only for ambipolar diffusion). We also performed additional simulations with increased and decreased resolutions to study the convergence of our results for our fiducial case ($\mu_{\trm{}}=5$, aligned case). The various run parameters are given in Table~\ref{table_runs}.

\begin{table*}
\centering
\footnotesize
\caption{Summary of the initial conditions for the different simulations}
\begin{tabular}{r  cc cc  cc cc }
\hline
Alignment and MHD type  & \multicolumn{2}{c}{niMHD Aligned}  & \multicolumn{2}{c}{iMHD Aligned}  & \multicolumn{2}{c}{niMHD Misaligned}  & \multicolumn{2}{c}{iMHD Misaligned}  \\
\hline\hline
magnetisation ($\mu$) & $2$ & $5$ & $2$ & $5$ & $2$ & $5$ & $2$ & $5$ \\
Ambipolar diffusion                        & yes & yes & no & no & yes & yes& no & no \\
Angle                                      & 0   & 0   & 0  & 0  & 40  & 40 & 40 & 40 \\
Rotational support ($\beta_{\trm{rot}}$)   & 0.02& 0.02& 0.02& 0.02& 0.02& 0.02& 0.02& 0.02 \\
Mass of the core (in $\msol$) & 1 & 1 &1 & 1& 1 & 1 &1 & 1 \\
\hline
\end{tabular}
\label{table_runs}
\end{table*}

\section{Early evolution\label{sec:Results}}

\subsection{Aligned case}\label{sec:AlignedCase}

In this section, we
study the general properties of the
collapsing system by comparing the iMHD case and a case with ambipolar diffusion in the fiducial aligned case. 
We first focus on the $\mu_{\trm{}}=5$ case, which we describe in detail. We then
report the differences or similarities with the $\mu_{\trm{}}=2$ case.

\subsubsection{First Larson core}\label{sec:AlignedFirstCore}

We determined the onset of the first core formation as the point
when the density in the computational domain reaches $10^{-12}$~g~cm$^{-3}$. We subsequently defined the first core itself by the cells fulfilling the criterion $\rho > 10^{-12}$~g~cm$^{-3}$. 
The core radius, mass, mass-to-flux ratio, and peak magnetic field value were
measured 200 years after its formation. The properties of the first core are summarised in Table~\ref{table_fc}.

The first core is oblate with a radius ranging from 8 to 9~au, independently of the initial magnetisation ($\mu_{\trm{}}=2$ or $\mu_{\trm{}}=5$). 
While pure hydrodynamical simulations \citep[e.g.][]{2011MNRAS.417.2036B,2010ApJ...725L.239T} and some radiation iMHD (RMHD) calculations \citep[][]{2010ApJ...714L..58T} found a similar result with a flattened first core, other RMHD studies \citep{commercon10} did not find flattened cores in the case $\mu=5$. 
As a result of the additional magnetic support, the free-fall (thus core formation) timescale is twice longer  for $\mu_{\trm{}}=2$  than for $\mu_{\trm{}}=5$ . The additional support also hinders accretion, and the first core is significantly less massive for $\mu_{\trm{}}=2$  than for $\mu_{\trm{}}=5$ . 
Similarly, for the stronger magnetisation ($\mu_{\trm{}}=2$), the core mass is lower for niMHD than for iMHD.
Indeed, the strongest magnetic field inside the core
remains of the order of $10^{-1}$~G in niMHD, while it is ten times stronger in iMHD.
This effect of ambipolar diffusion on the magnetisation during the core formation is illustrated by the differences in the mass-to-flux ratio $\mu$ at $10$ and $100$~au
between the ideal and non-ideal MHD cases. While $\mu$ is higher than 10 at $10$~au and at $\lesssim 4$ at $100$~au in niMHD, it is at most $\lesssim 5$ in iMHD. 

\begin{table*}
\centering
\footnotesize
\caption{Summary of the main properties of the first core for fiducial cases with solid-body rotation}
\begin{tabular}{r  cc cc  cc cc }
\hline
Alignment and MHD type  & \multicolumn{2}{c}{niMHD Aligned}  & \multicolumn{2}{c}{iMHD Aligned}  & \multicolumn{2}{c}{niMHD Misaligned}  & \multicolumn{2}{c}{iMHD Misaligned}  \\
magnetisation  & $\mu_{\trm{}}=2$ & $\mu_{\trm{}}=5$ & $\mu_{\trm{}}=2$ & $\mu_{\trm{}}=5$  & $\mu_{\trm{}}=2$ & $\mu_{\trm{}}=5$ & $\mu_{\trm{}}=2$ & $\mu_{\trm{}}=5$ \\
\hline\hline
Formation time ($\times 10^{3}$ years)            & 49.2  & 24.3 & 50.5 & 24.3 & 52.2  & 24.6 & 54.6  &  24.6     \\
Radius (au)                                     & 8-9   & 8-9  & 9-10 & 7-9  &  8-9  & 8-9  & 9-10  &  8-9      \\
Mass ($\text{M}_{\odot}$)          & 0.016   & 0.036  & 0.025  & 0.03  & 0.016   & 0.058  & 0.024   &  0.058      \\
$\mu(r=10 \ \trm{au})$                          & 14.1  & 21.2 &  3.5   & 3.9  & 14.1  & 24.8   & 3.4   &  3.9      \\
$\mu(r=100 \ \trm{au})$                         & 2.8     & 4.2    & 1.8  & 4.6  & 2.8     & 3.5    & 1.6   &  2.8        \\
Magnetic field strength B (G)                     & 0.09  & 0.1  & 0.8  & 0.98 & 0.09  & 0.21 & 0.77  &  4.4     \\
\hline
\end{tabular}
\label{table_fc}
\end{table*}

\subsubsection{Magnetic field repartition}

\begin{figure}
    \begin{center}
            \label{fig_1_1_b}
        \includegraphics[width=0.33\textwidth, angle=270]{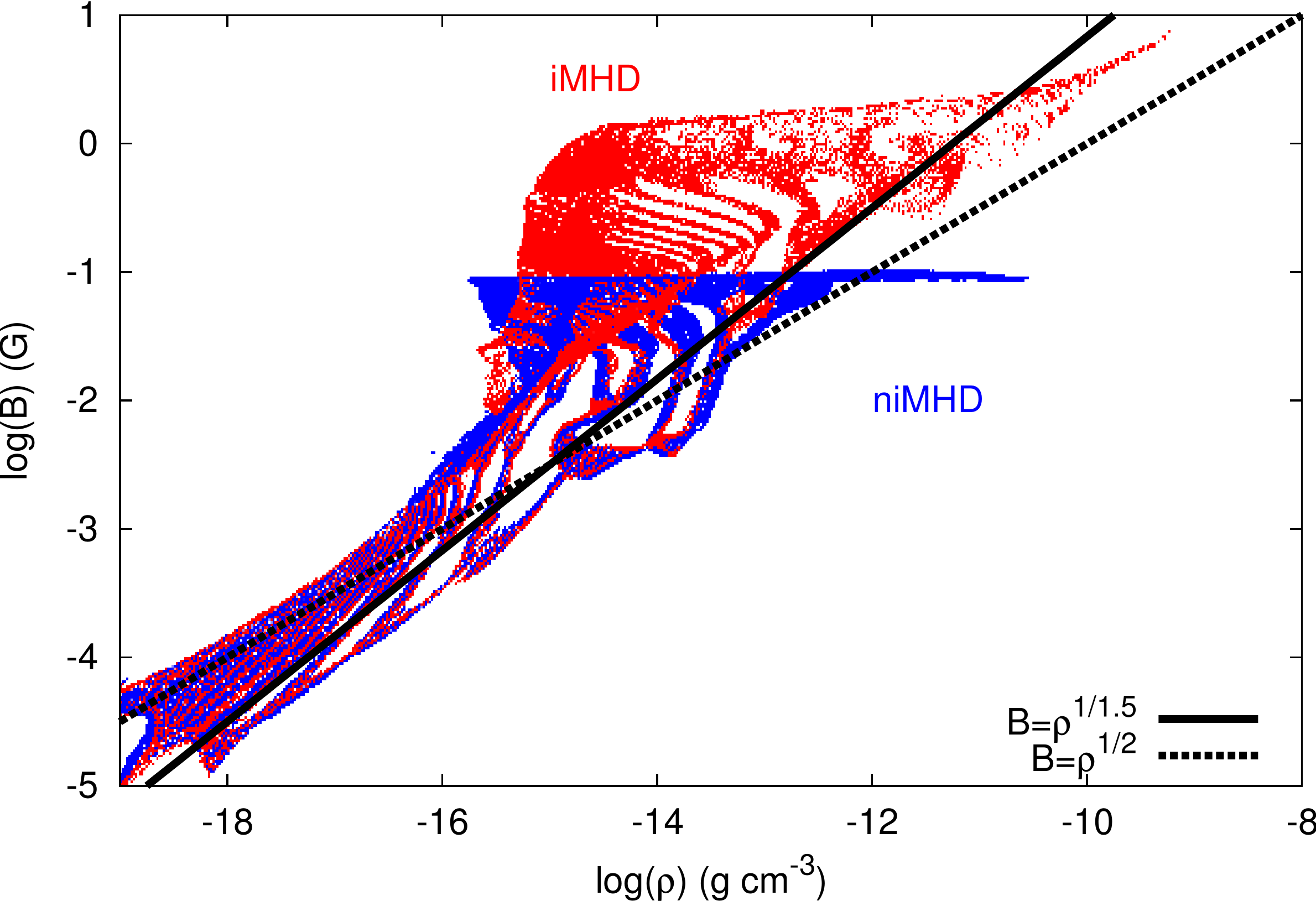}
        \caption{Magnetic field distribution as a function of gas density for $\mu_{\trm{}}=5$ aligned, using the iMHD (red) and the niMHD (blue) formalisms, $200$~years after first core formation. The solid black line shows the scaling of the magnetic field $B\propto \rho^{2/3}$ and the dashed black line the scaling $B\propto \rho^{1/2}$.}
        \label{fig_1_1}
    \end{center}
\end{figure}

Ambipolar diffusion enables neutral particles to overcome magnetic field lines and redistribute the magnetic field. We first focus on the magnetic field intensity and topology
to understand its consequences for the dynamics of the collapsing core.
\refig{fig_1_1} shows the magnetic field repartition as a function of gas density for the iMHD and niMHD runs. The flux-freezing behaviour, $B \propto \rho^{1/\xi}$ with $\xi \sim 3/2$, is obvious for iMHD (red) with a highest magnetic field value
of $\sim$1~G. The niMHD simulation (blue)
shows an identical initial evolution for densities below
$\sim10^{-14}$~g~cm$^{-3}$. After this point, ambipolar diffusion starts to dominate the later evolution of the core, and a very well defined diffusion plateau forms that is usually referred to as the decoupling stage in star formation \citep[]{DeschMouschovias}. In both the iMHD and niMHD calculations, the generation of outflows
can be identified by the
high values of the magnetic field at high density, compared with the expected perfect flux-freezing result. 


As seen in \refig{fig_1_1}, a scaling relation $B \propto {\rho}^{2/3}$ seems to represent the evolution of the magnetic field during the collapse better than the
 traditional $B \propto \sqrt{\rho}$ relation \citep[see the differences in the scaling for the different components of the field in][]{HennebelleFromang2008}. Further details
 about this scaling relation and the analytical derivation of an estimate of the field saturation value are given in Appendix A.

\refig{fig_1_2} illustrates the comparison between  $\mu = 2$ and $\mu = 5$ niMHD. Both simulations show similar initial evolutions (\refig{fig_1_2_a}) when flux-freezing is still dominant ($\mu =2$  naturally shows a higher magnetic field intensity at a given density), but have slightly different saturation values.
At later stages (\refig{fig_1_2_b}) a plateau forms in both low magnetisation and high magnetisation runs with $B_{\trm{plateau}} \simeq 0.1$~G.

This plateau arises when ambipolar diffusion becomes efficient enough to overcome the dynamical effects. The non-linearity of the effective diffusion coefficient, $\eta_{\trm{AD}} \propto \frac{B^2}{\gamma \rho_{\trm{i}} \rho}$, explains the self-regulation and the formation of the diffusion plateau. Indeed,
for any non-potential ($\rot{B}\ne 0$) field configuration, an increase in the magnetic field leads to
a rise
of the local diffusion coefficient and thus in turn to a decrease in field intensity. The plateau first forms when the dynamical evolution of the magnetic field, $\vect{v}\times \vect{B}$, no longer dominates the diffusion term, $\eta_{\trm{AD}} \rot{B}\times \frac{\vect{B}}{||B||} \times \frac{\vect{B}}{||B||}$ (with $||~||$ standing for the the L2 norm in this paper). Assuming flux-freezing holds during the first isothermal phase of the collapse (i.e. $B \propto \rho^{1/\xi}$), we can estimate the saturation value for the magnetic field
\begin{equation}
B_{\trm{sat}}=\left(c_{\trm{s}}^2 C \gamma_{\trm{AD}} \sqrt{\frac{3}{2 \pi G}} \right)^{\frac{1}{2-\xi}}\left( \frac{B_0}{\rho_0^{1/\xi}} \right)^{\frac{-\xi}{2-\xi}},
\end{equation}
where $\gamma_{\trm{AD}}$ is the drag coefficient (\refeqp{dragcoef}), $C$ the ionisation fraction such that $\rho_{\trm{i}} = C \sqrt{\rho_{\trm{n}}}$, $G$ the gravitational constant, $B_0$ and $\rho_0$ the typical initial conditions for the magnetic field and gas density, and $\xi$ the power index for the proportionality law
$B(\rho) \propto \rho^{1/\xi}$.
The initial thermal support and mass-to-flux ratio are linked to the values of $B_0$ and $\rho_0$. Further details on the derivation of $B_{\trm{sat}}$ can be found in Appendix~\ref{A1}.
The values we obtain for $B_{\trm{sat}}$ with $\xi = \frac{3}{2}$ for various values of thermal support $\alpha$, that is, the ratio of thermal over gravitational energy, 
are shown \refig{bsat} (thick black solid line). For $\xi = \frac{3}{2}$, which corresponds to a homogeneous sphere permeated by a uniform magnetic field, we find that the saturation value does not depend on $\alpha$. In this case, the saturation value estimate for $\mu_{\trm{}}=5$ is $B_{\trm{sat}}= 0.13$~G, very close to the numerical value $B \lesssim 0.1$~G (see \refig{fig_1_1}). According to this simple order-of-magnitude calculation, the saturation value should even be lower in more magnetised cores, with $\mu < 5$. For example, the $\mu_{\trm{}}=2$ case yields $B_{\trm{sat}}= 8 \times 10^{-3}$~G. In our numerical simulation, we do not observe such a strong diminution, but the saturation value {\it \textup{is}} lower for $\mu_{\trm{}}=2$ than for $\mu_{\trm{}}=5$ (see \refig{fig_1_2}).
The discrepancy between the analytical prediction and the numerical value for $B_{\trm{sat}}$ in the strongly magnetised case originates from the fact that in the latter case
the magnetic field distribution departs appreciably from a simple power-law parametrisation $B\propto \rho^{2/3}$ (see \refig{fig_1_2}). Using a slightly different power index that better fits the effective $B(\rho)$ repartition in the $\mu_{\trm{}}=2$ case, for instance $\xi=1.73$, as shown on the inset in \refig{bsat}, we obtain $B_{\mathrm{sat}} = 0.06$ (coloured lines in \refig{bsat}), which
is closer to the value obtained in the simulation.

\begin{figure}
     \begin{center}
        \subfigure[Early time, when flux freezing is still relevant.]{
            \label{fig_1_2_a}
            \includegraphics[width=0.325\textwidth, angle=270]{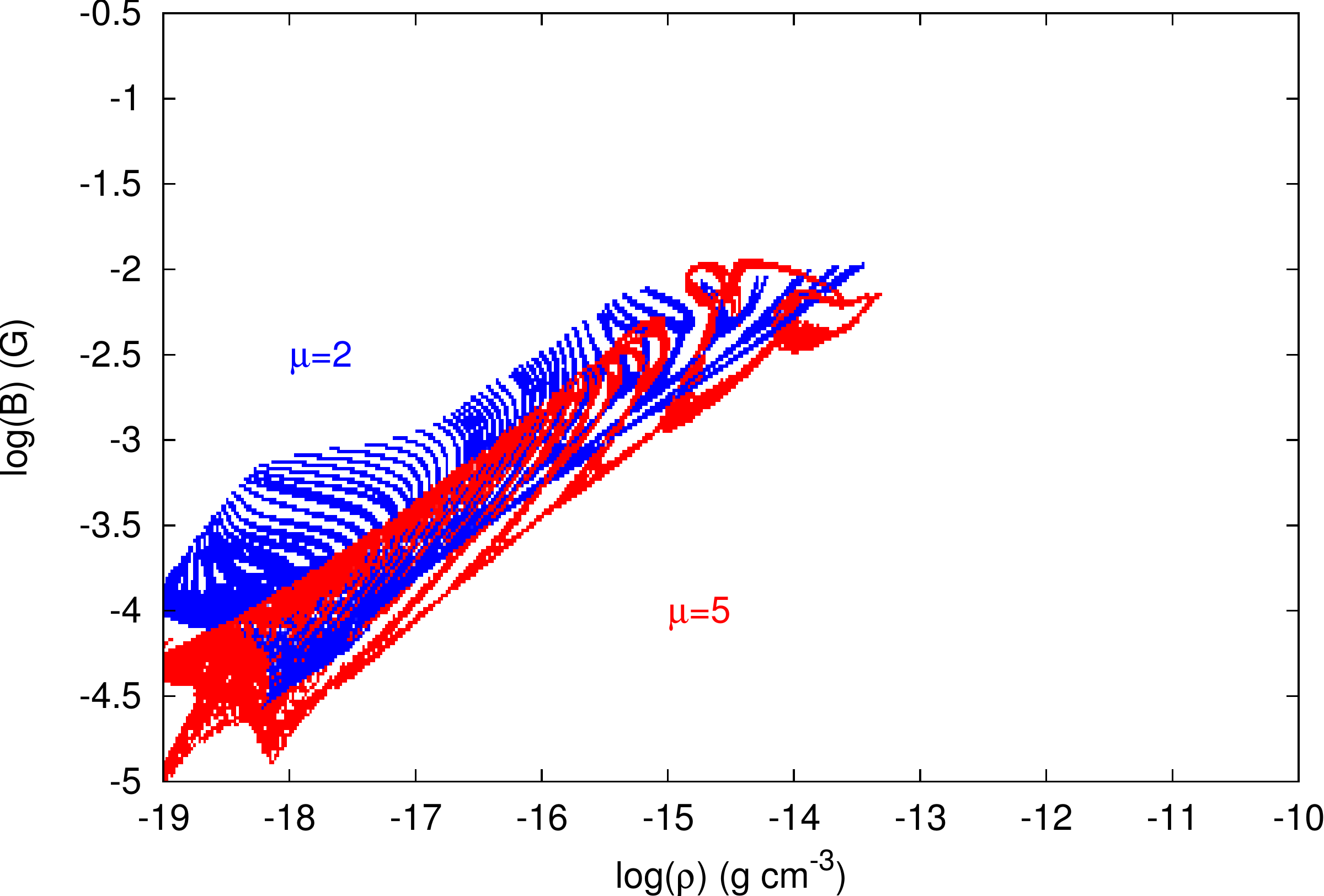}
        }\\ 
        \subfigure[After the decoupling stage.]{
            \label{fig_1_2_b}
           \includegraphics[width=0.325\textwidth, angle=270]{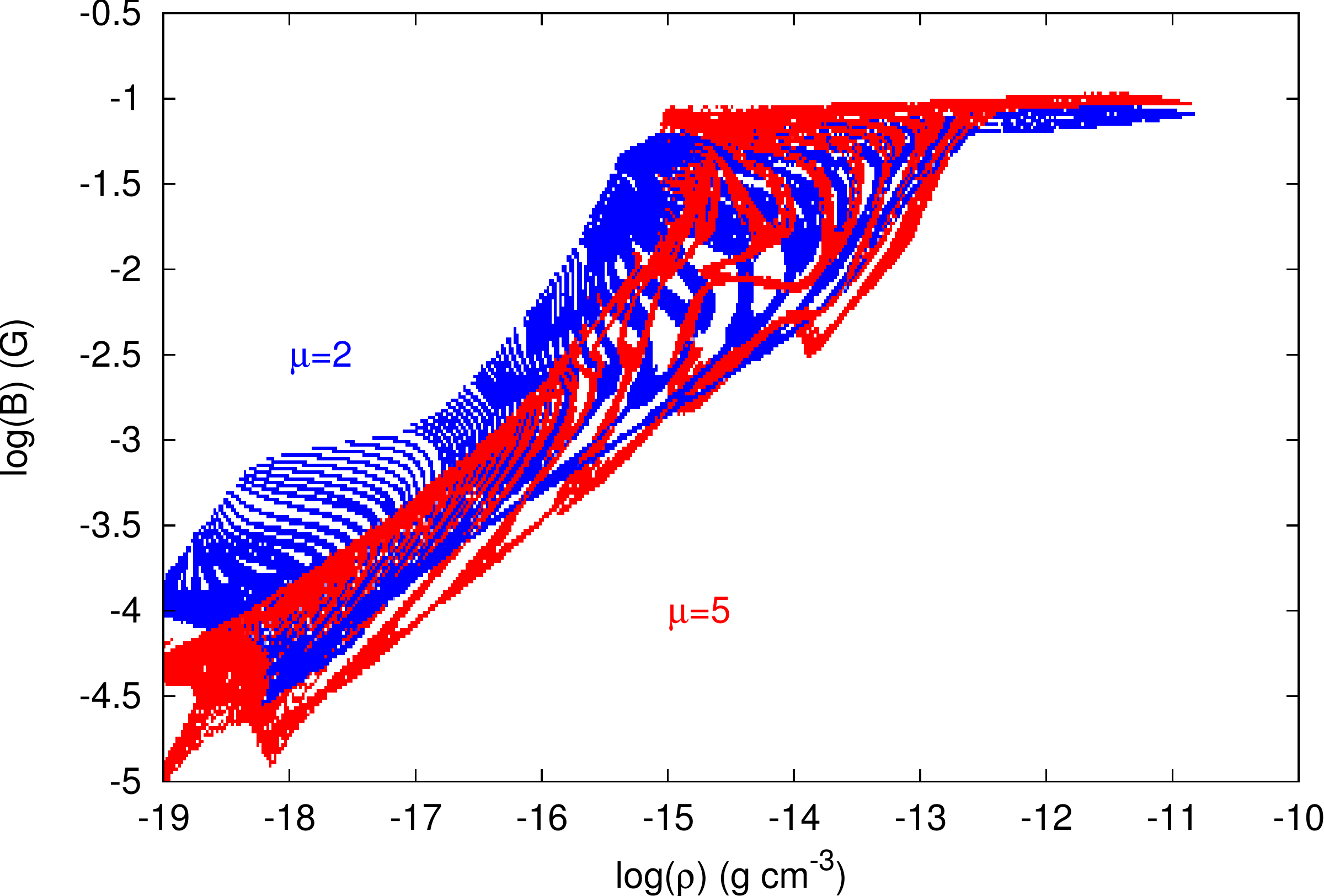}
       }
    \end{center}
    \caption{$B(\rho)$ for $\mu_{\trm{}}=5$ (red) and $\mu_{\trm{}}=2$ (blue).}
   \label{fig_1_2}
\end{figure}

\begin{figure}
     \centering
     \includegraphics[width=0.49\textwidth]{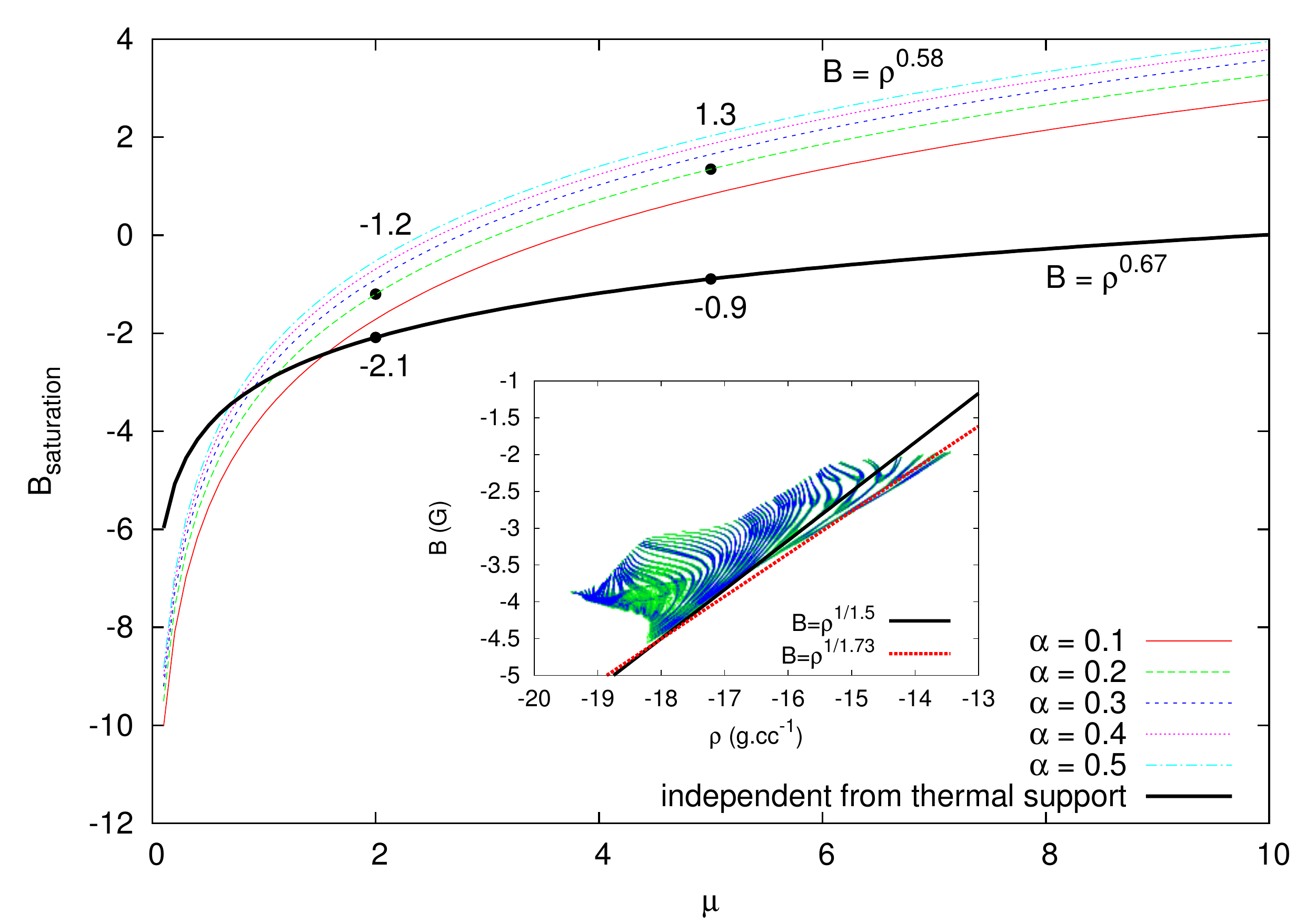}

        \caption{Saturation value as a function of $\mu$ for various values of thermal support $\alpha$. The points corresponding to $\mu=2$ (lower values) and $\mu=5$ (upper values) are marked in the figures.}
        \label{bsat}
\end{figure}

\begin{figure}
    \begin{center}
        \includegraphics[width=0.33\textwidth, angle=270]{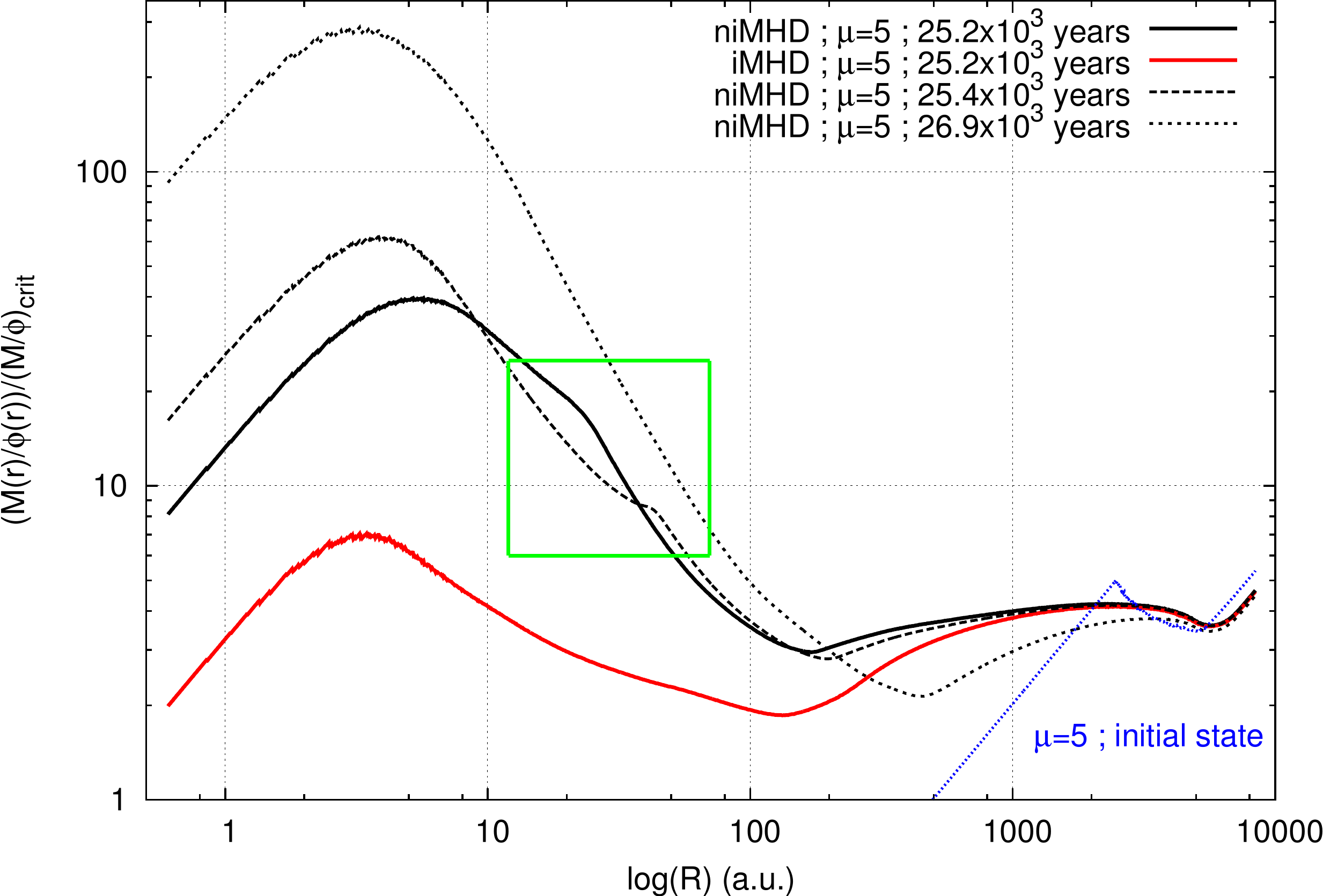}
        \caption{$\mu_{\trm{}}=5$, aligned case. Mass-to-flux ratio ($\mu_{\trm{}}$) after first core formation (at $\sim 24$~kyears) for the niMHD (black) and IMHD (red) cases (solid lines). We also display the initial condition (dotted blue line) and two later outputs for the niMHD case (dashed and dotted solid black lines). The decoupling stage region due to magnetic field pile-up is emphasised with a green square.}
        \label{mur_admhd}
    \end{center}
\end{figure}

\refig{mur_admhd} shows snapshots of the mass-to-flux ratio $\mu_{\trm{}}$, as defined~\refeq{mumumu}, as a function of radius at different times in the simulations.
At large radii ($r \gtrsim 1000$~au), the magnetisation is similar for iMHD (red) and niMHD (black), confirming the fact that
at these scales the AD timescale is orders of magnitude longer than the dynamical timescale. There is a kink in the AD case at a few tens of au (highlighted by the green square in the figure) that slowly propagates outwards (compare the blue dashed and solid curves); this corresponds to the efficient diffusion of the field at these densities.

\subsubsection{Outflows}

\begin{figure*}
    \centering
        \includegraphics[width=\textwidth]{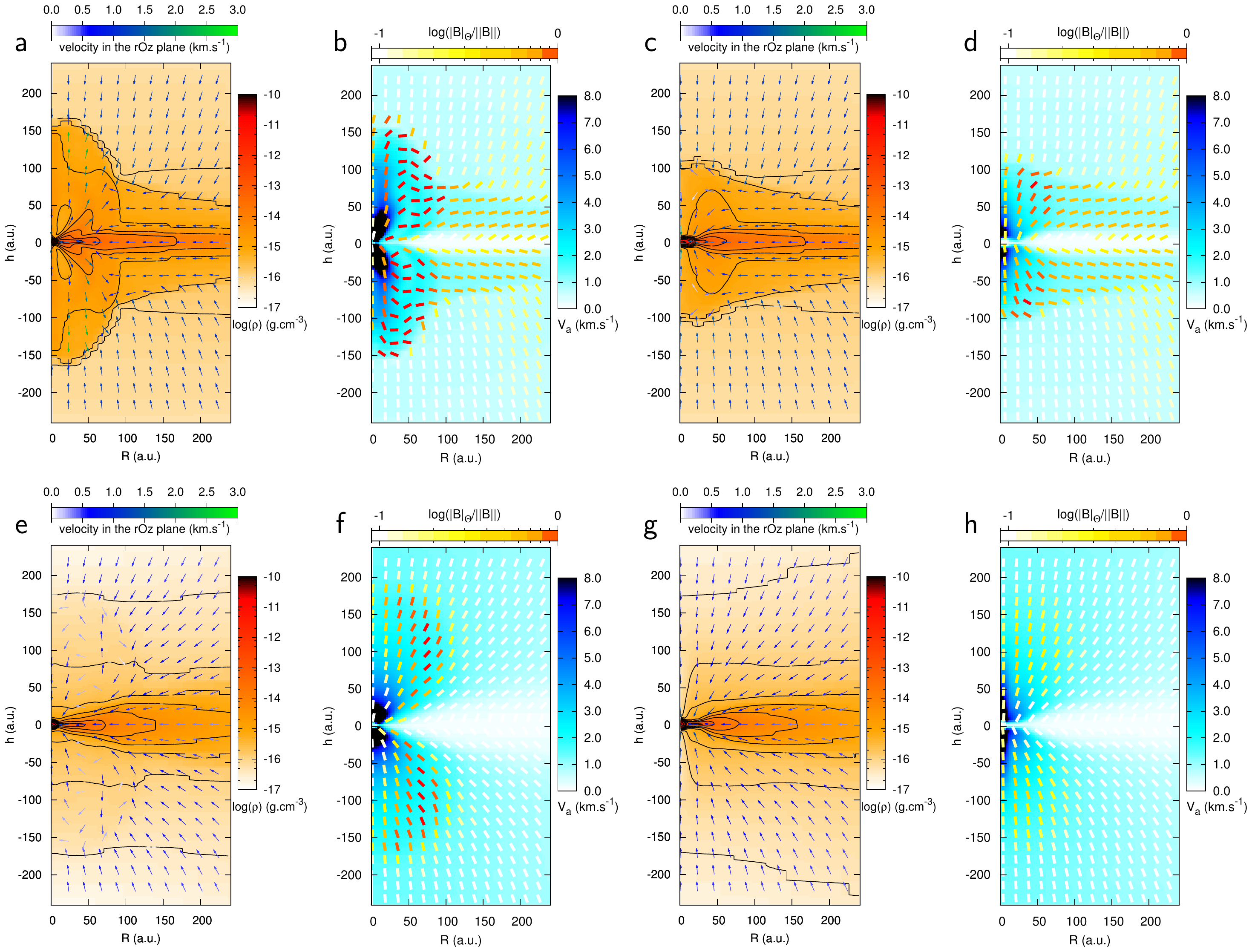}
        \caption{Top row: $\mu_{\trm{}}=5$, aligned case, snapshots at a time $t=1.02~\tff$ (500 years after first core formation). Panel (a) shows a map of the density in the ideal MHD simulation. The black solid lines are logarithmically spaced density contours. The arrows represent the velocity field where blue to green corresponds to the increasing magnitude of the velocity in the ($r,z$) plane. Panel (b) is a map of the Alfv\'en speed (blue), with the magnetic field direction in the ($r,z$) plane indicated by the segments. Yellow to red in these segments corresponds to increasing relative magnitude of the toroidal component of the field compared to the total field. Panels (c) and (d) are the same as (a) and (b), respectively, but for the niMHD case. Bottom
row: same as the top row for $\mu = 2$, and at the same time $t=1.01~\tff$ (500 years after first core formation). Every quantity is azimuthally averaged.}
        \label{outflow1}
\end{figure*}

The piling-up of the toroidal component of the magnetic field during the collapse of the dense core ultimately leads to the formation of a growing vertical structure, called magnetic tower. We note that we use the generic term magnetic tower to describe any outflowing structure. In reality, there are several ways to launch outflows in a magnetised environment, as studied initially by \citet{1994MNRAS.267..146L} and in contemporary studies by \cite{HennebelleFromang2008,2010MNRAS.409L..39C}.
This outflow is launched shortly ($\lesssim 1$~kyear) after the formation of the first core. \refig{outflow1} (top row) displays the azimuthally averaged density and velocity fields (panels a and c) and Alfv\'en speed with the magnetic field direction (panels b and d) for the iMHD and niMHD $\mu = 5$ runs. The simulations are compared $500$~years after the formation of the first Larson core, but the same conclusions hold during all the time evolution. In all panels, the $z$-axis is aligned with the direction of the average angular momentum vector computed in a sphere of radius 100 au, centred around the densest cell in the simulation. The velocity vectors are overlaid on the density maps, while the direction of the magnetic field in the $(r,z)$ plane is superimposed on the Alfv\'en speed colour maps. The colour coding for the magnetic field illustrates the intensity of the toroidal component $B_{\phi}$. 

The magnetically driven outflow is reinforced in the iMHD case by the
more important magnetic field pile-up.
This accumulation stems both from
the radial bending of the field lines, as clearly seen in \refig{outflow1}b,
and from the increasing toroidal field (of prime importance for magnetic braking, as shown by \refeqp{eqfluxmb})
that is generated by the rotation, as clearly seen in \refig{outflow1}b and d, especially close to the first Larson core where the toroidal support is significantly stronger in (b). While the gas inside the magnetic tower is slightly less dense in the AD case, the lower magnetic field intensity
(compared to the iMHD case)
yields an overall Alfv\'en velocity
that is lower by one order of magnitude. This in turn explains the weaker magnetic braking, since the angular momentum is carried away by slower Alfv\'en 
waves\footnote{In a simple representation of bent field lines, it is possible to derive the angular momentum transport equation.
Angular momentum
propagates along the field lines and follows a wave equation with a velocity defined by the local Alfv\'en speed and the topology of the field. For more details see \citet{kejathese} or the original article by \citet{gillis2}.}.

In the iMHD simulation, the velocity field in the tower (in the r-z plane) is vertical and the gas mainly flows out, the launching mechanism taking root in the strong toroidal field and differential rotation close to the Larson core. The field is dominantly vertical, and neutral matter follows the field lines. When AD is included, the growth of the magnetic tower is still magnetically regulated, but the growth is slower and the velocity field in the tower is almost null, or slightly directed towards the protostar. A detailed analysis shows that there is no real outflowing gas motion of gas per se, but that the magnetic tower structure or
growth is supported by magnetic pressure. We also note that close to the core ($r < 50$~au), while the field lines are pinched in the iMHD case (split monopole topology), field re-distribution has operated in the niMHD run because of the ambipolar diffusion (the magnetic field vectors are much more vertical).

\paragraph{\underline{\smash{$\mu =2$} }:}

In a more magnetised case, the
picture is very different. The bottom row of \refig{outflow1} shows the structure of the outflow for $\mu =2$ . Magnetic braking is much more effective (the magnetic field is overall stronger, yielding a stronger magnetic braking, as seen Sect.~\ref{magbrak}),
causing the field topology to come closer to
a split monopole configuration at large scale, with strong pinching of the lines in the equatorial plane.
The enhanced braking weakens the outflow-launching mechanism because the pile-up of the toroidal field is less effective.
For iMHD, there is still an outflowing feature, but its velocity field is either null in the r-z plane or even falls back onto the core. The boundary of the outflow region is less well defined,
but is characterised by a discontinuity in the velocity field.
In the AD case, the braking is stronger for $\mu =2$  than for $\mu =5$  at the early stages of the collapse (see Sect.~\ref{regionAD}),
which results in a weaker magnetic field (the diffusion plateau is lower) and
reduces the pile-up of toroidal field close to the core. This weakens the launching
process
to the point that no outflow
is produced
in this case. The Alfv\'en speed close to the core is very similar to the weak field $\mu =5$ run (\refig{outflow1}d),
as expected from the field magnitude distribution as a function of density (\refig{fig_1_2}),
where $B(\rho> 10^{-12}$~g~cm$^{-3})$ distributions are almost identical for the strongly and weakly magnetised cases.

\subsubsection{Regions of active ambipolar action \protect\label{regionAD}}

Figure~\ref{outflow1}
also shows that for $R>$100~au
the iMHD and AD runs look similar. This resemblance
arises from the strong dependence of ambipolar diffusion efficiency upon density and the sharp transition between the flux-freezing and AD dominated regimes. The initial isothermal phase of the collapse thus remains very similar in both cases. 
To examine the effect of flux dissipation in detail, we now focus on the regions where ambipolar diffusion dominates.

\begin{figure*}
    \centering
        \subfigure[$\mu = 5$, aligned]{
            \label{fig_act1_a}
            \includegraphics[width=0.44\textwidth]{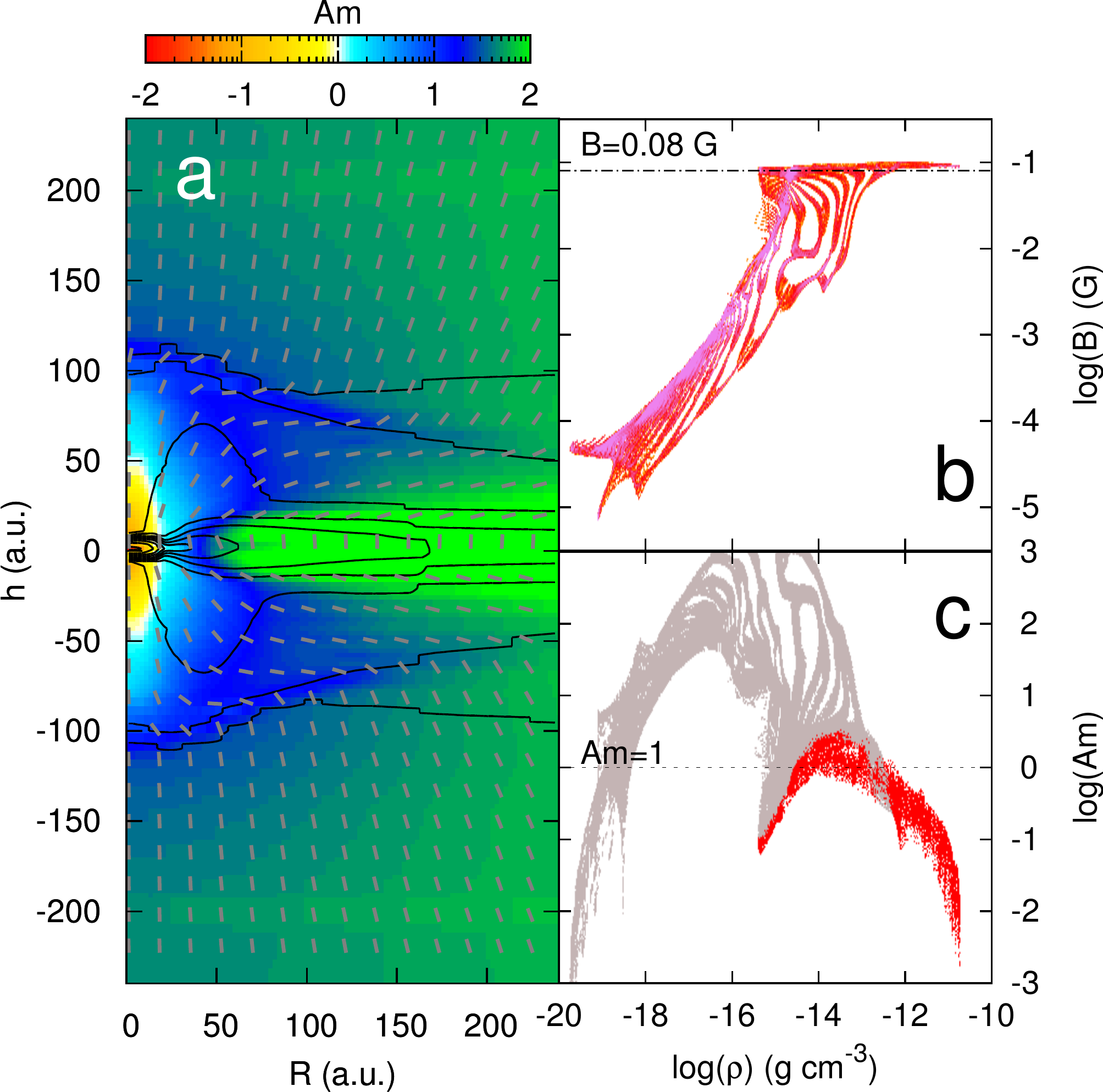}
        }
        \subfigure[$\mu = 5$, misaligned]{
            \label{fig_act1_b}
            \includegraphics[width=0.44\textwidth]{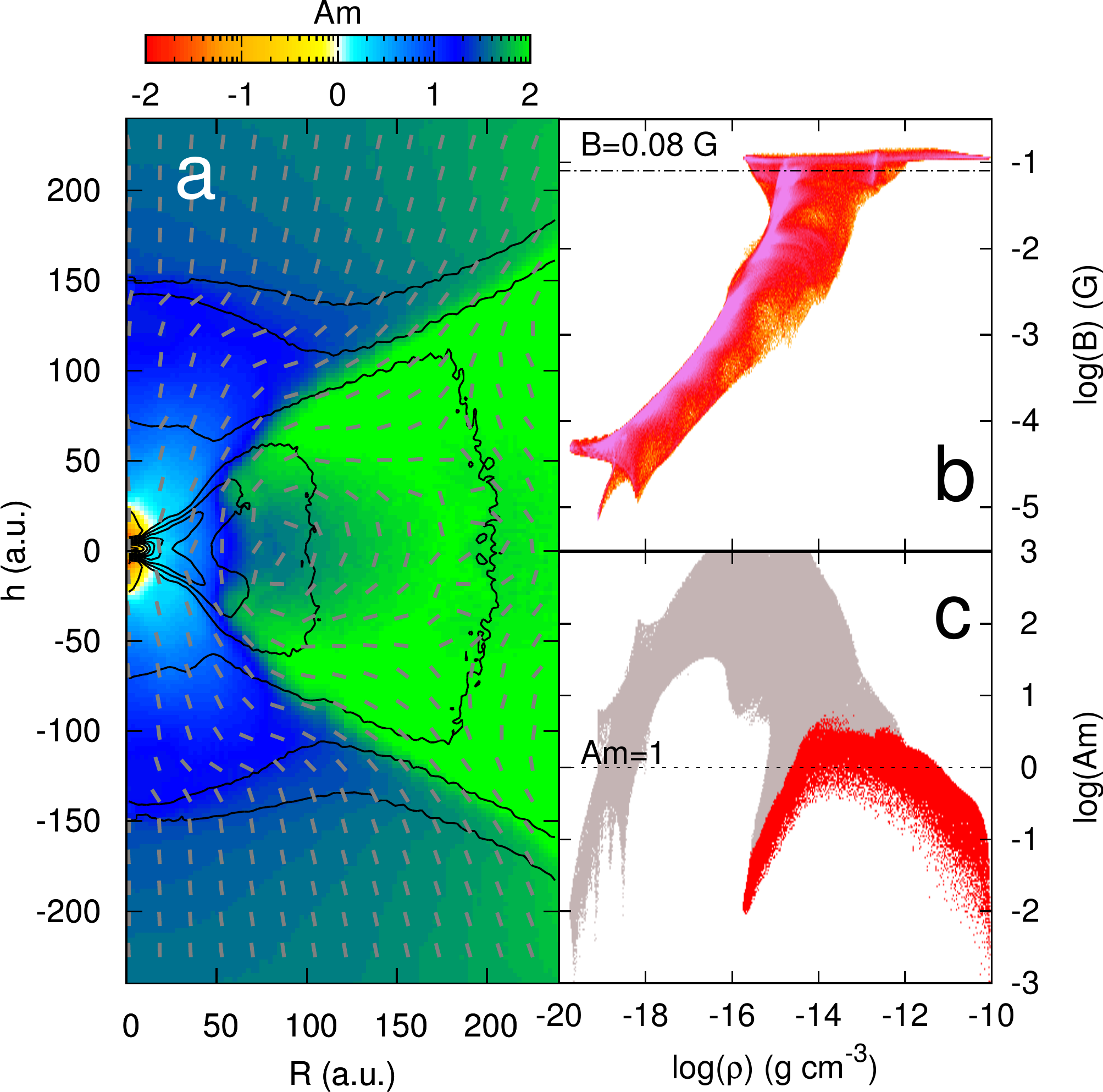}
        }\\ 
        \subfigure[$\mu = 2$, aligned]{
            \label{fig_act1_c}
            \includegraphics[width=0.44\textwidth]{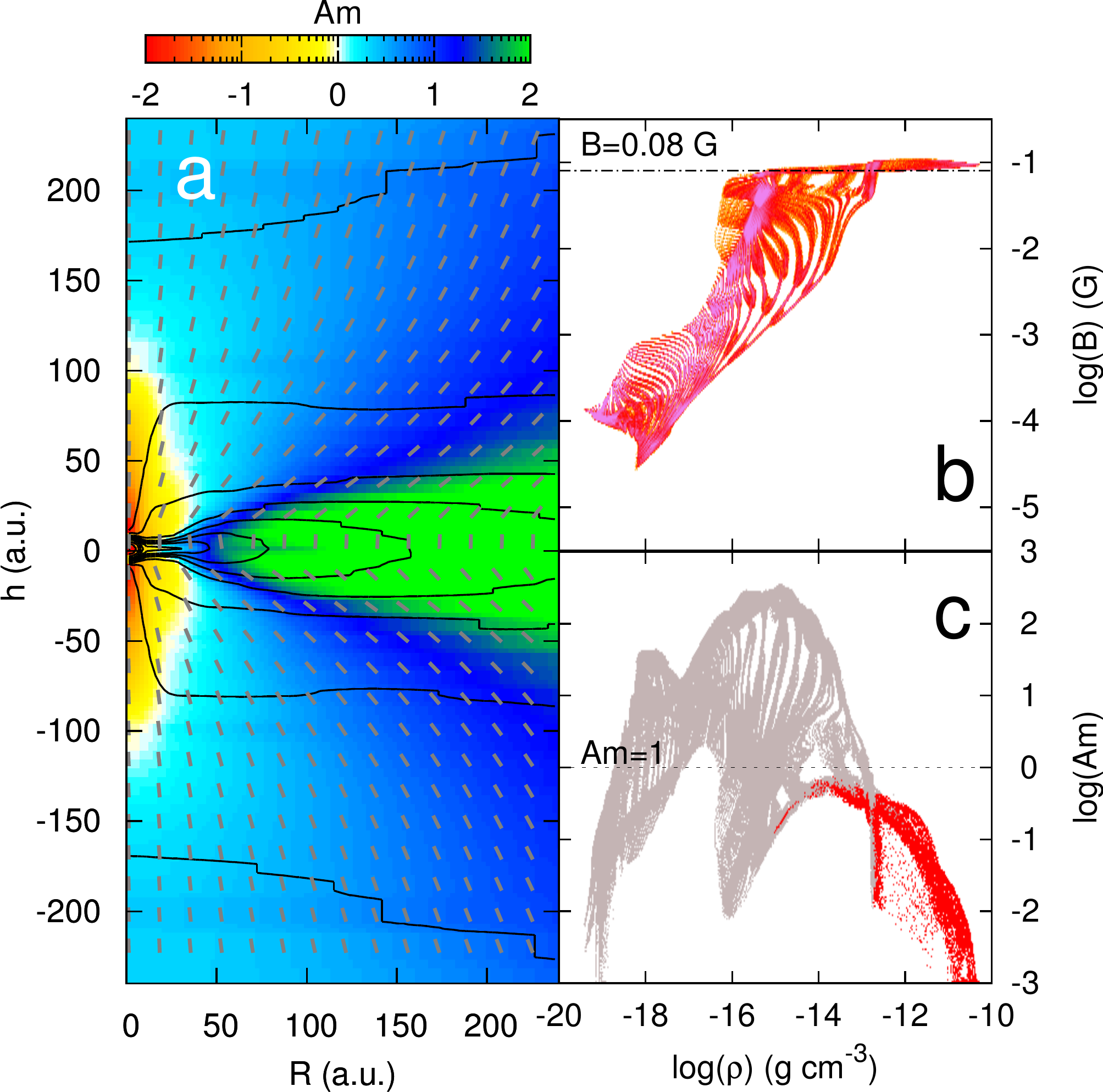}
        }
        \subfigure[$\mu = 2$, misaligned]{
            \label{fig_act1_d}
            \includegraphics[width=0.44\textwidth]{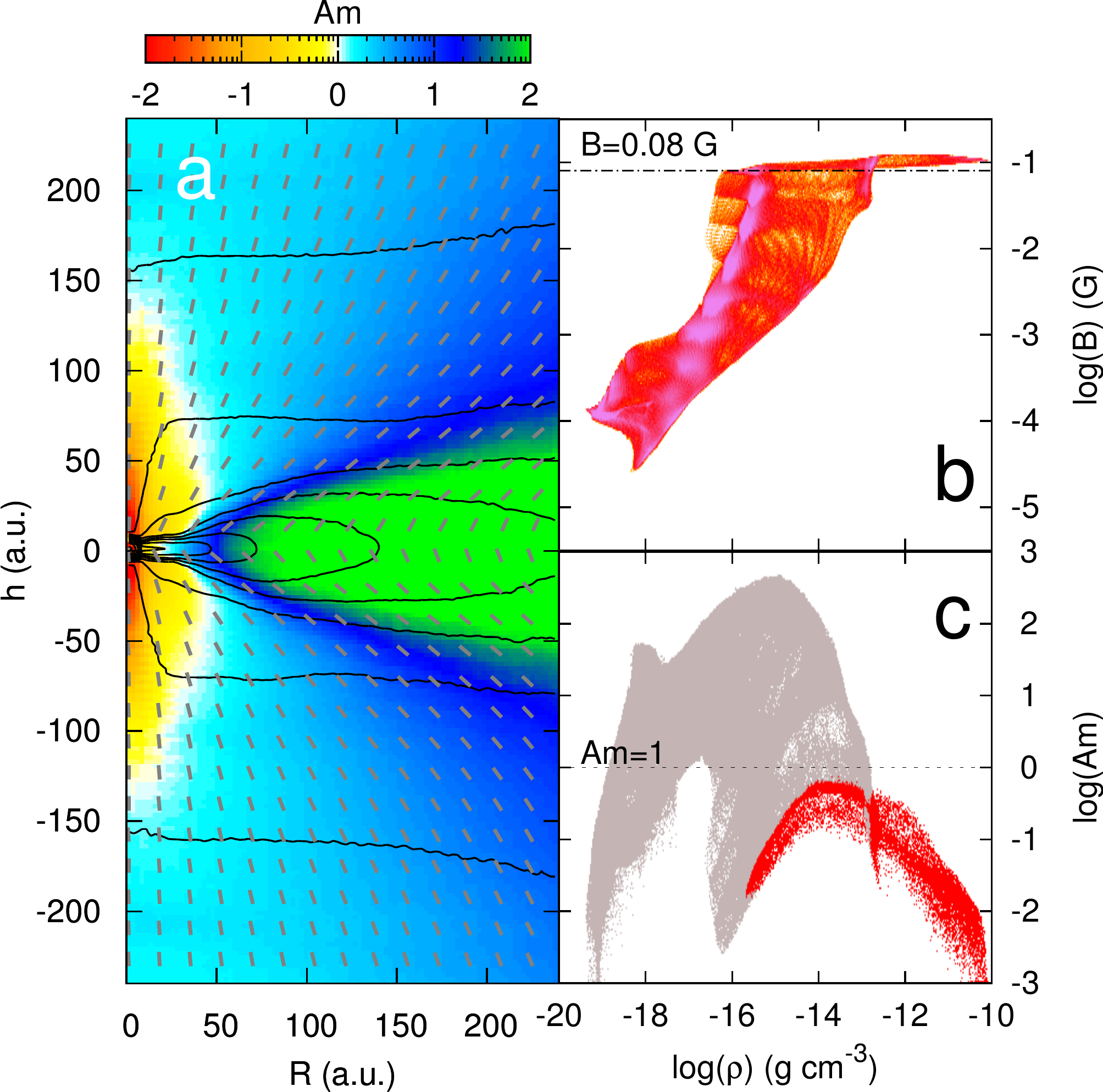}
        }
        \caption{Four main panels show the $\mu_{\trm{}}=5$ aligned (\ref{fig_act1_a}), $\mu_{\trm{}}=5$ misaligned (\ref{fig_act1_b}), $\mu_{\trm{}}=2$ aligned (\ref{fig_act1_c}), and $\mu_{\trm{}}=2$ misaligned (\ref{fig_act1_d}) cases 500 years after first core formation. In each main panel, subpanel (a) shows a colour map of the azimuthal average of $\trm{Am} = \frac{V L}{\eta_{\trm{AD}}}$. Green to blue means that dynamical processes (collapse, Keplerian disk) dominate, while red to yellow means that ambipolar diffusion dominates. The grey segments show the direction of the magnetic field, and black solid lines are isodensity contours. Subpanels (b) and (c) display the magnetic field magnitude $B$ and the value of Am as a function of density. Red cells in the Am scatter plot are cells where $B(\rho)> 0.08$~G, while grey cells have $B(\rho) < 0.08$~G.}
        \label{act1}
\end{figure*}

These regions are highlighted in \refig{act1}a, which shows maps of Am, an adimensional number that characterises the efficiency of the diffusion process compared to the dynamical ones, defined as
\begin{equation}
\trm{Am} = \frac{V L}{\eta_{\trm{AD}}} ~.
\end{equation}
The typical length-scale $L$ is taken as the distance to the protostar and the velocity $V$ as the local velocity along the field lines,
$V = (\vect{v} \cdot \vect{B})/||\vect{B}||$.
Ambipolar diffusion essentially plays no role in regions where $\trm{Am}>1$, which is the case for $r > 100$~au, where niMHD and iMHD calculations produce identical structures.
The region of the outflow, in contrast, is a region of very active ambipolar diffusion ($\trm{Am} < 1$) because of a lower density that corresponds to a higher resistivity $\eta_{\trm{AD}}$ (see \refeqp{cavient}). Dissipative effects are also very strong
around the mid-plane for $r<50$~au.
This reduces
magnetic braking by
relaxing the split monopole configuration close to the dense core.
The right panels in \refig{act1} show the magnetic field distribution as a function of density, along with the Am number (right axis).
All cells with $B>0.08$~G (horizontal dot-dashed line) have Am values
coloured in red (compared to non-flagged cells, which are grey). We note that almost every cell belonging to the magnetic field plateau is indeed dominated by ambipolar diffusion effects rather than dynamical effects ($\text{Am} \lesssim 1$).

\paragraph{\underline{\smash{$\mu =2$} :}}

The same diagram for the more magnetised case is shown in \refig{act1}b. Dynamical motions are less dominant because of the additional magnetic support. As a consequence, the active ambipolar diffusion region
is stretched into an hourglass shape with a pinched equatorial waist in the mid-plane
for $r<50$~au, of about the same size as for $\mu=5$ . The pinching of the field lines is also weaker in this central region than
for iMHD. However, compared to $\mu_{\trm{}}=5$, the split monopole configuration remains, with a more radial field in the midplane, especially in the domain $50<r<150$~au.
This eventually leads to increased magnetic braking, which significantly hampers the formation of
large Keplerian disks. Cells belonging to the diffusion plateau are, as previously, dominated by ambipolar diffusion effects and characterised by $\trm{Am} \lesssim 1$.

\subsection{Misaligned case}\label{sec:MisalignedCase}

The effects of tilting the rotation axis of the molecular cloud dense core with respect to the orientation of the magnetic field has consequences on the properties of the first Larson core and the accretion disks that can form around it \citep[see][]{joos}. We have repeated this analysis in MHD calculations including ambipolar diffusion.

\subsubsection{First Larson core}

The properties of the
first Larson core in the misaligned case are summarised
in Table~\ref{table_fc}. As in Sect.~\ref{sec:AlignedFirstCore}, the listed
quantities were measured 200 years after peak density reached $10^{-12}$~g~cm$^{-3}$. 

In the misaligned case, the regulation of the magnetic flux occurs as in the aligned case, yielding similar properties for the mass-to-flux ratio repartition and the value of the strongest magnetic field. For the collapse phase, the additional magnetic tension leads to a slightly longer free-fall time.
For the weakly magnetised $\mu_{\trm{}}=5$ , the mass of the core is increased in the misaligned case. Indeed, the mass-to-flux ratio reaches the same value, but the magnetic field is twice stronger than in the aligned case, yielding a core mass almost twice higher.

Therefore, misalignment of the magnetic field and rotation axis does not change the properties of the first core
significantly. It affects the formation and properties of rotationally supported structures, however, because of the weaker magnetic breaking, as we examine in Sect.~\ref{rot_mis}

\subsubsection{Magnetic field repartition}

\begin{figure}
     \begin{center}
        \subfigure[$\mu = 5$]{
            \label{figm_1_1_a}
            \includegraphics[width=0.325\textwidth, angle=270]{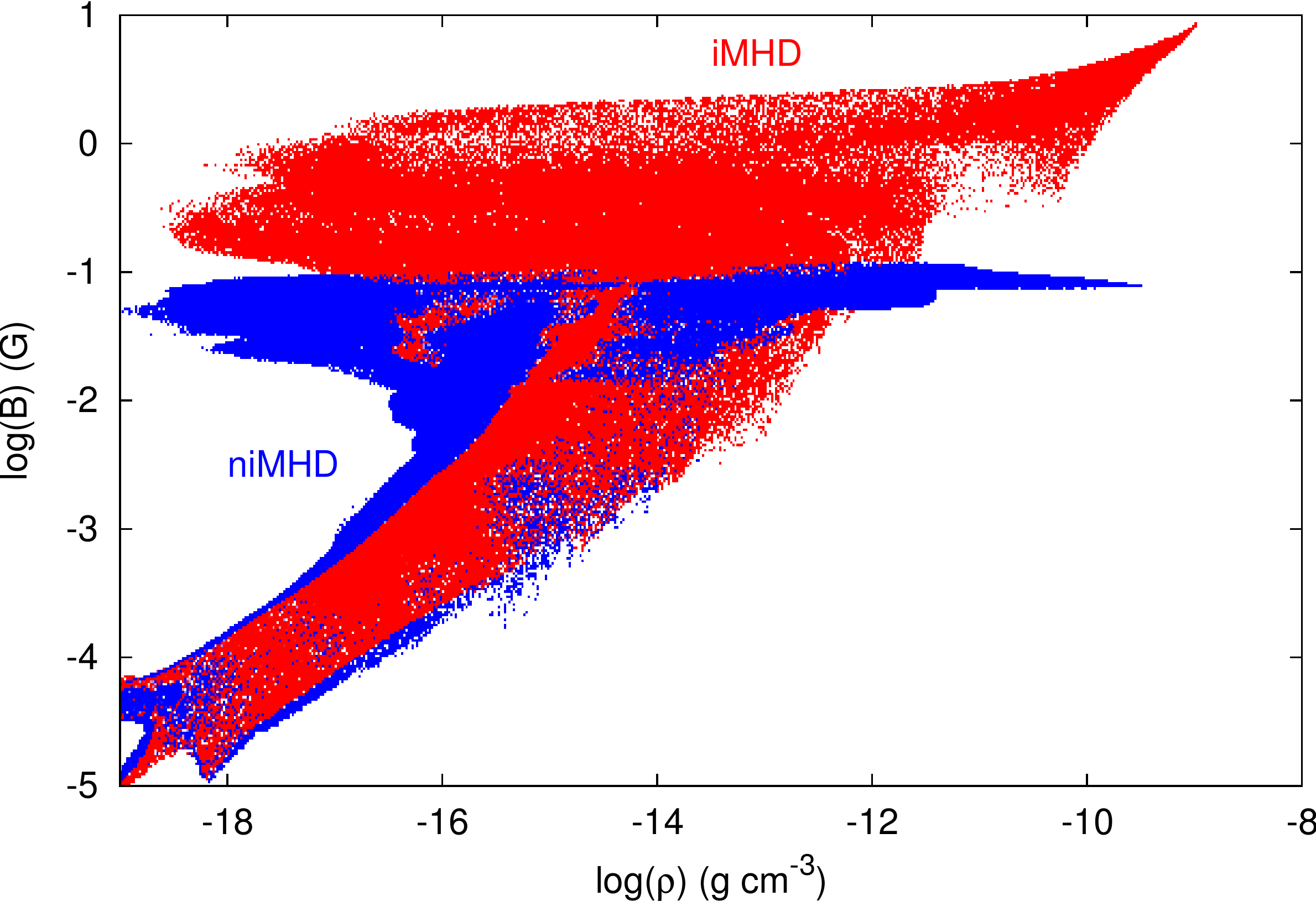}
        }\\ 
        \subfigure[$\mu = 2$]{
            \label{figm_1_1_b}
            \includegraphics[width=0.325\textwidth, angle=270]{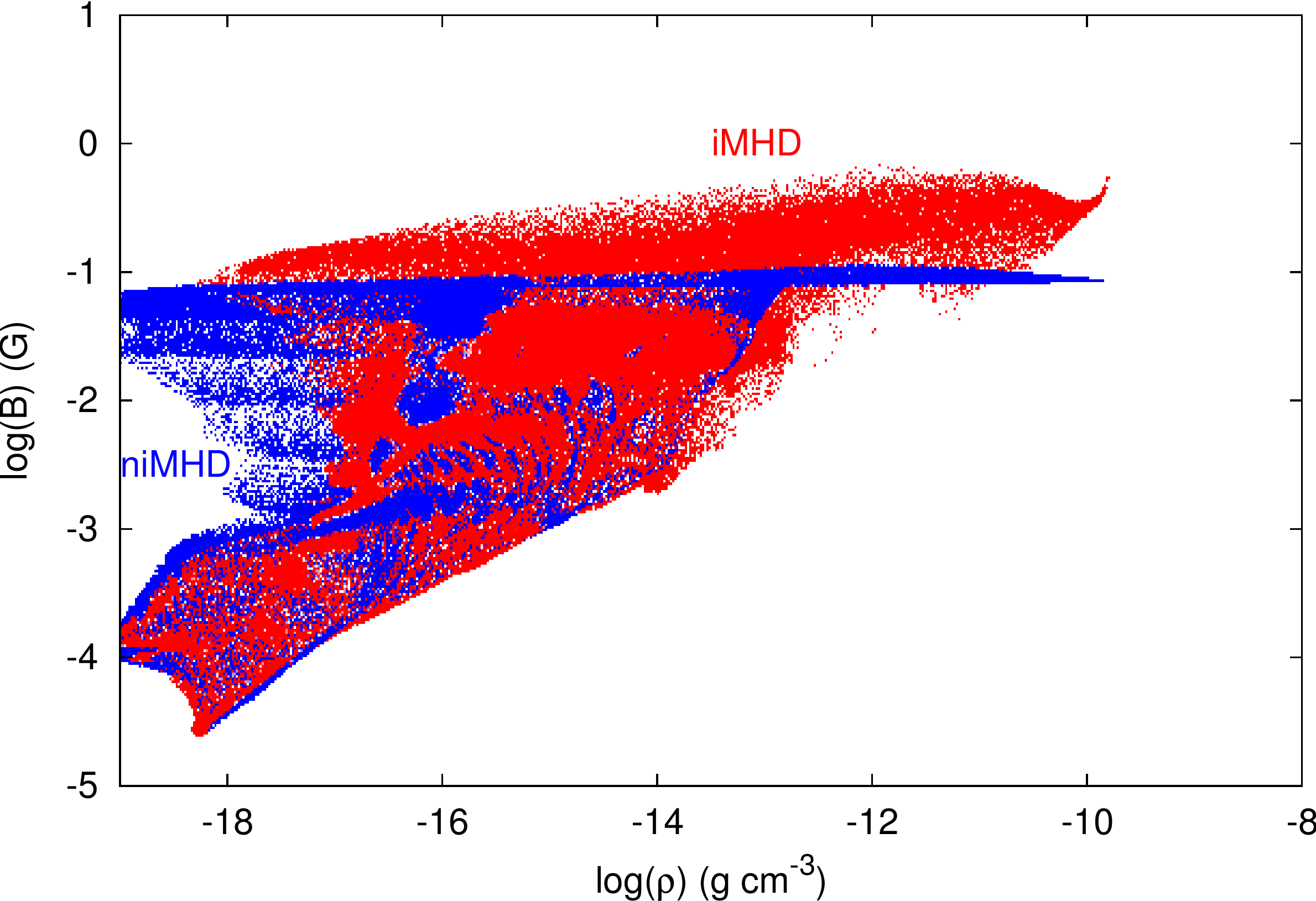}
        }
    \end{center}
    \caption{Same as \refig{fig_1_1} for the misaligned case.}
   \label{figm_1_1}
\end{figure}

Magnetic field repartition as a function of density is qualitatively unchanged compared to the aligned case, as seen in \refig{figm_1_1}. The same non-linear self regulation process yields a diffusion plateau. The region above and below the core is strongly depleted, leading to a highly magnetised and low-density region ($\rho < 10^{-16}$~g~cm$^{-3}$, $B > 10^{-2}$~G) that corresponds to the polar cavity mentioned in \cite{2015ApJ...801..117T}. It is more extended in the more magnetised case (see \refig{figm_1_1_a} compared to \refig{figm_1_1_b}), as in the aligned configuration (see Figs~\ref{act1}c and \ref{act1}d).
In iMHD, flux freezing still yields high peak magnetic field values, but the mixing due to the misalignment increases the numerical reconnection. As a result, a plateau somewhat similar to the niMHD case starts to develop (especially in the $\mu = 2$ case). We will discuss this in further detail in a forthcoming paper (Paper II), which will include the effect of turbulence.

\subsubsection{Outflows}

We consider only niMHD in this paragraph. Because of the misalignment angle, outflows, when present, exhibit a wider opening angle in the misaligned case than in the aligned one, and their growth is hindered. Misalignment yields less pile-up of the toroidal field, which weakens the outflow-launching mechanism. As for the aligned case, we do not see any outflowing motions in the $\mu = 2$ calculations.

We conclude that the presence or absence of outflowing gas is closely linked to conservation of angular momentum and to the regulation of magnetic field accumulation (and thus magnetic braking). Low magnetisation ($\mu = 5$) allows for enough toroidal field accumulation to grow a magnetically and density-enhanced tower, while higher magnetisation ($\mu = 2$) produces too effective braking, preventing the formation of magnetically launched and supported outflows.

\subsubsection{Region of active ambipolar action}

\refig{act1}c shows the efficiency of the ambipolar diffusion along with density contours and magnetic field orientation for $\mu_{\trm{}}=5$. The results are
 similar to the aligned case (cf. \refig{act1}a), with a larger pseudo-disk\footnote{The pseudo-disk is the overdensity region resulting from the loading of pinched field lines in a collapsing core that resembles a disk. See \cite{GalliShu1993} for the first study of the pseudo-disk.}. At these mid- to large scales ($>50$~au), ambipolar diffusion has little influence
($\trm{Am} \gg 1$). The polar cavities above and below the core are still present and well defined, and the region of ambipolar action in the mid-plane leads to the formation of a rotationally supported structure. Again, the magnetic plateau that forms at $B \lesssim 0.1$~G corresponds to regions where $\trm{Am} \lesssim 1$, dominated by ambipolar diffusion.

The $\mu_{\trm{}}=2$ misaligned case is almost indistinguishable from the aligned one, as can be seen when comparing Figs~\ref{act1}b and \ref{act1}d. This suggests that in more magnetised cases the final
configuration of the protostellar system results essentially from a saturation process
(self-regulation from the ambipolar diffusion) and is independent of the alignment or misalignment. Another way to formulate this result is that all the relevant
regions, including the diffusion plateau and most of the simulated box, are dominated by ambipolar diffusion ($\trm{Am} \lesssim 1$).

\section{Long-term evolution in non-ideal MHD: disk formation}\label{sec:disks}

In this section, we examine the rotationally supported structures that can form around the first Larson core. We focus on niMHD calculations; iMHD calculations are discussed in Sect~\ref{sec:longtermevo}. 

\subsection{Aligned case}

Conservation of angular momentum during the collapse yields a
very high specific angular momentum close to the protostar, which eventually leads to the formation of rotationally dominated structures with a strong toroidal velocity component and in some cases to
genuine disks with Keplerian velocity profiles. To define a "disk" in
the simulations, we used the criteria defined in \citet{joos_turb}.
Structures in which the magnetic pressure either dominates or is at least not negligible, however, may
satisfy these criteria while being very different from
the flat rotationally supported structures characteristic of Keplerian disks.
It is important to clearly identify and distinguish these two types of structures, since the existence of flat disks around Class-0 objects remains a subject of heated debate, and this issue is addressed below.
Following \citet{joos_turb}, a piece of fluid belongs to the disk if it fulfils all the following constraints (using $f=2$):
\begin{itemize}
    \item it is close to hydrostatic equilibrium:
    $v_{\theta} > f v_z$~;
    \item it has strong rotational support: $v_{\theta} > f v_r$~;
    \item its rotational support is stronger than the thermal support: $\frac{\rho v_{\theta}^2}{2} > f P$~;
    \item it has a high density: $\rho > 3.8 \times 10^{-15}~\text{g~cm}^{-3}$.
\end{itemize}

\begin{table}
\centering
\footnotesize
\caption{Properties of the disks that are formed in the non-ideal MHD simulations.}
\begin{tabular}{r cc cc cc cc}
\hline
Alignment  & \multicolumn{2}{c}{Aligned}  & \multicolumn{2}{c}{Misaligned}  \\
magnetisation  & $\mu_{\trm{}}=2$ & $\mu_{\trm{}}=5$ & $\mu_{\trm{}}=2$ & $\mu_{\trm{}}=5$ \\
\hline
\hline
Disk age (k years)                                            & 5.1  & 2.1              & 7.6  & 2.5 \\
Disk (aspect ratio)                                          &  core (2)  & spiral (8)  & spiral (5)  & warped (5) \\
Outer radius (au)                                          & 15  & 80               & 30  & 45 \\
M$_{\trm{first core}}$ (M$_{\soll}$)         & 0.17  & 0.23              & 0.17  & 0.27 \\
M$_{\trm{disk}}$ (M$_{\soll}$)               & 0.029 & 0.14         & 0.027  & 0.19 \\
$\beta(\trm{inner radius})$                                  & $>10^3$  & $>10^3$          & $>10^3$  & $>10^2$ \\
$\beta(\trm{outer radius})$                                  & $\lesssim 1$  & $\sim 1$         &  $\sim 1$  & $\lesssim 1$ \\
\hline
\end{tabular}
\label{table_disk}
\end{table}

\begin{figure*}
    \centering
       \includegraphics[width=0.90\textwidth]{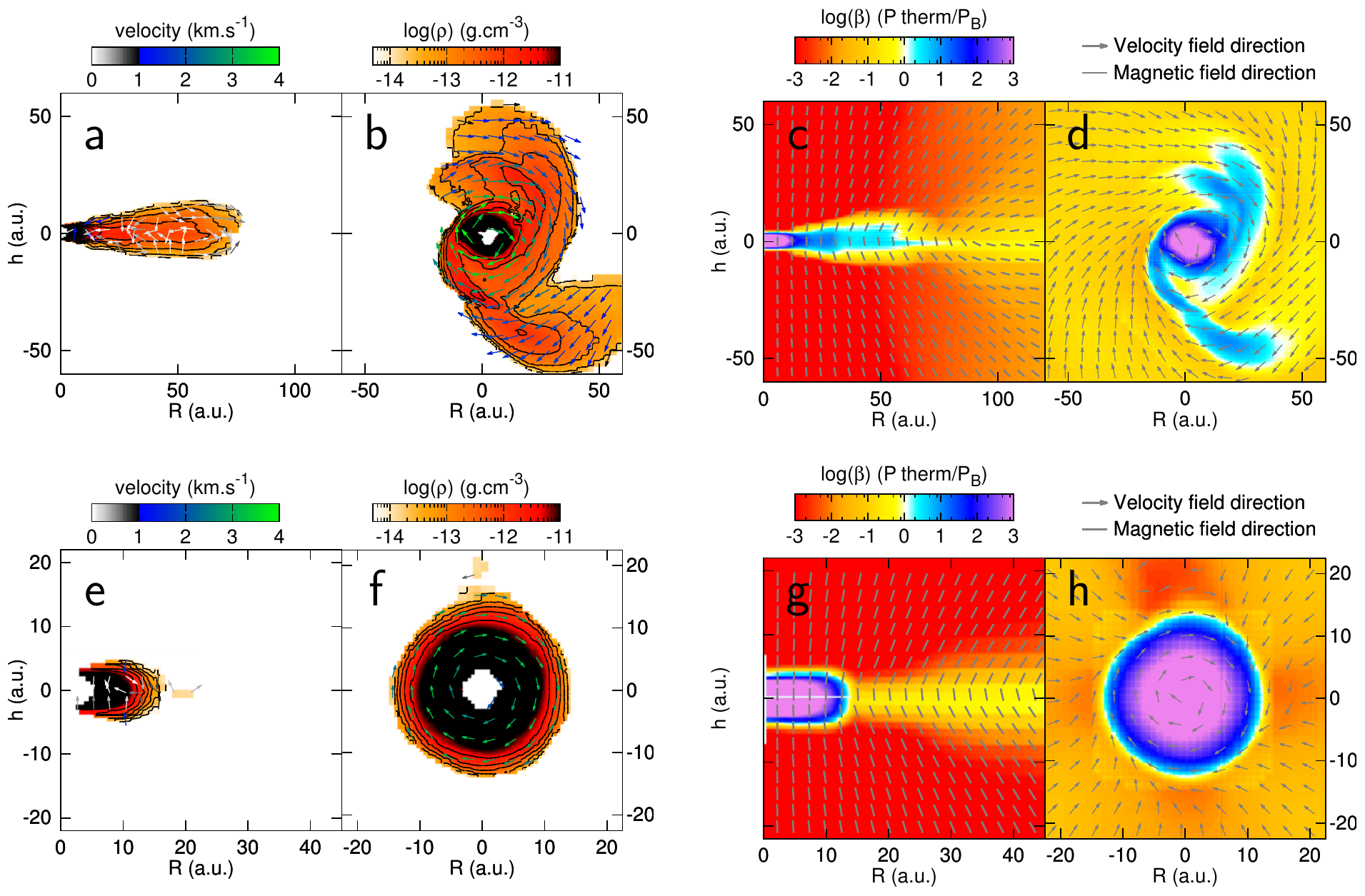}
       \caption{Visualisation of the disk structures. Top
row: Panels (a) and (b) show side and top views, respectively, of the disk density (orange), with the velocity vectors superimposed for  $\mu = 5$ . Panels (c) and (d) are the same side and top views, but showing the plasma $\beta$ parameter (colour map), over which we plot the magnetic field direction for the side view (c) and the velocity vectors for the top view (d). Bottom row: Same as for the top row, but for the $\mu = 2$ simulation. Each row has a different spatial scale.}
        \label{fig:disks_aligned}
\end{figure*}

\begin{figure}
    \begin{center}
       \includegraphics[width=0.33\textwidth, angle=270]{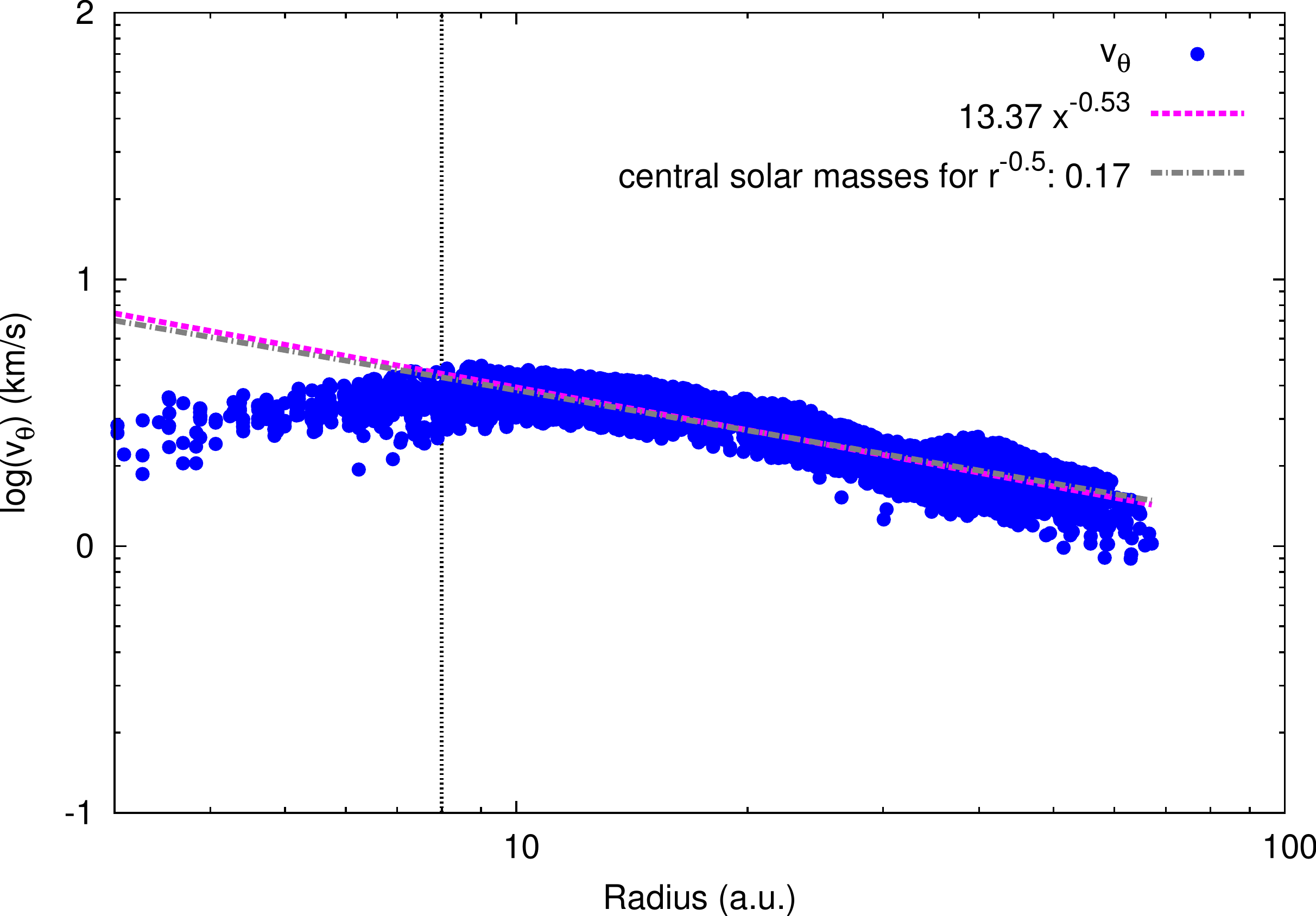}
       \caption{$\mu_{\trm{}}=5$, aligned case. Toroidal ($v_{\theta}$) velocity as a function of radius for disk cells. The pink dashed line is the fit (in the log/log space) of the blue points by a power law $a x^b$. The grey dashed line is the same fit but by enforcing $b=-1/2$. The vertical dotted line at 7~au is the inner radius for the fit.}
        \label{fig_19_2}
    \end{center}
\end{figure}

In our fiducial aligned case with $\mu_{\trm{}}=5$,
we do observe the formation of a disk according to the above criteria that strongly resembles a Keplerian disk.
Its characteristic properties are given in Table~\ref{table_disk} and the results of the simulations are shown
in \refig{fig:disks_aligned} (top row).
The disk is well defined in density, beginning just at the edge of the core and exhibiting sharp edges at its periphery. The breaking of symmetry observed in the top view (\refig{fig:disks_aligned}b) is due to the evolution over several orbital periods at the core radius, with small numerical errors leading to a bar-like instability that evolves into two spiral arms. 
The disk then grows to a radius of $r\sim 80$~au and is flat. The velocity profile in the disk is displayed \refig{fig_19_2} and is very close to a Keplerian one, with an estimated mass for the central object of $0.17\msol$ compared to $0.23\msol$ (Table~\ref{table_disk})
(see Appendix C for an important remark about assuming a Keplerian profile to estimate the mass of the central object).

\begin{figure}
     \begin{center}
        \includegraphics[width=0.33\textwidth, angle=270]{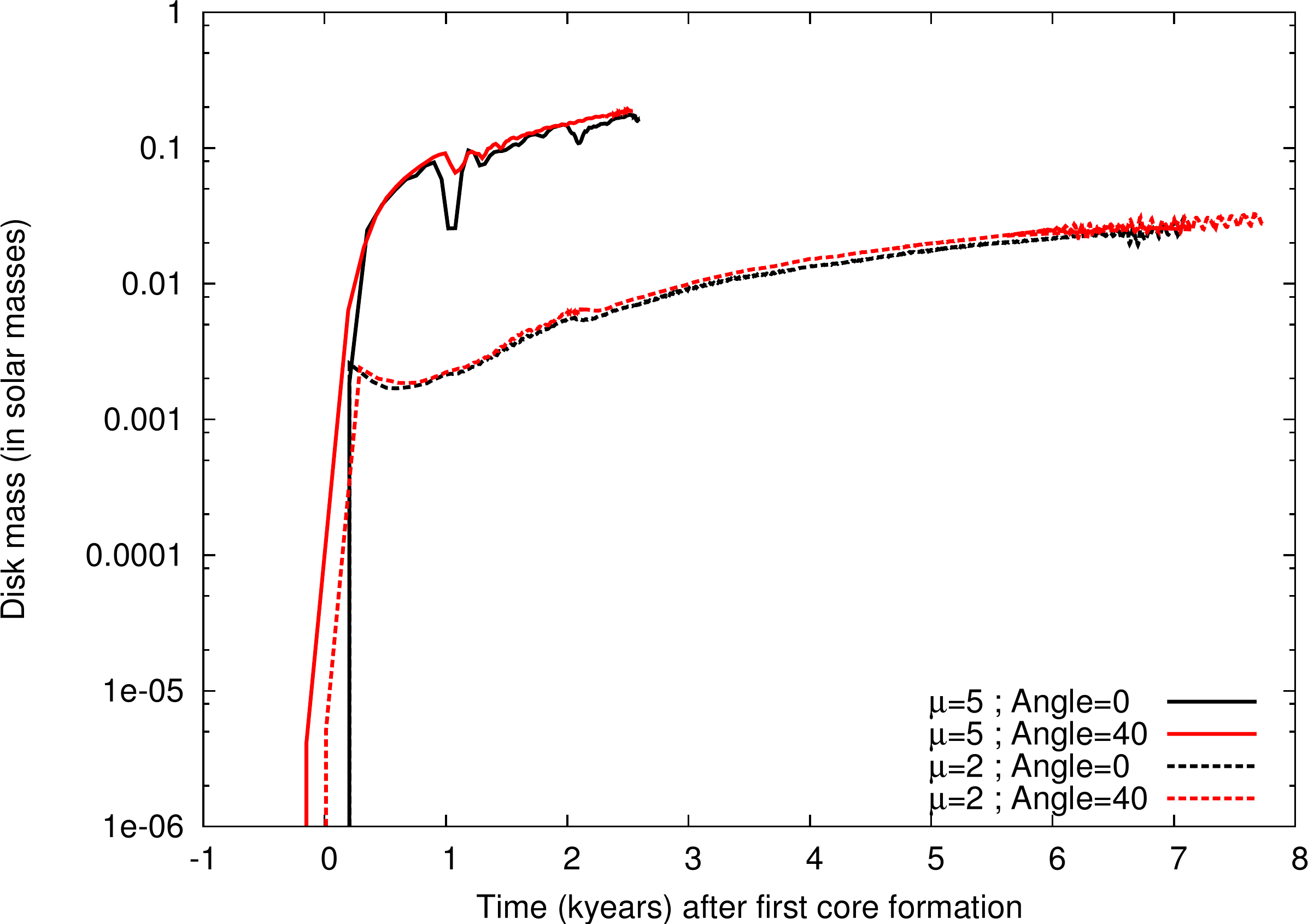}
    \end{center}
    \caption{Disk mass evolution. Solid lines: $\mu = 5$; dashed lines: $\mu = 2$. Black: aligned case. Red: misaligned case. $t=0$ corresponds to the first core formation to allow direct comparison between cases.}
    \label{figm_24_2_a}
\end{figure}

The evolution of the disk for our two values of magnetisation is displayed in \refig{figm_24_2_a} (solid lines for $\mu =5$, dashed lines for $\mu=2$).
In niMHD, the pinching of the field lines the midplane is greatly reduced and the radial component of the field is almost null. In this case, there is almost no magnetic flux accreted onto the central object in the radial direction.

Figures~\ref{fig:disks_aligned}c and \ref{fig:disks_aligned}d
show the plasma $\beta$, defined as the ratio of
thermal over magnetic pressure, $\beta=P/P_{\trm{mag}}$, in the disk and its surroundings.
We note that everywhere inside the disk
and the core, $\beta \gg 1$, meaning that the thermal pressure dominates magnetic pressure in these regions.

These results are of prime importance both from a physical and numerical point of view. Physically, they bear major consequences on our understanding of
fragmentation and angular momentum transport in disks. In ideal MHD calculations, magnetic fields have been shown to inhibit fragmentation (\cite{HennebelleTeyssier2008} and \cite{commercon10}. The
fact that in more realistic non-ideal MHD calculations $\beta \gg1$ suggests that magnetic fields have less impact than expected according to iMHD results and that fragmentation may be facilitated. A second important consequence of the present calculations, of the characterisation of plasma $\beta,$ and of the detailed topology of the field is the major role of these quantities in determining viscous transport in disks, in particular for the magneto-rotational instability \citep{balbus2001}. A better knowledge of these quantities will enable us to define more accurate initial conditions in such studies \citep[see e.g.][]{2014A&A...566A..56L}.

Form the numerical point of view, the present studies are very important for the use of sink particles in simulations. 
Current sink particle models do not treat magnetic flux transport at the sink-gas interface in a consistent way. The flux is not accreted, which creates an effective barrier of infinite diffusion and spurious numerically driven interchange instabilities. As mentioned above, in the AD case, the radial component of the magnetic field is almost non-existent, yielding a configuration of zero net radial flux, allowing the accurate characterisation of the accreted (or in this case not accreted) magnetic flux onto the sink particle.

\paragraph{\underline{\smash{$\mu =2$} }:}

In the more magnetised case, a disk-like structure around the first core is still observed, according to the above criteria, but the disk mass is an order of magnitude lower than in the less magnetised $\mu_{\trm{}}=5$ case.
Figures~\ref{fig:disks_aligned}e and \ref{fig:disks_aligned}f
represent the disk cells in an edge-on and in a top
view. As for the less magnetised case, the disk
inner radius coincides with
the outer layers of the core,
and the disk slowly
grows as matter is accreted. The final radius is about $20$~au with a mass of $0.03 \msol$. The aspect ratio remains close to unity, and the structure exhibits the same characteristic properties as
in the previous case: it has
a close-to-Keplerian velocity field with a high plasma $\beta$ in the rotationally dominated structure, while in contrast, the regions outside the disk are magnetically dominated.
No spiral arms have developed in this case, even at the end of the simulation, but steady state has not been reached yet as the disk is still slowly accreting.

As an important global remark, we note that since values of $\beta \gg 1$ in the disk are found in all our simulations in niMHD for any initial condition, a $\beta$-based criterion could thus be used as a robust disk criterion in niMHD simulations\footnote{This criterion encompasses the core itself, which needs to be removed from the disk
by considering the thermal to kinetic energy ratio, as done in \cite{joos_turb}.}.

\subsection{Misaligned case\label{rot_mis}}

\begin{figure*}
    \centering
       \includegraphics[width=0.90\textwidth]{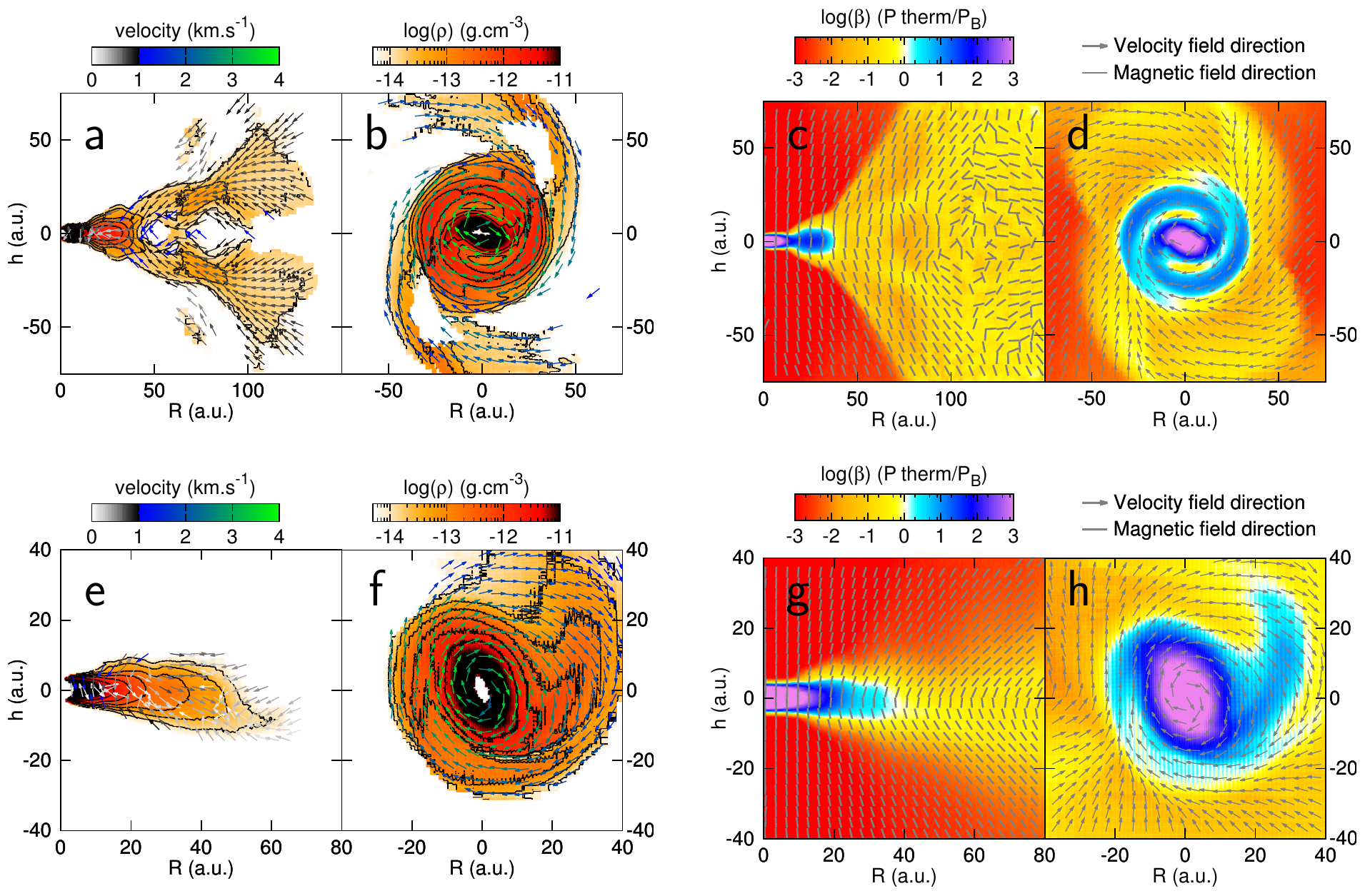}
       \caption{Same as \refig{fig:disks_aligned}, but for the misaligned configuration. Each row has a different spatial scale.}
        \label{fig:disks_misaligned}
\end{figure*}

In the misaligned case, a disk can form more easily
and is more massive, as already noted in previous studies \citep{joos,2013ApJ...774...82L}. Indeed, misalignment yields a thicker pseudo-disk, which reduces the magnetic braking.
\refig{fig:disks_misaligned} (top row) shows the disk
structure at the end of the simulation. It exhibits a flat inner morphology, with a Keplerian velocity profile and
high plasma $\beta$ values.
In this case we note large accretion spiral arms (with significant infalling motions) above and below the disk plane, with $\beta \ll 1$. These spiral arms, however, do not represent a significant part of the disk mass. We note that using a $\beta$-based disk selection criterion would exclude these arms from the disk (see Fig.~\ref{fig:disks_misaligned}d compared to \ref{fig:disks_misaligned}b).

For $\mu_{\trm{}}=2$ (Figs.~\ref{fig:disks_misaligned}e-h), the early evolution is similar to the aligned case, the disk is at
first hardly distinguishable from the core (outer radius is about 10 to 15~au) and then grows by accretion. Significant differences compared to the aligned case, however, occur at later stages, as seen in the
figures. In this case, the disk is massive enough to trigger the formation of spiral arms, with a mean radius of 20 to 50~au. As for the previous cases, the disk is very well defined by the region of high $\beta$ plasma.

\subsection{Emerging consistent picture}

\begin{figure}
     \begin{center}
         \includegraphics[width=0.38\textwidth]{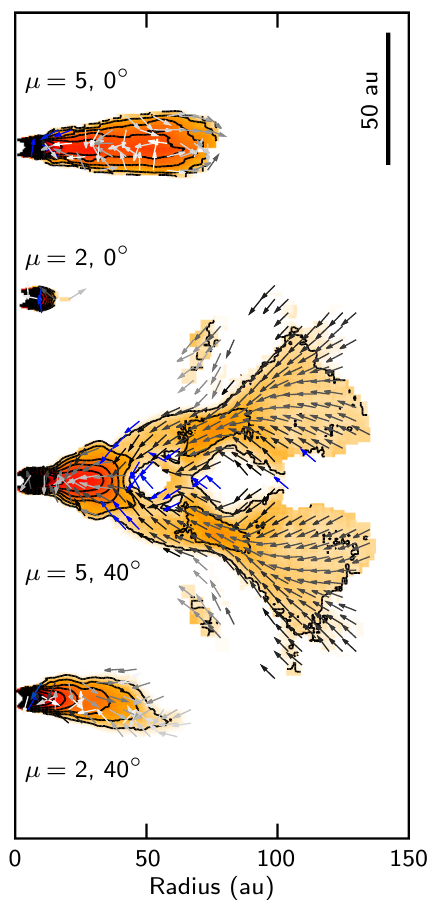}
    \end{center}
    \caption{Representation of disk sizes and shapes for every niMHD simulation. The density colour map and contours are the same as in Figs.~\ref{fig:disks_aligned} and \ref{fig:disks_misaligned}.}
    \label{fig:alldisks}
\end{figure}

\refig{fig:alldisks} shows a direct visual comparison of disks in each simulation. The disk formation and evolution during the collapse in the aligned and misaligned cases is presented in \refig{figm_24_2_a}. In the misaligned cases, the disk forms at the same time as or even before the first core forms. This confirms, as discussed above, that the disk emerges from the outer layers of a distorted first core embryo. After about a hundred years, the disk is well defined, with a mass $\gtrsim 10^{-3}\msol$ in all cases. At this stage,
the disk growth histories for the aligned and misaligned cases are essentially indistinguishable.

\begin{figure}
     \begin{center}
        \includegraphics[width=0.33\textwidth, angle=270]{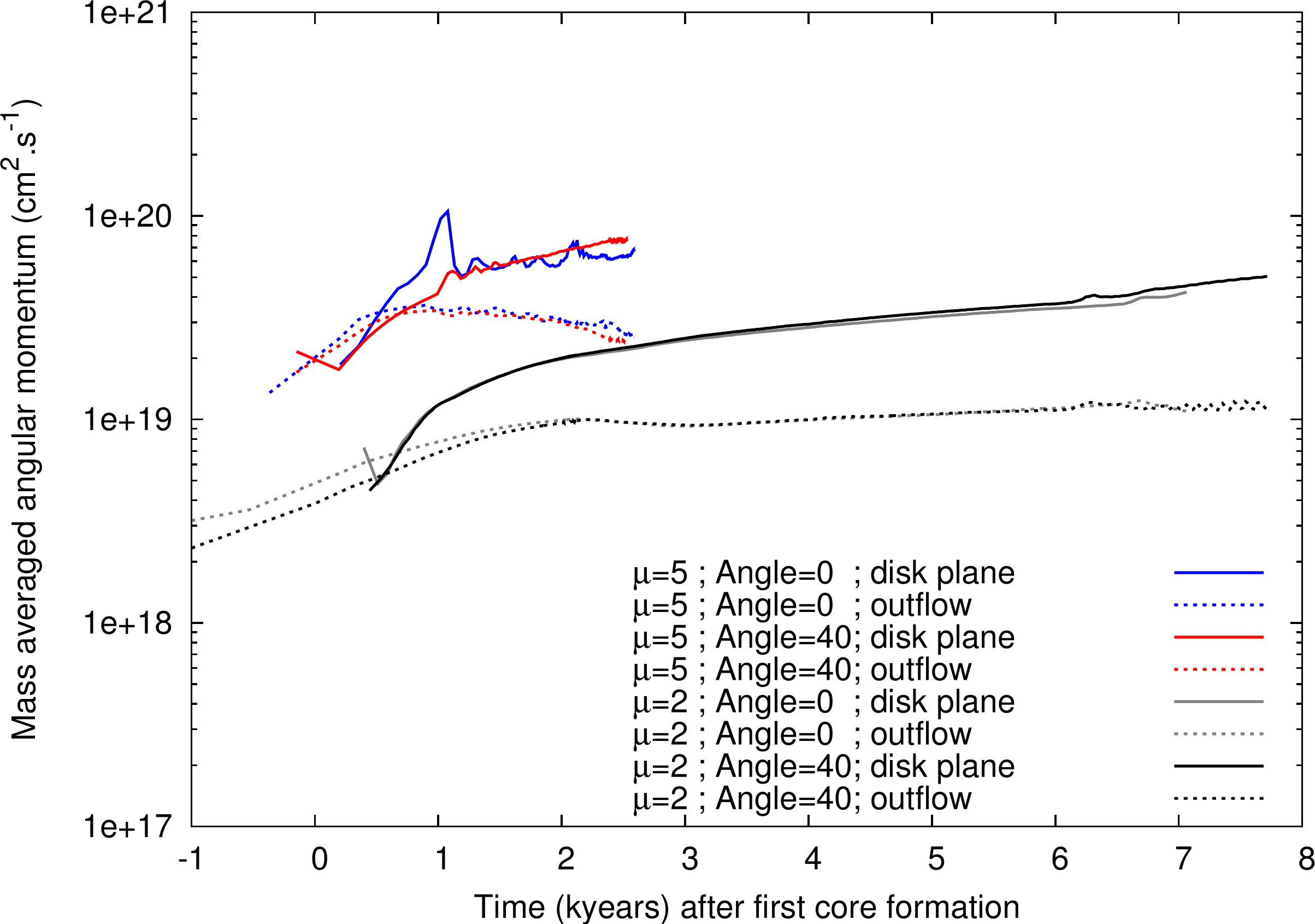}
    \end{center}
    \caption{Angular momentum evolution. Blue and red: $\mu = 5$; grey and black: $\mu = 2$. Solid lines: angular momentum in the disk plane; dashed lines: angular momentum in the outflow (see text). $t=0$ corresponds to the formation  of the first core.}
    \label{figm_24_2_b}
\end{figure}

\refig{figm_24_2_b} displays the angular momentum evolution in spatially restricted regions, separating the disk and outflow components in the system. The disk plane includes every cell fulfilling the disk criteria inside a sphere of radius $100$~au. Conversely, the outflow component corresponds to every other cell in a sphere of radius $100$~au. The $\mu = 5$ and $\mu = 2$ cases exhibit the same pattern.
The mass-averaged angular momentum is slightly larger in the outflow region in the aligned case (dashed blue line) than in the misaligned one (dashed red line)
during the first thousand years after the core formation. In the disk plane, the definition of the disk and the presence of spiral arms introduce variability, but the trend remains similar for the aligned and misaligned configurations. The misaligned case in iMHD releases the strong gradients in the pseudo-disk (strongly pinched magnetic field lines) found in the aligned case and thus enables larger disks to form.
 Non-ideal MHD produces the same smoothing of gradients, yielding eventually similar structures in the aligned and misaligned cases.

This similarity between aligned and misaligned simulations in niMHD shows that the formation and evolution of a disk is independent of the initial misalignment. Ambipolar diffusion, when it is efficient, regulates the angular momentum transport and the magnetisation during the collapse.
A stronger
magnetic braking will operate in the aligned case, leading to more angular momentum being evacuated in the vertical direction and ultimately yielding the same disk mass and final angular momentum once a physical equilibrium has been reached. The larger the ambipolar diffusion, the more efficient the regulation mechanism, leading eventually to similar structures.
This is well illustrated by the even more pronounced
similarity between the aligned and misaligned cases for $\mu = 2$.

\section{Limits of ideal MHD \label{sec:longtermevo}}

In this section, we highlight several limits of the ideal MHD approximation in the context of prestellar magnetised collapse and disk formation, justifying in passing the fact that we did not discuss iMHD simulations in Sect.~\ref{sec:disks}.

\subsection{Counter-rotation}

\begin{figure}
    \begin{center}
        \includegraphics[width=0.48\textwidth]{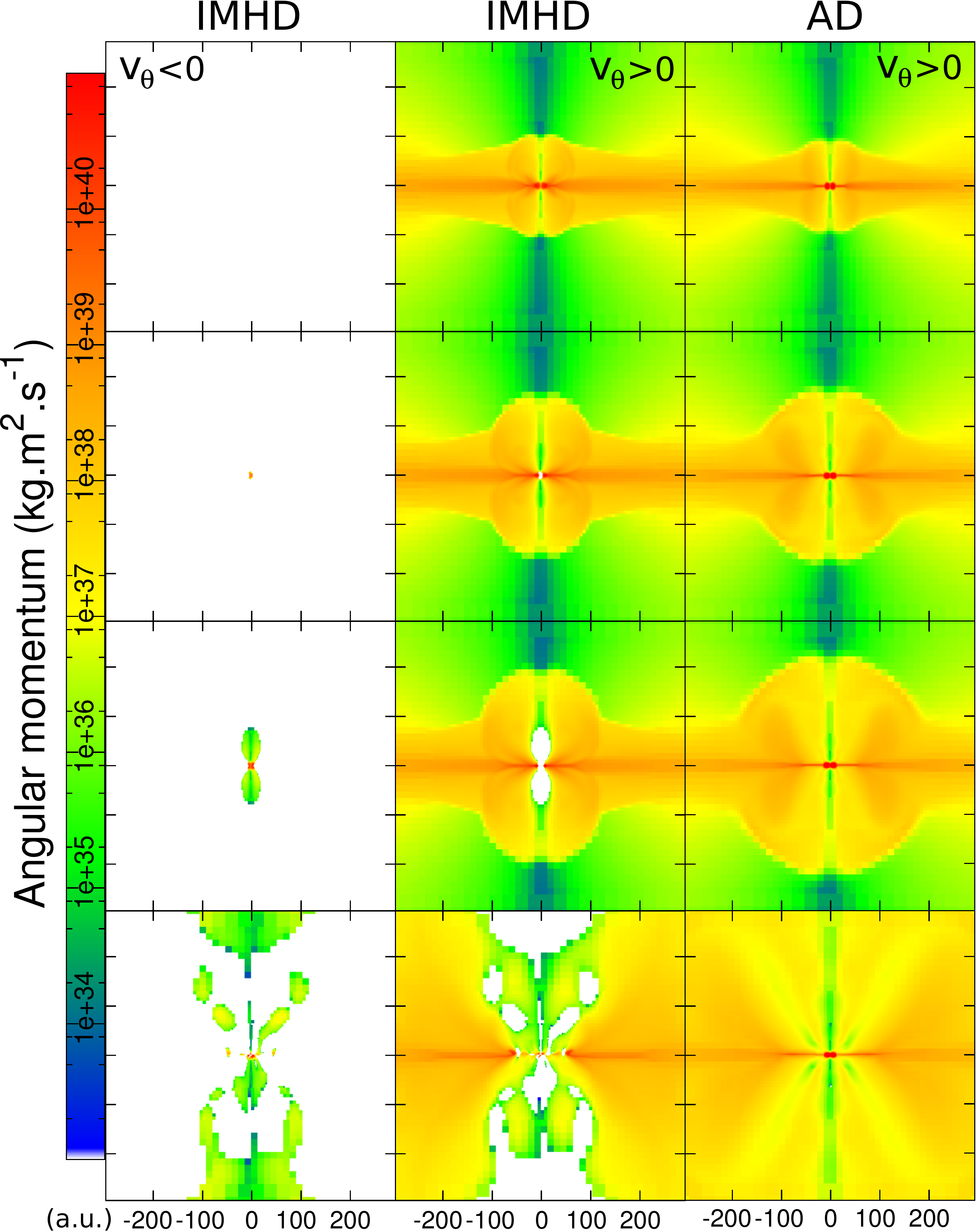}
        \caption{$\mu = 5$, aligned case. Four
        snapshots (time increases from top to bottom) from an ideal MHD simulation (two left columns) and the ambipolar diffusion case (right column) with the magnetic field initially aligned with the rotation axis. For the iMHD simulation,
        the angular momentum is plotted in the left or right panel, depending on the sign of the azimuthal speed $v_{\theta}$ in cylindrical coordinates.}
        \label{cntr}
    \end{center}
\end{figure}

In ideal MHD, the increase in the toroidal field component is due to the rotation of the gas around the forming protostar and is only hindered by numerical reconnection.
The resulting magnetic
braking can be efficient enough to completely stall
the rotation of the core.
During the time the information propagates outwards, the core can start to counter-rotate,
creating a new braking,
until eventually co-rotation with its surrounding is reached again. This is illustrated in \refig{cntr}, where we slightly increased the initial rotation in our fiducial case to better illustrate our purpose:
$\mu=5$, $\alpha=0.25$, $\beta_{\trm{rot.}}=0.03$.
Time is evolving from top to bottom. To trace the rotation, we calculated the angular momentum for each cell
with respect to the mean direction of rotation in a sphere of radius $200$~au.
The first and second columns show the negative and positive toroidal velocity ($v_{\theta}$) maps, respectively, in a iMHD simulation. The third column displays the positive toroidal velocity for niMHD \footnote{We do not display the counter-rotating $v_{\theta} < 0$ for AD since there are almost no counter-rotating cells except very close to the rotation axis at the very end of the simulation.}.
Strong counter-rotation occurs in the iMHD simulation, starting near the core (second row), and propagates into the outflow (third row).
There is no sign of counter-rotation in the niMHD run. The mechanism continues and generates
zones that extend to very large radii
with alternate positive and negative toroidal velocities, producing a butterfly-shaped outflow (bottom row).
This has important consequences on the expansion of the outflow, which is
narrower for iMHD (this is most visible in the third row) because the
counter-rotation hampers the pile-up of the toroidal component of the magnetic field.

We carried out similar studies 
in the misaligned case and found out that
counter-rotation still develops after formation of the first core, although with a more limited spatial extent
than in the aligned case.
We also examined lower initial magnetisations and found out that counter-rotation is greatly reduced when $\mu > 5$.

\subsection{Flux redistribution}

\begin{figure}
    \begin{center}
        \includegraphics[width=0.33\textwidth, angle=270]{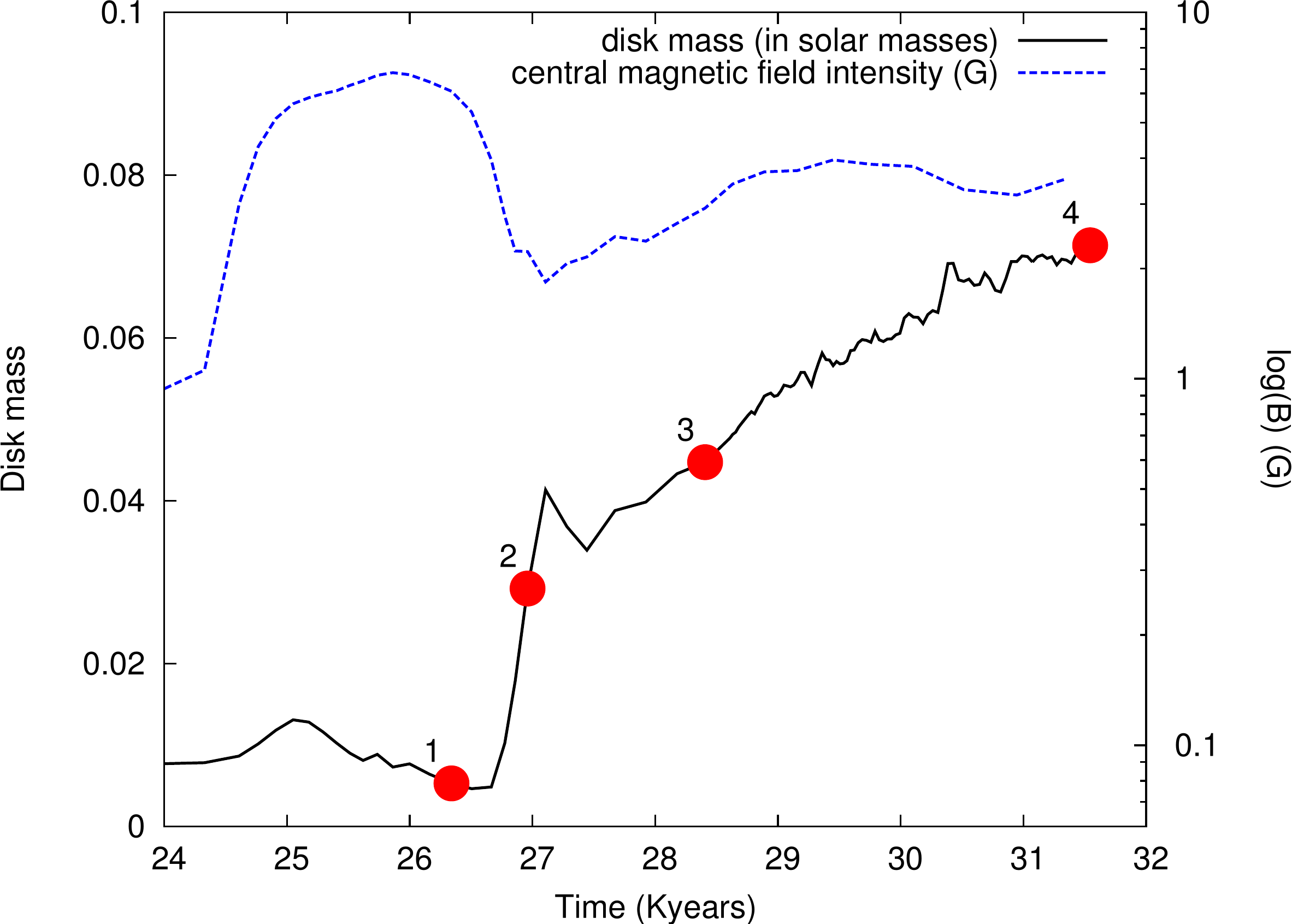}
        \caption{Evolution of the disk mass (solid line) and the peak magnetic field value (dashed line) for the $\mu_{\trm{}}=5$, aligned case, in iMHD. The four red dots marked
        1 to 4
        indicate the times at which the snapshots in \refig{plein} were taken. We recall that the formation of the first Larson core occurs at $t=24300$~years (see Table~\ref{table_fc}). The number 1 indicates the beginning of the release of magnetic flux from inside the core.}
        \label{evol2}
    \end{center}
\end{figure}

\begin{figure}
     \begin{center}
         \subfigure[Evolution of the density/velocity map for the four times labelled 1-4 in \refig{evol2}.]{
            \label{plein_a}
        \includegraphics[width=0.48\textwidth]{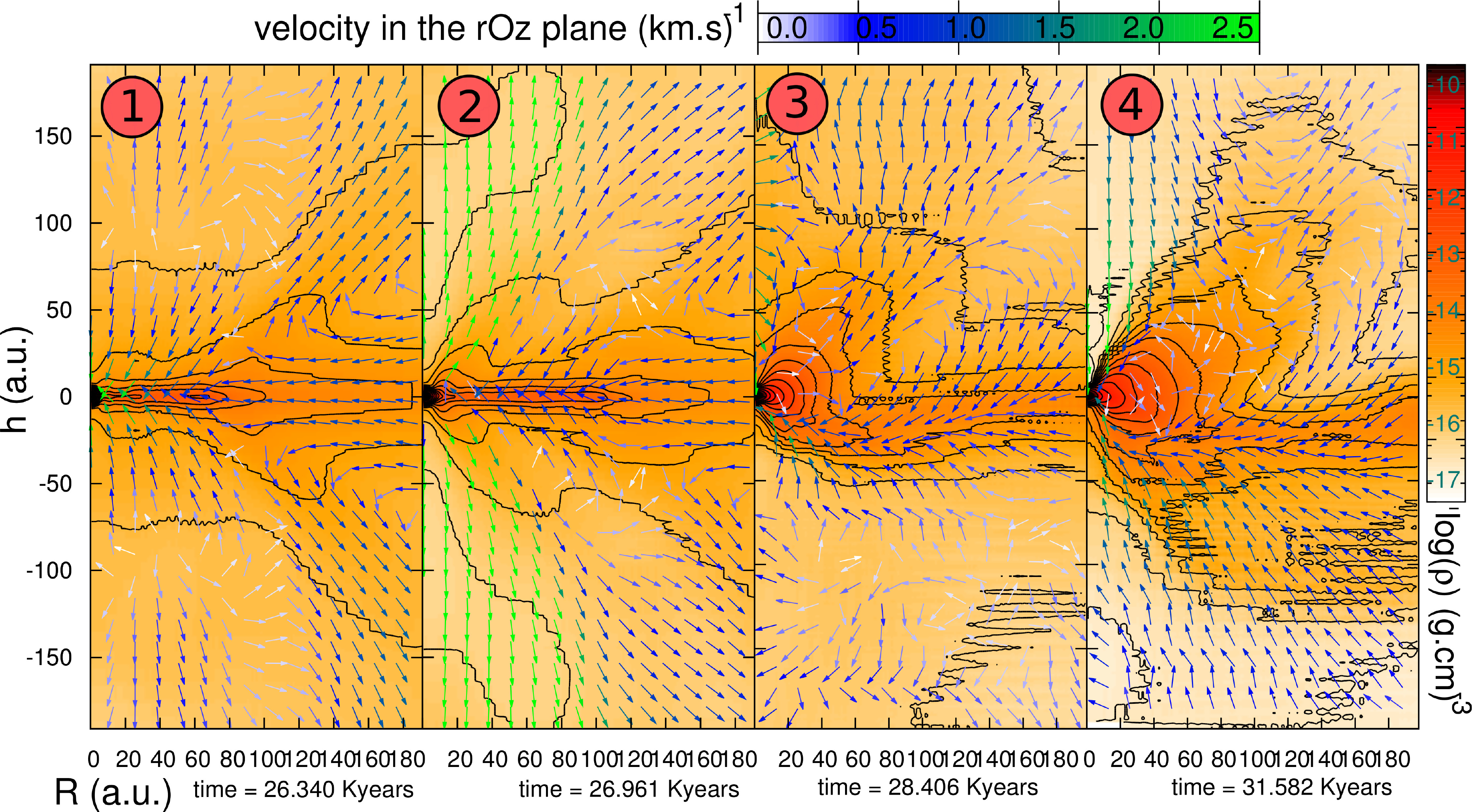}
        }\\ 
         \subfigure[The three first columns represent the relative importance of each component of the magnetic field. The colour scale shows the value of each component as follows: black for high values ($\gtrsim 1$), white for low values ($\lesssim 10^{-1.5}$); blue and red (in-between $\gtrsim 10^{-1}$). The rightmost column displays the plasma parameter $\beta=P/P_{\trm{mag}}$.]{
            \label{plein_b}
        \includegraphics[width=0.48\textwidth]{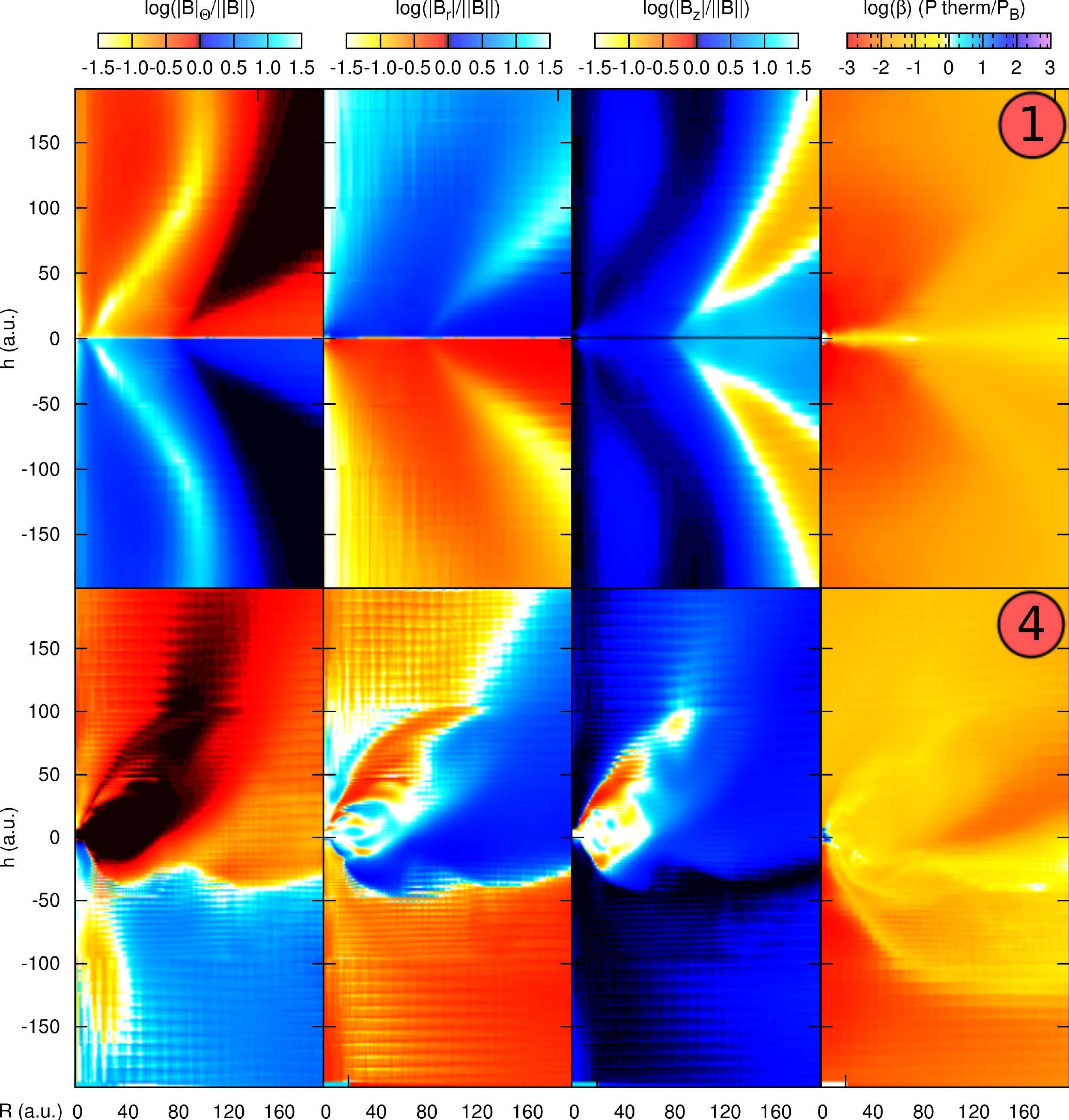}
        }
    \end{center}
        \caption{Snapshots for different times of interest in the iMHD $\mu=5$ aligned case, as labelled \refig{evol2}.}
        \label{plein}
\end{figure}

\begin{figure}
    \begin{center}
         \subfigure[iMHD]{
            \label{disk_compa_a}
        \includegraphics[width=0.24\textwidth]{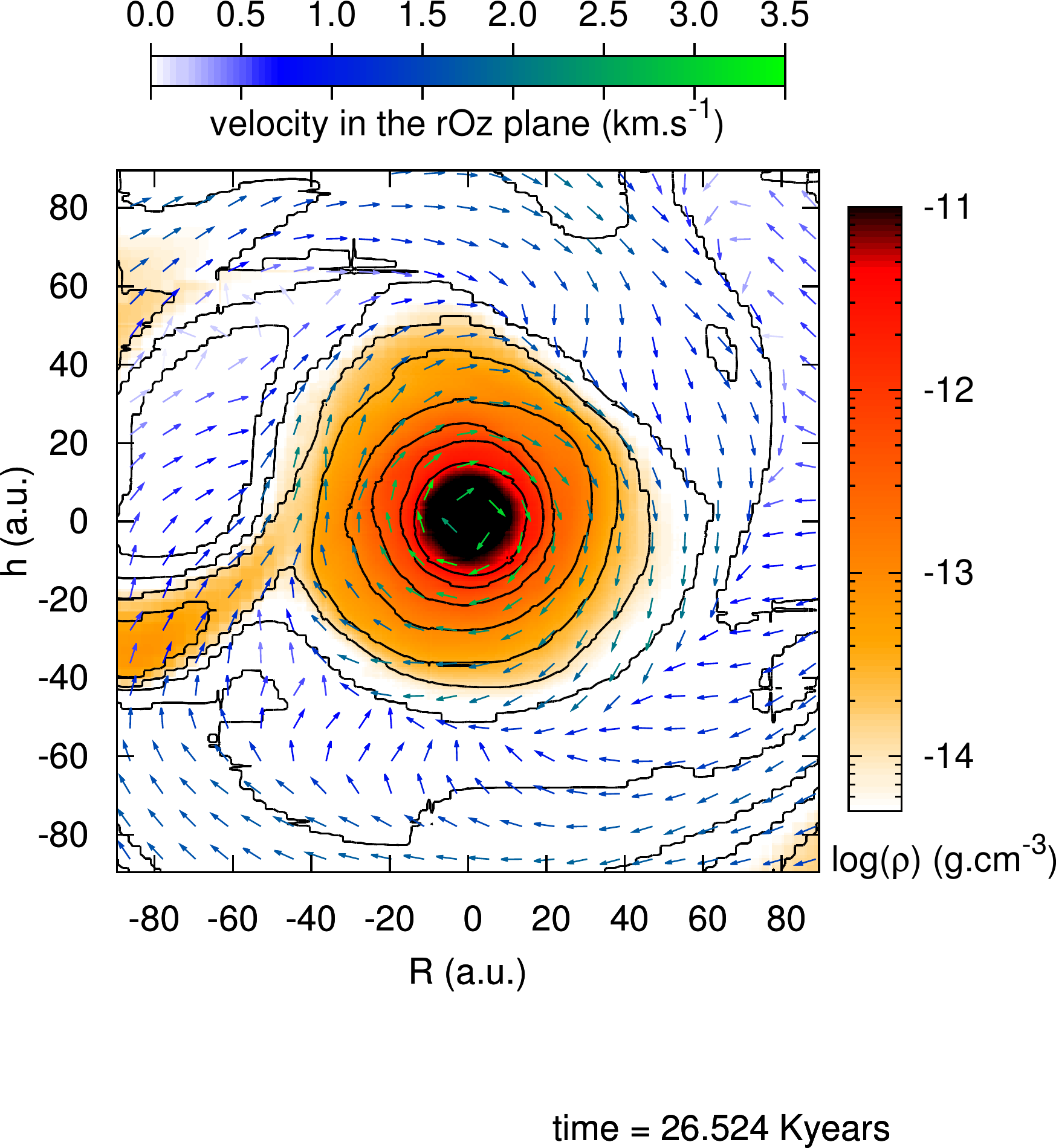}
        \includegraphics[width=0.24\textwidth]{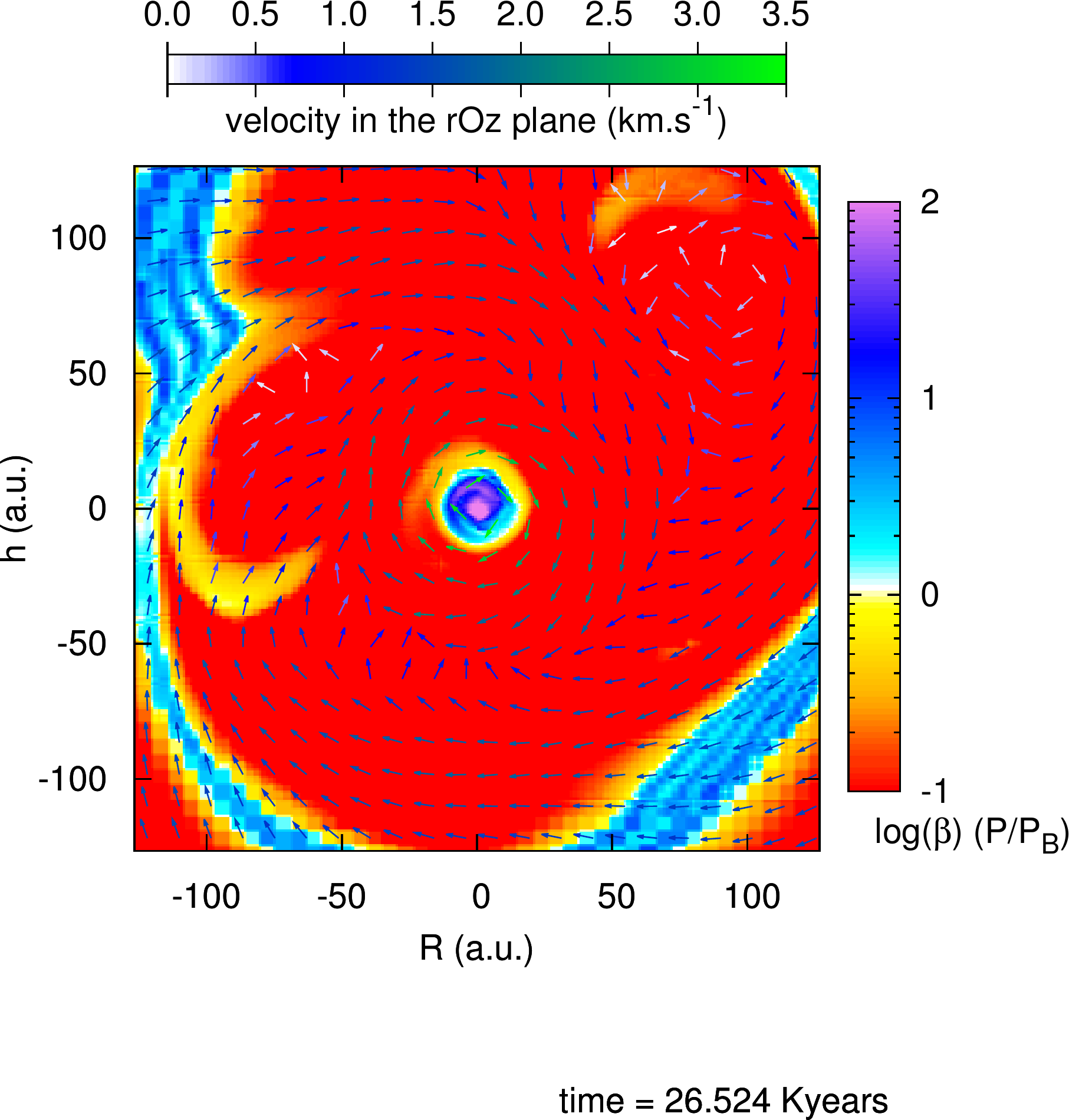}
                     }\\
         \subfigure[niMHD]{
            \label{disk_compa_b}
        \includegraphics[width=0.24\textwidth]{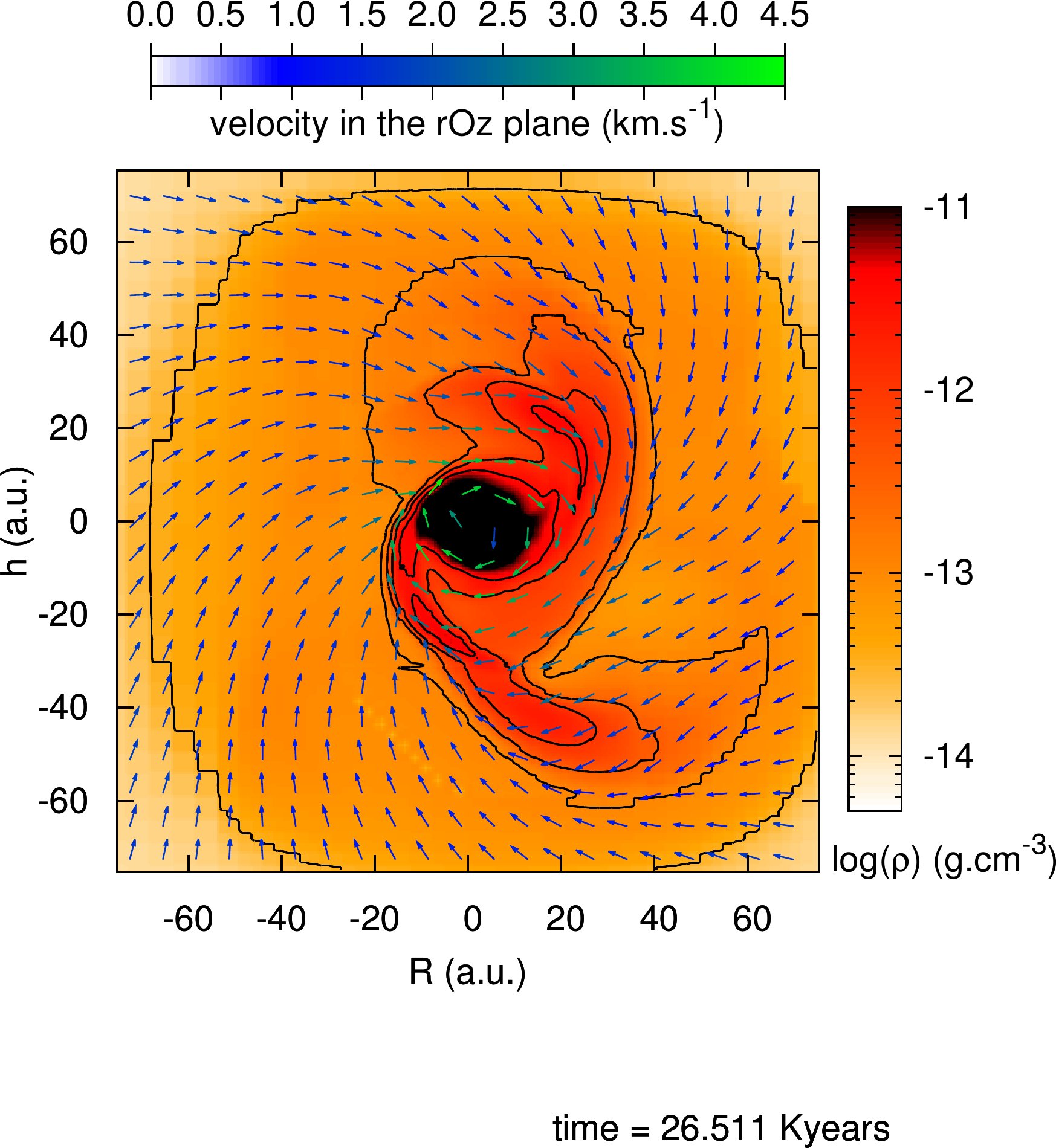}
        \includegraphics[width=0.24\textwidth]{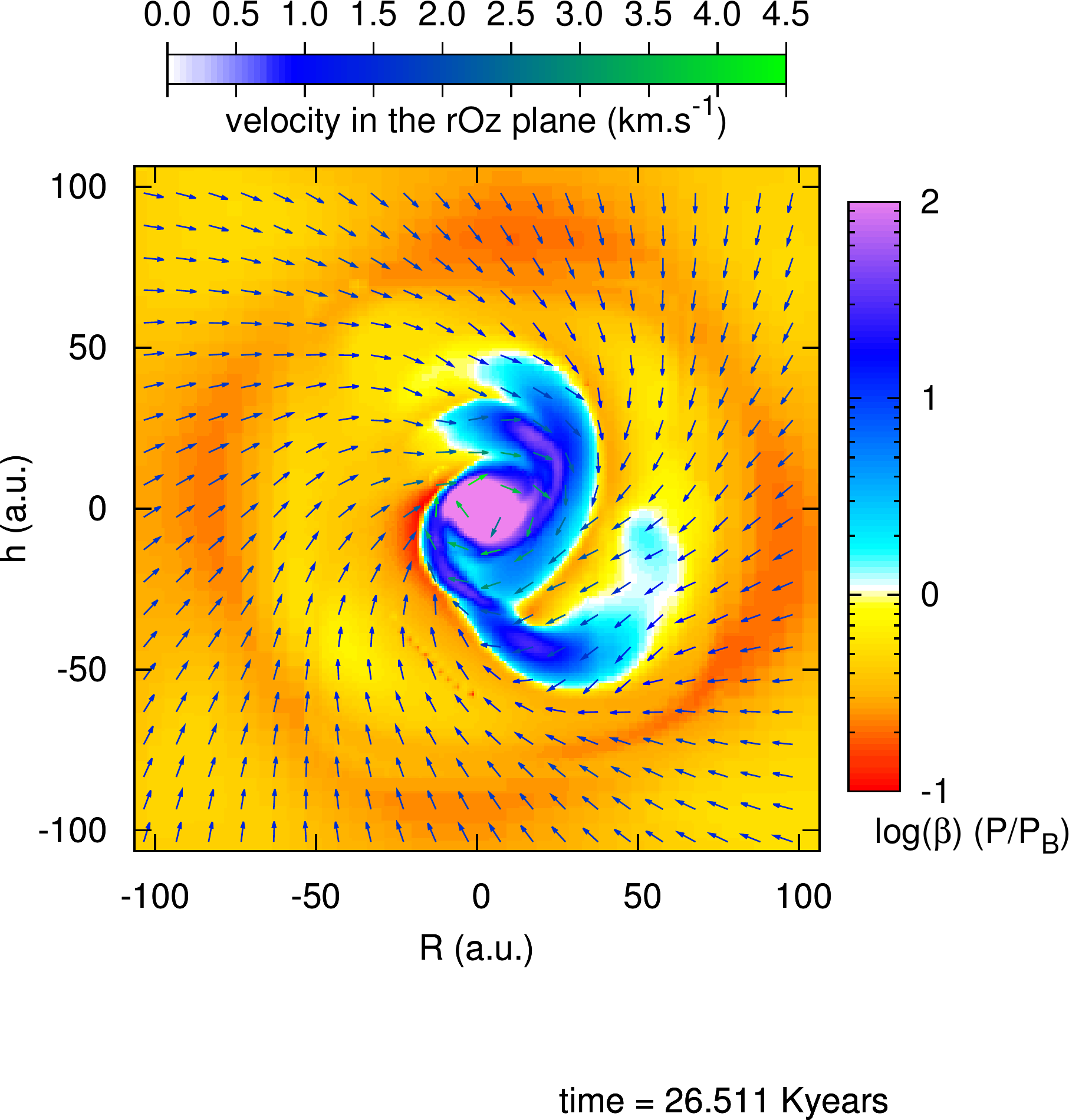}
                     }
        \caption{$\mu_{\trm{}}=5$, aligned case. View of the disk plane in a cylindrical volume of height h=$20$~au. Left and right: density/velocity field and plasma $\beta$, respectively, with the velocity field.}
        \label{disk_compa}
    \end{center}
\end{figure}

\refig{evol2} illustrates the evolution of the disk mass and central magnetic field for iMHD . At $t \gtrsim 26.5$ k years, the field intensity decreases by almost an order of magnitude, producing a drastic increase of the disk mass. This stems from the flux accumulation at the centre of the collapsing system 
because the magnetic flux is frozen with the flow, which produces a major interchange instability. After this event, the core environment is weakly magnetised and quite disorganised (looking similar to turbulent runs), enabling the formation of a massive rotationally supported disk.
The evolutionary sequence is portrayed in \refig{plein} (top),
where the times corresponding to snapshots 1 to 4 in \refig{evol2} are indicated in \refig{evol2}. The thin disk-like structure in 1 and 2, with a mass
$M_{\trm{disk}}~\lesssim 5~\times 10^{-2}~\trm{M}_{\odot}$,
evolves into a thick structure (3 and 4) with strong magnetic support, although it does not fulfil the disk criteria we defined
above. The violent release of magnetic flux produced by the interchange instability between epochs 2 and 3 breaks the top-down symmetry of the system and displaces the core by several tens of au (this is not visible in \refig{evol2} since we have integrated the maps by always using the densest cells as the origin).
Detailed maps of the different magnetic components and the plasma $\beta$ are displayed in \refig{plein} (second and third rows) for snapshots 1 and 4. 
The symmetry breaking, as well as the toroidal support, are apparent in the figure.
Throughout most of the structure,
$\beta$ is lower than unity, showing again that magnetic support is ubiquitous in the system. As examined in the previous section, for AD the pile-up of the field occurs outside the core and is much less extended (see \refig{mur_admhd}, at $r \sim 30$ au), and is controlled by the physical, rather than numerical, resistivity. Furthermore, it does not lead to the growth of the pseudo-disk\footnote{Growth of the pseudo-disk due to pile-up of the field has been studied in \citet{HennebelleFromang2008} in the appendix with a model based on quasi-equilibrium growth of a magnetised self-gravitating rotationally dominated structure in iMHD.}. \refig{disk_compa} compares the density/velocity and plasma $\beta$ maps, viewed from above, in the iMHD (top) and AD (bottom) cases. We clearly see the contrast between the highly magnetised ($\beta \ll 1$) disk-like structure that arises in iMHD and the rotationally supported disk that develops spiral arms with $\beta \gg 1$ when ambipolar diffusion is taken into account.

\subsection{Numerical convergence}

Numerical convergence is always a problem in numerical simulations. We carried out a resolution study in the aligned case and, while good agreement was found between the various AD runs, convergence in iMHD was not satisfactory.
Numerical diffusion effects, as well as the choice of Riemann solver to compute fluxes at the cell interfaces, prevent a clear identification of the source of this problem.
This issue has been highlighted in \cite{commercon10,HennebelleFromang2008} for the influence of different solvers, and explored in detail by \cite{2014ApJ...793..130L}, to which we refer to for further information and discussions.
As part of our resolution study, we performed new runs of our benchmark aligned $\mu = 5$ case (hereafter denominated \texttt{case-0}), with a mesh resolution increased by a factor 2 (16 cells per Jeans length) and a resolution decreased to six cells per Jeans length (this latter is only applied for densities above $10^{-12}~\text{g~cm}^{-3}$, while keeping the original Jeans length sampling elsewhere).

By comparing the high-resolution iMHD calculation to \texttt{case-0}, we found a good agreement until $t \sim 25$~kyears (the disk mass is slightly lower by a factor $1.5$ at this stage, but this probably stems from the uncertainties of the criteria used to define the disk
when no clearly identifiable rotationally supported structures are present; see Sect.~\ref{sec:disks}), at which point magnetic field oscillations at the centre of the core start to appear
due to the flux freezing condition (earlier in the high-resolution run). This is reassuring in the sense that it implies that the resolution we have used in \texttt{case-0} is high enough for numerical resistivity effects to be negligible, which agrees with the absence of a strong field accumulation at the boundary of the core. The higher resolution, however, further diminishes the numerical resistivity, and thus flux freezing and flux accumulation are increased,
facilitating the development of instabilities, which now appear earlier in the simulation.

In the low-resolution run, the numerical resistivity is increased, and we observe the appearance of a diffusion plateau, similar to the one in AD runs, but developing at a slightly higher value ($B \simeq 1$~G). In this case, the disk mass is overestimated but the symmetry is preserved longer than in the high-resolution runs. We observe transient ejections similar to those produced by the interchange instability, however, that are due to a strong pile-up of the magnetic field
close to the core. These ejections ultimately lead to a similar quasi-steady final state.

\begin{figure}
    \begin{center}
        \includegraphics[width=0.48\textwidth]{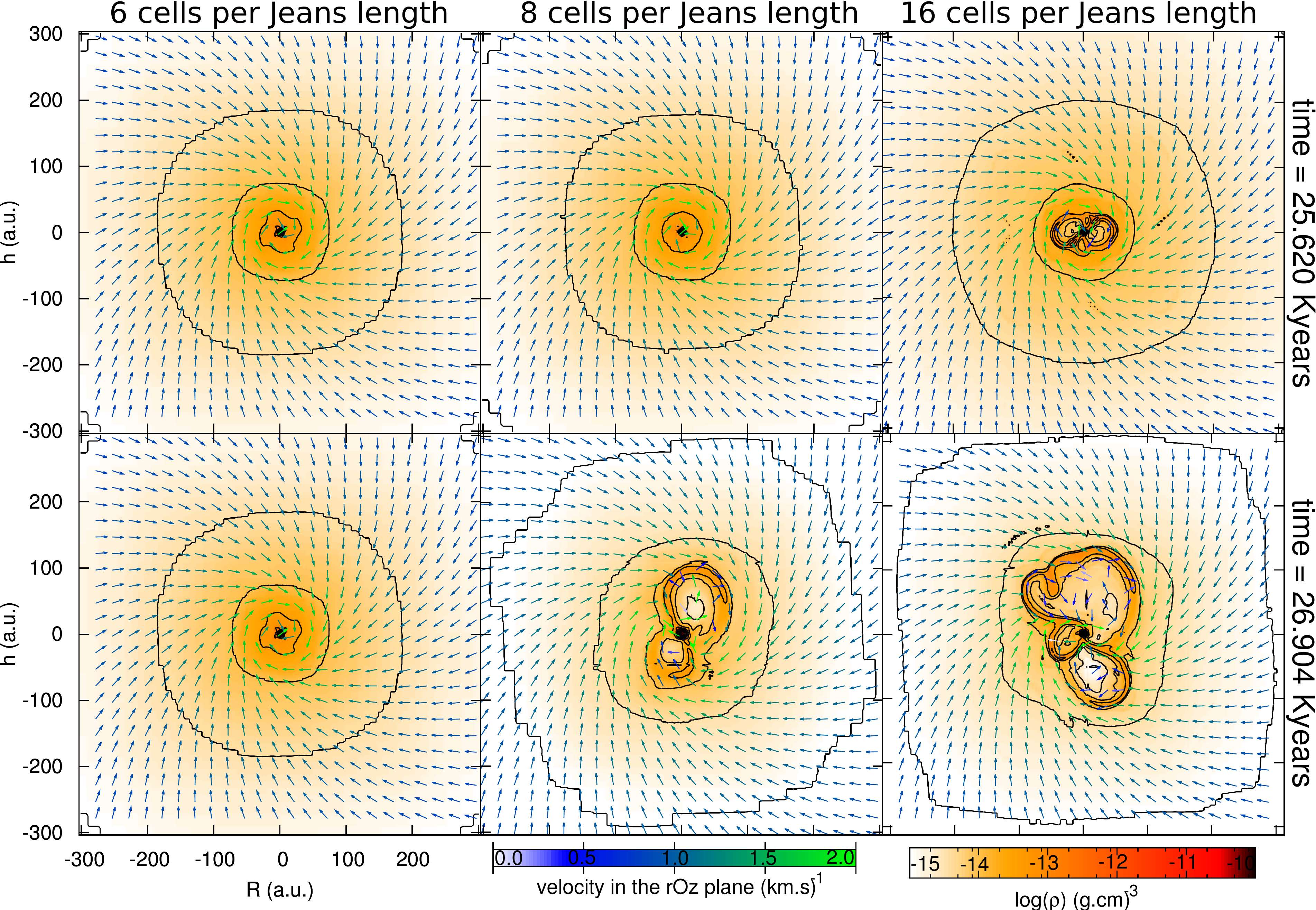}
        \caption{iMHD, $\mu_{\trm{}}=5$, aligned case. View of the disk plane in a cylindrical control volume of height h=$20$~au. First row: $t\sim25.6$~k years. Second row: $t\sim 27$~k years. From left to right, the same simulation with increased resolution: 6, 8, and 16 cells per Jeans length.}
        \label{compa_res}
    \end{center}
\end{figure}

A visual comparison of the different resolutions (6, 8, and 16 cells per Jeans length) at two different times is shown in \refig{compa_res}. As seen, the resolution strongly affects the simulation output when numerical resistivity is not negligible. This is of prime importance in turbulent simulations, where it is never clear that the turbulent reconnection accurately describes the sub-grid unresolved physics. 
Another parameter that leads to numerical diffusivity is the numerical method chosen
to solve the induction equation. More details on this issue are given in Appendix~\ref{part_solver}.

\begin{figure}
     \begin{center}
        \includegraphics[width=0.33\textwidth, angle=270]{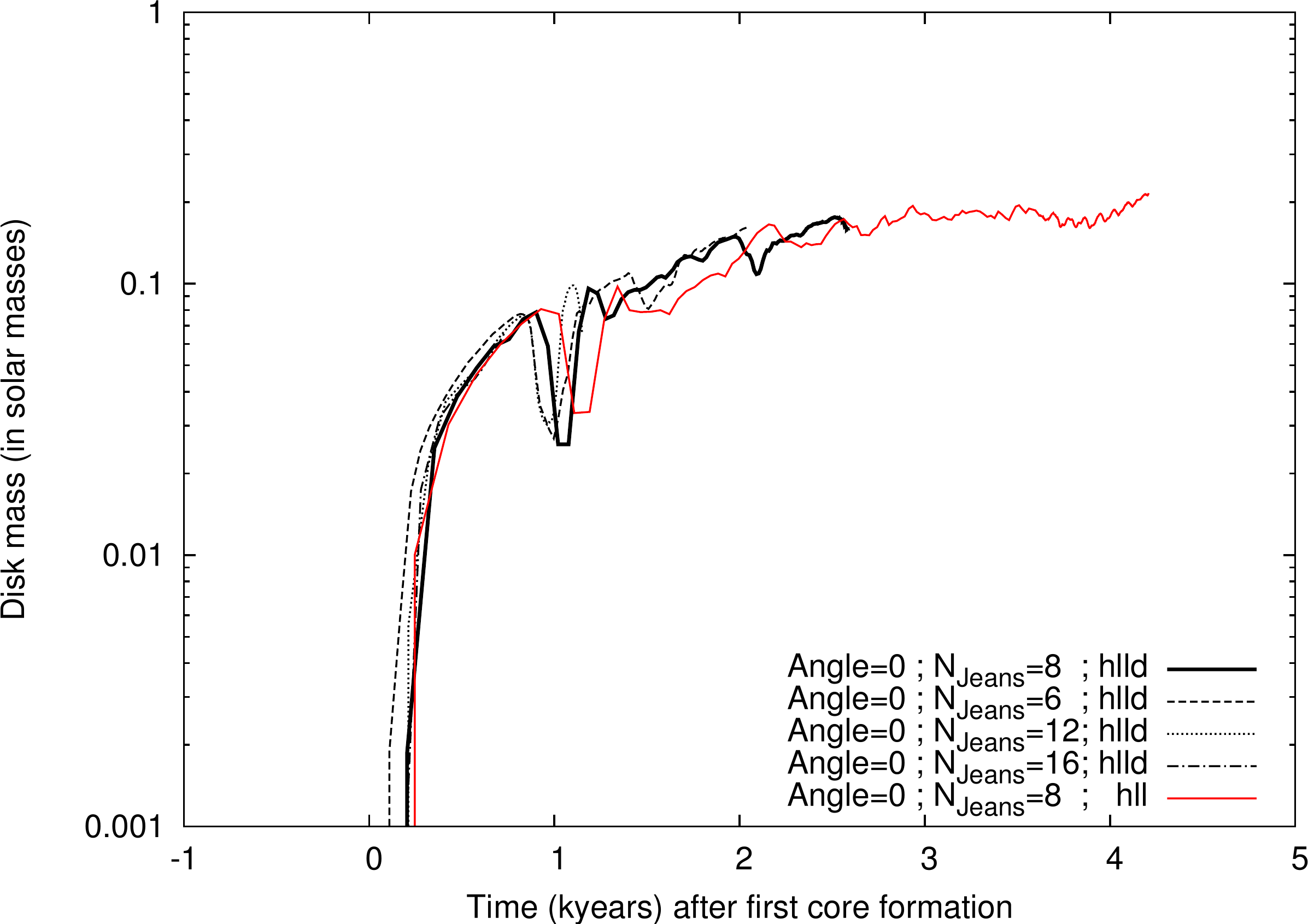}
        \caption{niMHD runs. Comparison of the disk mass in \texttt{case-0} for different numerical resolutions using the HLLD solver (black) and HLL solver (red).
        }
    \label{figm_25_2}
    \end{center}
\end{figure}

The numerical convergence is much better for niMHD. \refig{figm_25_2} shows the disk mass evolution using different mesh resolutions and Riemann solvers. The long-term evolutions are similar, with both the same global variations and final disk mass. 
The differences
also remain small
when changing the solver from HLLD to HLL, as long as the resolution is $\gtrsim 8$ cells per Jeans length. We used the HLLD solver for all our simulations to ensure that ambipolar diffusion dominates the numerical one. 

The conclusion of our convergence study is as follows. 
While in ideal MHD the isothermal first phase of the collapse is accurately described, with the formation of the first core and the early growth of the magnetic bubble/tower, the later evolution strongly depends on the numerical resolution. The key issue is the sudden release of the magnetic energy accumulated in the core that leads to oscillations in the peak magnetic field value (see \refig{evol2} dashed blue line). The release of this energy, depending on the numerical tools, ultimately leads to either a quasi-steady state in which a massive disk-like rotationally and magnetically supported structure forms (see \refig{disk_compa} in the top panels) or, for HLLD, to the appearance of a high-velocity jet and high Alfv\'en speeds that are due to depleted cavities above and below the core (see Appendix \ref{part_solver} for a discussion of the Riemann solver). The long-term evolution of these structures, which directly inherit their properties from the angular momentum and magnetic flux repartition close to the core, is thus of highly questionable validity. 
In misaligned configurations, the pseudo-disk is thicker and the spurious effects due to very high gradients are less acute \citep[see e.g.][]{joos}. The interchange instability is less visible than in \refig{compa_res}, but the flux release and the growth of a rotationally supported structure (as in \refig{evol2}) is still present in the simulation. 

In niMHD, in contrast, neither the resolution (as long as it is good enough) nor the choice of the solver have significant effects on the final outputs. This brings confidence to our global study and reinforces our general conclusion that ambipolar diffusion in the protostellar first collapse plays a major role in regulating angular momentum transport and magnetic flux diffusion, not to
mention preventing spurious instabilities found in ideal MHD.


\section{Comparison with previous work}\label{sec:Con-parisons}

In this section, we compare our results with those of four previous studies that are representative of the exploration of star formation in non-ideal magnetohydrodynamics. 

\cite{Krasnopolsky} first explored the impact of enhanced resistivities on disk formation. Their study was motivated by the fact that in ideal MHD, magnetic fields prevent the formation of large Keplerian disks, which were ubiquitous in hydrodynamical simulations. Resistive MHD was identified as a physical process able to limit this effect. Their setup uses an isothermal equation of state, does not account for self-gravity, and assumes axisymmetry. Their main results, compared to ours, are the following:
\begin{itemize}
    \item in their Figs. 5-8, the pinching of the field lines in the equator and therefore the split monopole configuration is released as the resistivity is increased.
    \item they found that a resistivity $\gtrsim 10^{19}$~cm$^2$~s$^{-1}$ is needed to form a substantial disk.
\end{itemize}
While they did not directly link the release of the split-monopole configuration to the formation of rotationally supported disks, they concluded that increasing the resistivity helps both to form disks and to reduce numerical artefacts through physical rather than numerical dissipation, and it also changes the topology of the field that  is prone to reconnect close to the star where the pinching was strong. We agree with these conclusions and note that the resistivity above which they formed disks corresponds to or is slightly stronger than the typical values of ambipolar resistivity we used \citep[see][]{chimie} that lead to the formation of disks. However, these authors assumed a constant enhanced resistivity in the induction equation in the Laplace operator. Consequently, they could not grasp any non-linear effect, which, as we showed, can lead to a saturation of the magnetic field and further facilitate the formation of rotationally supported structures. 

\cite{2014ApJ...793..130L} continued the work started by \cite{Krasnopolsky} and studied the mechanisms of disk formation. While they focused on iMHD, they raised many of the questions we developed in Sect.~\ref{sec:longtermevo}. In particular, they insisted on the role of the warping and the reduced reconnection close to the protostar, and they emphasized that a correct treatment of magnetic flux accretion on the protostar (especially if a sink particle is used) and reconnection at the boundary of the disk/pseudo-disk are mandatory and are lacking in many previous studies. The problem of turbulent reconnection in iMHD was also discussed and placed in perspective. We agree with their findings and insist on the fact that numerical questions regarding flux conservation and reconnection are crucial for a proper study of disk formation.

\cite{2015MNRAS.452..278T} presented the first second core calculations with both Ohmic dissipation and ambipolar diffusion. They performed three-dimensional simulations using an SPH code with a Godunov module and divergence cleaning methods for the induction equation. Their main conclusions are the following:
\begin{itemize}
    \item 1) a value $\beta_{\trm{plasma}} \gg 10^4$ in a flattened first core.
    \item 2) formation of a circumstellar disk (with a radius of $1$~au) at the formation of the protostar.
    \item 3) occurrence of a saturation plateau at $B\sim {\rm a\,few} 10^{-3}$ G in the evolution of the magnetic field at densities $10^{-15} < \rho < 10^{-14}$ g cm$^{-3}$ followed by a sheet-like collapsing phase (in which $B\propto \rho^{1/2}$) until the formation of the first core.
    \item 4) they did not observe disks around the first Larson core.
\end{itemize}
Their values for $\beta_{\trm{plasma}} $ (point 1) are similar to ours (see Table~\ref{table_fc}). So far, we cannot comment on point 2 since we did not address the second core formation in this paper. 
When studying the evolution of the central magnetic field as a function of density, they only plotted the data at the centre of the system (while we showed every cell in the simulation), and they thus failed to report a saturation of the magnetic field that is seen in our \refig{fig_1_1}. Through private communication, we have obtained the confirmation that they indeed observed a diffusion plateau that was due to ambipolar diffusion, but its origin or consequences are not discussed in their work.
We note that they assumed a sheet-like accretion through the disk, whereas in our simulations accretion onto the first core mostly occurs in channels driven by the hourglass configuration. There is, however, no real evidence for their assumption. Finally, the lifetime of the first core in their simulations is comparable to the time of integration after the first core formation in our runs (of the order of a thousand years for $\mu=4$). However, they did not form any mid-scale (a dozen to a few dozen au) rotationally supported structure around the first core. Instead, in their Ohmic and ambipolar+Ohmic calculations, the flattened first core itself evolves into a rotationally dominated structure that they called a disk. These differences in the dynamics of the collapse highlight the difficulty of conducting reliable numerical calculations during prestellar core collapse, 
which involves a wide range of temporal and spatial scales, and the necessity of including all relevant physical processes and of using reliable numerical schemes. These studies, and the
present ones, open the route to such explorations.

\cite{2015ApJ...801..117T} performed a study similar to ours, focusing on the first core formation and the surrounding structures, with the addition of radiation transfer instead of a barotropic equation of state. They used only one value of magnetisation
that is close to our low-magnetisation case. Their model with ambipolar diffusion also included Ohmic dissipation, and they did not study the effects of ambipolar diffusion alone. We agree with the following of their results:
\begin{itemize}
    \item a flattening of the first core, especially in their OA (Ohmic+ambipolar diffusion) run.
    \item depleted polar cavities where the ambipolar action is dominant.
    \item in the simulation labelled I (ideal MHD), they found counter-rotation of the core (as visible in their Fig.~$4$), as in the present \refig{cntr}, but explained it by the interchange instability mixing the physical quantities. While we also found counter-rotation in iMHD, it occurs {\it \textup{before}} the interchange instability develops. Following \cite{GalliShuLizano2006}, we explain this counter-rotation by a strong pile-up of the magnetic field yielding enough magnetic braking (or torque) to completely stop the rotation of the core and spin it backwards. The general conclusion, however, remains similar: iMHD models are unable to accurately describe angular momentum transport after the instability has developed.
    \item an increase of the mass-to-flux ratio up to or above a factor 10 due to Ohmic and Ohmic+ambipolar diffusion action, especially at the first core scales.
    \item no splitting of the field lines in the midplane.
    \item the use of a genuine physical dissipation scale helps numerical convergence and thus assesses the reliability of the results.
\end{itemize}
Some of their results are more difficult to discuss. For instance, their explanation for a higher first core mass in the resistive cases where the magnetic pressure and the thermal and rotational supports are higher. In our experience, the interplay between magnetic braking (yielding stronger or weaker rotational support) and removal of magnetic flux (yielding stronger or weaker magnetic support) strongly affects the mass of the core. Although it is possible that in their set of simulation, the iMHD models lead to less massive first cores, in our runs we found that the first core mass is higher in the iMHD framework for $\mu=2,$  but somewhat similar to or lower than the niMHD value for $\mu=5$  (see Table~\ref{table_fc}). The physical explanation remains the same: less support in total, but it is not straightforward to predict the results over a broad range of initial conditions. Although their discussion of the disk is not expanded enough to allow detailed comparison with our results, 
the main result is similar: non-ideal MHD enables disk formation with no need to artificially relax the flux-freezing constraint or to invoke enhanced resistivities.

\section{Conclusions}\label{sec:Conclusions}

We have thoroughly explored the role of ambipolar diffusion in magneto-hydrodynamic collapse calculations of a dense molecular cloud core in the context of star formation.
Our setup involved a magnetised core of uniform density in solid-body rotation, which collapsed under its own gravity. The gas had a barotropic equation of state to mimic radiative transfer, and the magnetic resistivities entering the calculations of the non-ideal MHD terms were calculated using a reduced chemical network. We performed eight simulations, with two different magnetisations ($\mu = 5$ and 2) and two tilting values
of the rotation axis ($0^{\circ}$ and $40^{\circ}$) with respect to the magnetic field direction, both in ideal MHD and in non-ideal MHD with ambipolar diffusion. We paid particular attention to the problems of magnetic flux conservation, magnetic braking, and flux release due to diffusion processes.
Our main findings are summarised as follows.

\begin{itemize}

\item[\textbullet] Ambipolar diffusion creates a magnetic diffusion barrier at about the time the first Larson core forms, preventing the magnetic field from being amplified above 0.1~G. Flux freezing, however, still holds during the initial stages of the collapse, when resistivities are low. The mass and radius of the first Larson core remain rather similar between iMHD and niMHD models.
\vspace{0.5em}\item[\textbullet] The magnitude of $B$ at the diffusion plateau can be estimated by simple order-of-magnitude arguments. This saturation value appears to depend very weakly on the initial cloud magnetisation, suggesting a convergence of the final state once the initial conditions have been "forgotten" by the system.
\vspace{0.5em}\item[\textbullet] The occurrence of a diffusion plateau has crucial consequences on magnetic braking processes. Not only does it prevent a catastrophic amplification of the field $B$, which controls the braking efficiency, but it also reorganises the field topology, reducing the pinching of field lines close to the protostellar object, which also hinders the braking mechanism.
\vspace{0.5em}\item[\textbullet] Magnetic flux freezing and magnetic braking play a central role in the formation and development of the structures during the collapse. While in iMHD strongly amplified fields launch powerful magnetically supported outflows, these latter are much weaker or even disappear in niMHD. 
\vspace{0.5em}\item[\textbullet] Misalignment between the initial rotation axis and the magnetic field direction does not appear to affect the niMHD results, showing that the physical flux dissipation due to ambipolar diffusion dominates the effects of initial configuration or numerical diffusion. For iMHD models, the additional mixing due to a tilted configuration also produces a diffusion plateau, similar to the ambipolar calculations.
 
\vspace{0.5em}\item[\textbullet] 
The disks that form in iMHD, characterised by strong magnetic support and an inflated shape (due to toroidal field lines loaded by infalling matter) strongly differ from the flat disks that form in niMHD. Formation and long-term evolution of disks in ideal MHD, however, are of dubious validity. Furthermore, the excessive magnetic braking generates unphysical counter-rotation inside the outflow or the magnetic tower. Interchange instabilities develop at the interface between the core and the disk, producing a redistribution of the flux and displacement of the core and disrupting the top-bottom symmetry in the system. Mesh resolution is found to strongly affect the simulation results in iMHD. None of these effects is observed in the niMHD simulations if the
resolution is high enough.
\vspace{0.5em}\item[\textbullet] Disks with Keplerian velocity profiles around the protostar form in all our niMHD simulations for all different magnetisations and inclination angles. Their size and mass, however, is significantly reduced (by a factor $\sim 10$) in the more magnetised case ($\mu=2$) because of the increased braking. Such a magnetisation value seems to be typical of most molecular clouds. Aligned and misaligned initial configurations have no consequence on the disk properties in niMHD and yield disks with very similar properties.

\end{itemize}

Magnetic flux diffusivity due to ambipolar diffusion thus appears to play a dominant role during the first protostellar collapse and the formation of the first Larson core and its surrounding disk,
and it appears to dominate processes such as magnetic field and rotation axis orientations. Flux diffusion during the collapse allows the formation of quasi-Keplerian disks,
solving the "disk formation crisis" found in ideal MHD. The mass, size, and magnetic properties (plasma $\beta$) of these disks, however, strongly depend on the initial magnetisation ($\mu$ parameter) of the cloud, and in all cases the disks are significantly smaller and less massive than those found in pure hydrodynamics calculations. Characterisation of this plasma $\beta$ is of major importance to characterise viscous transport in disks, notably in the MRI.
To complete our study, we will explore the effect of turbulence on the first collapse of a magnetised cloud core in a forthcoming paper.

\begin{acknowledgements}
We thank the anonymous referee for the suggestions and remarks that contributed to improve the quality of this manuscript. The research leading to these results has received funding from the European Research Council under the European Community's Seventh Framework Programme (FP7/2007-2013 Grant Agreement no. 247060). BC gratefully acknowledges support from the French ANR Retour Postdoc program (ANR-11-PDOC-0031). We finally acknowledge financial support from the "Programme National de Physique Stellaire" (PNPS) of CNRS/INSU, France.
\end{acknowledgements}

\appendix

\section{Calculation of the saturation value \label{A1}}

In this section, we provide an analytical estimate of the value of $B$ at which the magnetic diffusion starts to overcome the dynamical MHD effects.
We rewrite the initial values for $B_0$ and $\rho_0$ in terms of the ratio of thermal to gravitational energy $\alpha$ and mass-to-flux ratio $\mu$, according to
\begin{align}
B_0 =&  \frac{1}{\msol \left( \frac{2G}{5c_{\trm{s}}^2}  \right)^2 \mu_{\trm{crit}}} \left( \frac{M}{\msol}  \right)^{-1} \left( \frac{\alpha}{\alpha=1}  \right)^{-2} \left( \frac{\mu}{\mu=1}  \right)^{-1} ~, \\
\rho_0 =& \frac{1}{\frac{4}{3}\pi \left( \frac{2 G}{5 c_{\trm{s}}^2} \right)^3 \msol^2} \left( \frac{M}{\msol}  \right)^{-2} \left( \frac{\alpha}{\alpha=1}  \right)^{-3} ~.
\end{align}
The relationship between the magnetic field and density is
assumed to follow $\rho \propto B^{\xi}$. We also
assume that in the early phases of collapse, the typical length scale is $L\simeq c_{\trm{s}} t_{\trm{free-fall}}$ and the velocity is $V\simeq c_{\trm{s}}$. 
We then seek the value of the magnetic field $B_{\mathrm{sat}}$ for which
\begin{equation}
    \vect{V} \times \vect{B} \simeq \eta_{\trm{AD}} \frac{\mathbf{B}}{||\vect{B}||}  \times \left[ (\nable \times \mathbf{B})\times \frac{\mathbf{B}}{||\vect{B}||} \right] ~,
\end{equation}
with $\eta_{\trm{AD}} \simeq \frac{B^2}{\gamma_{\trm{AD}} \rho_{\trm{i}} \rho}$. Using the fact that
$\rho_{\trm{i}}= C \sqrt{\rho}$ \citep[from][]{Shu_1992}, we can then write
\begin{equation}
B^2 = c_{\trm{s}}^2  t_{\trm{free-fall}} \gamma_{\trm{AD}} \rho^{3/2} C,
\end{equation}
which yields the saturation value for the magnetic field
\begin{align}
    B_{\trm{sat}}=\left(c_{\trm{s}}^2 C \gamma_{\trm{AD}} \sqrt{\frac{3}{2 \pi G}} \right)^{\frac{1}{2-\xi}}\left( \frac{B_0}{\rho_0^{1/\xi}} \right)^{\frac{-\xi}{2-\xi}}. \label{fuleq}
\end{align}

\section{Influence of the solver \label{part_solver}}

\begin{figure*}
    \begin{center}
        \subfigure[Lax-Friedrich]{
            \label{rs_llf}
        \includegraphics[width=0.2215\textwidth]{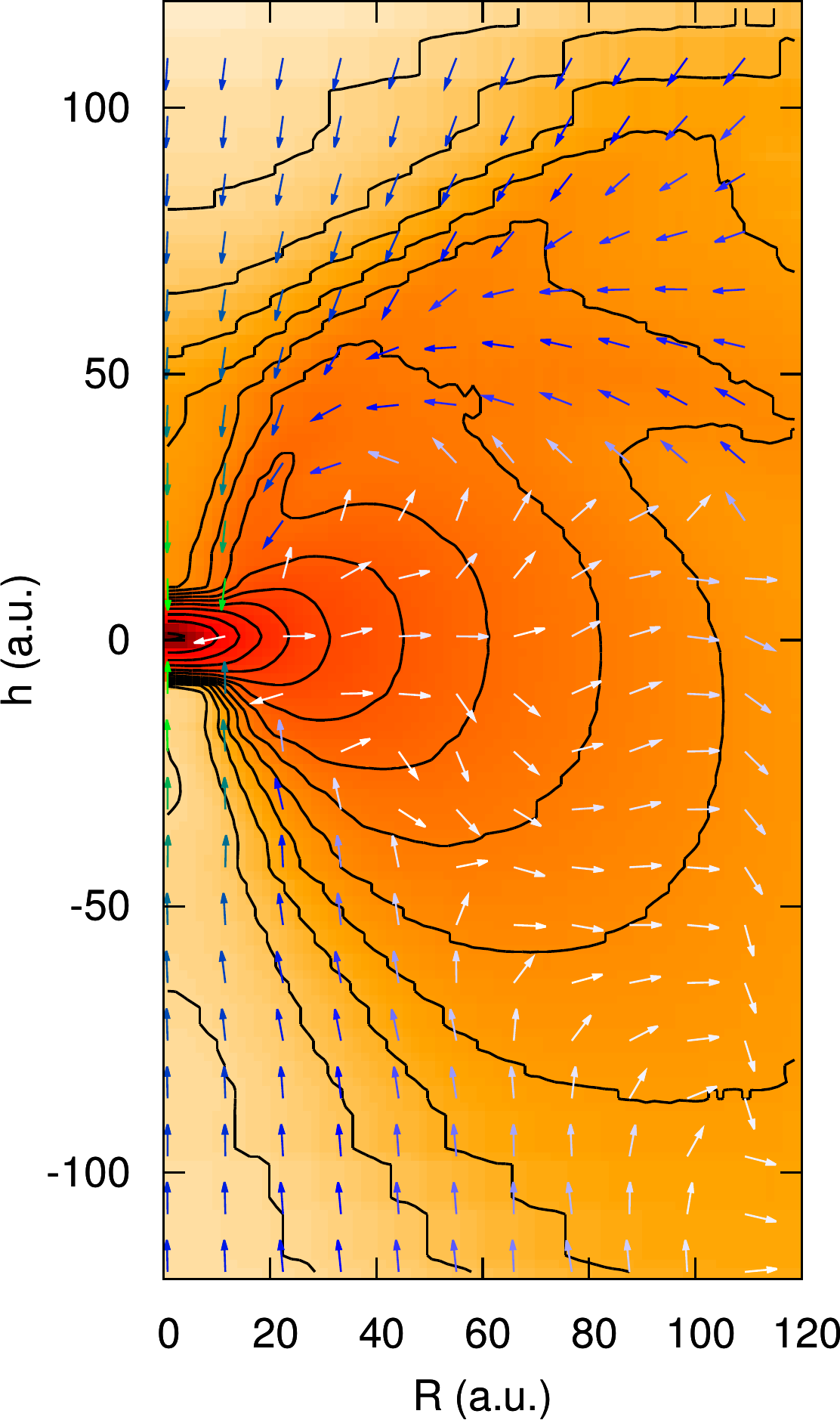}
    }
        \subfigure[HLL]{
            \label{rs_sshll}
        \includegraphics[width=0.18\textwidth]{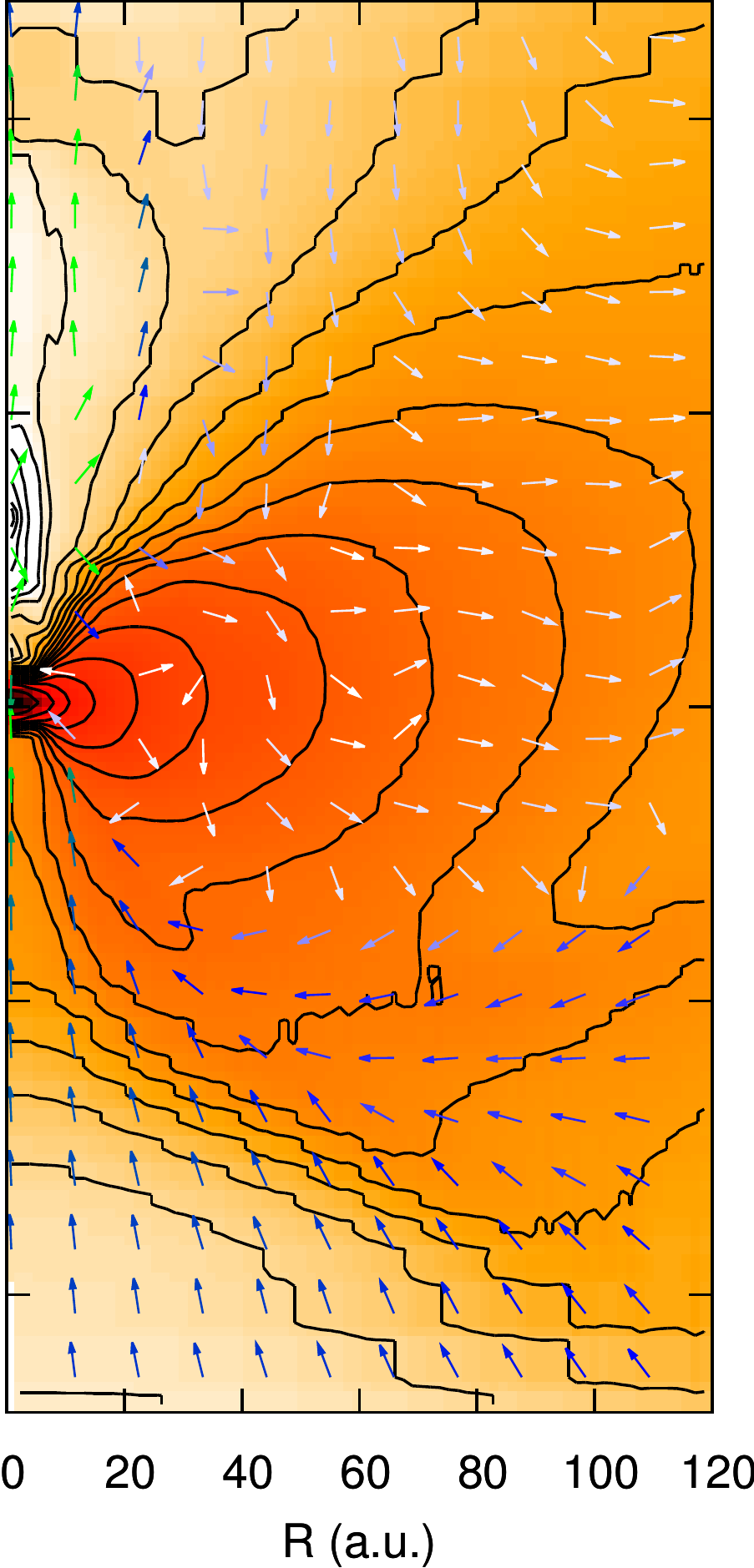}
    }
        \subfigure[HLLD+switch to LLF]{
            \label{rs_ashlld}
        \includegraphics[width=0.18\textwidth]{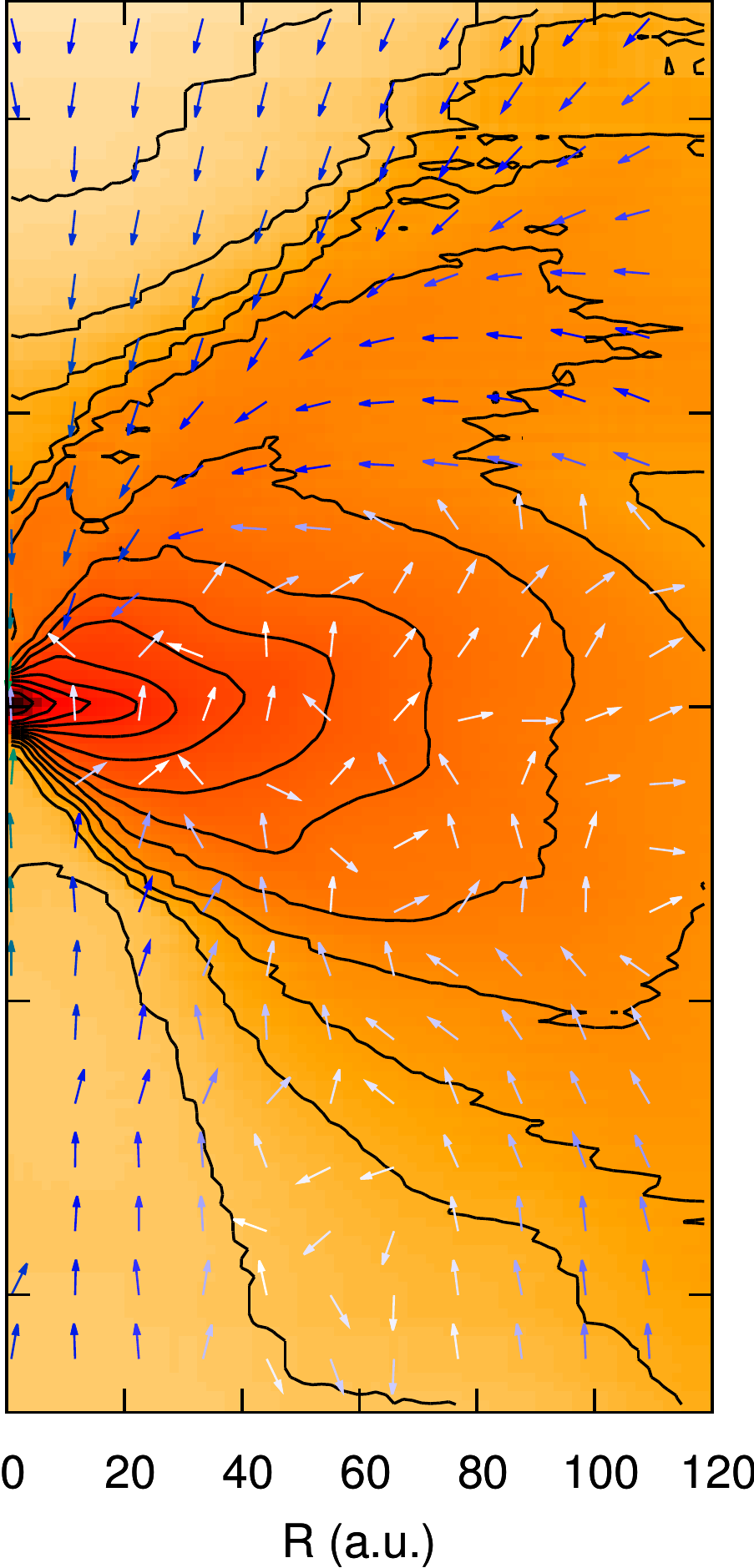}
    }
        \subfigure[HLLD]{
            \label{rs_sshlld}
        \includegraphics[width=0.18\textwidth]{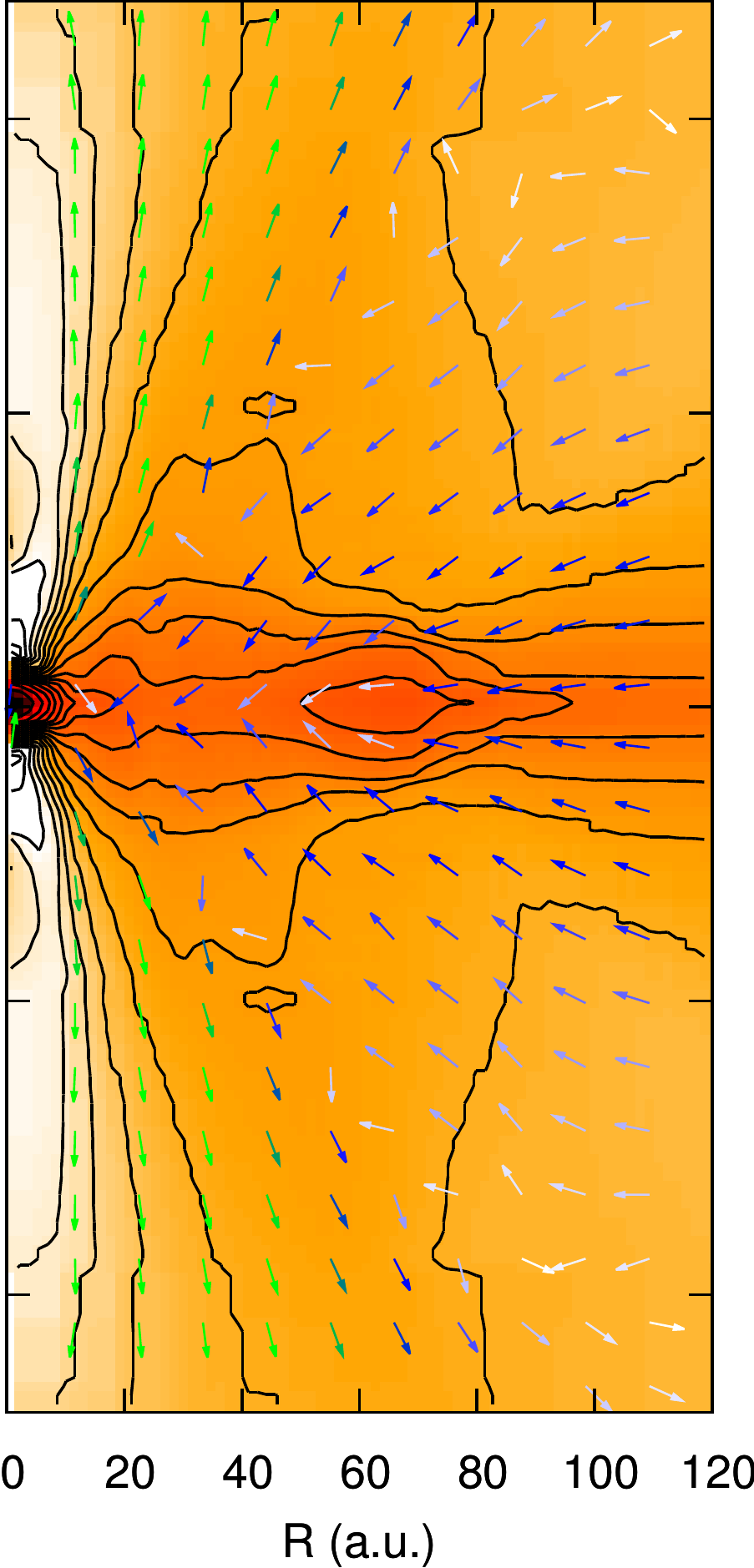}
        } 
        \subfigure{
            \label{rs_colorbar}
            \raisebox{1.8cm}{
                \includegraphics[width=0.15\textwidth]{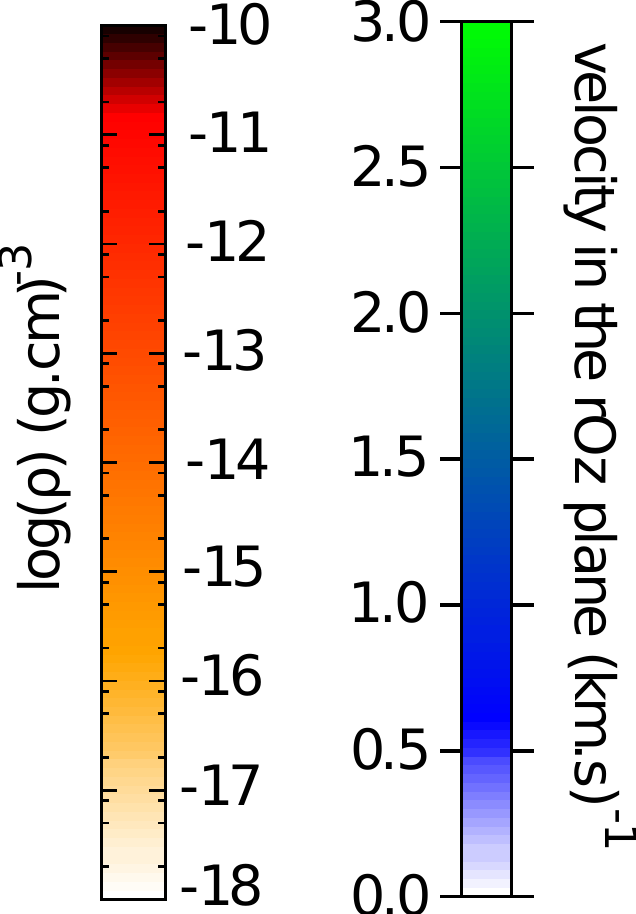}}
    }\\
        \caption{Density maps of the region around the first core using four different Riemann solvers: (a) Lax-Friedrich, (b) HLL, (c) HLLD+Switch, (d) HLLD. The velocity vectors are overlaid on density maps with green and blue arrows.}
        \label{RiemannSolvers}
    \end{center}
\end{figure*}

The choice of the Riemann solver to compute the conservative fluxes at the interfaces between domain cells can greatly affect the results of a simulation, especially when studying diffusion processes.
We compare here
four test simulations
that use different strategies to compute the fluxes. The resolution for each run is 8 cells per Jeans length. In \refig{RiemannSolvers} from left to right, the solver is increasingly less diffusive.

In the HLLD+switch simulation, we used the HLLD solver over most of the computational domain, but switched to a Lax-Friedrich solver whenever large discontinuities\footnote{Discontinuities either in density, or when the wave speed for the intermediate states for HLLD are ill defined due to a very small denominator.} are present in the flow variables between neighbouring cells.
This allowed us
to evolve the simulations for long periods of time, up to $\sim 1.5$ free-fall time. However, it
raises crucial problems on both the numerical reconnection in the vicinity of the core and the consequent interchange instability, as well as
a violent core displacement that accompanies an 
unphysical release of  magnetic energy at the centre.
The formation of large inflated disks is also questionable.

\begin{figure}
    \begin{center}
       \includegraphics[width=0.3\textwidth, angle=270]{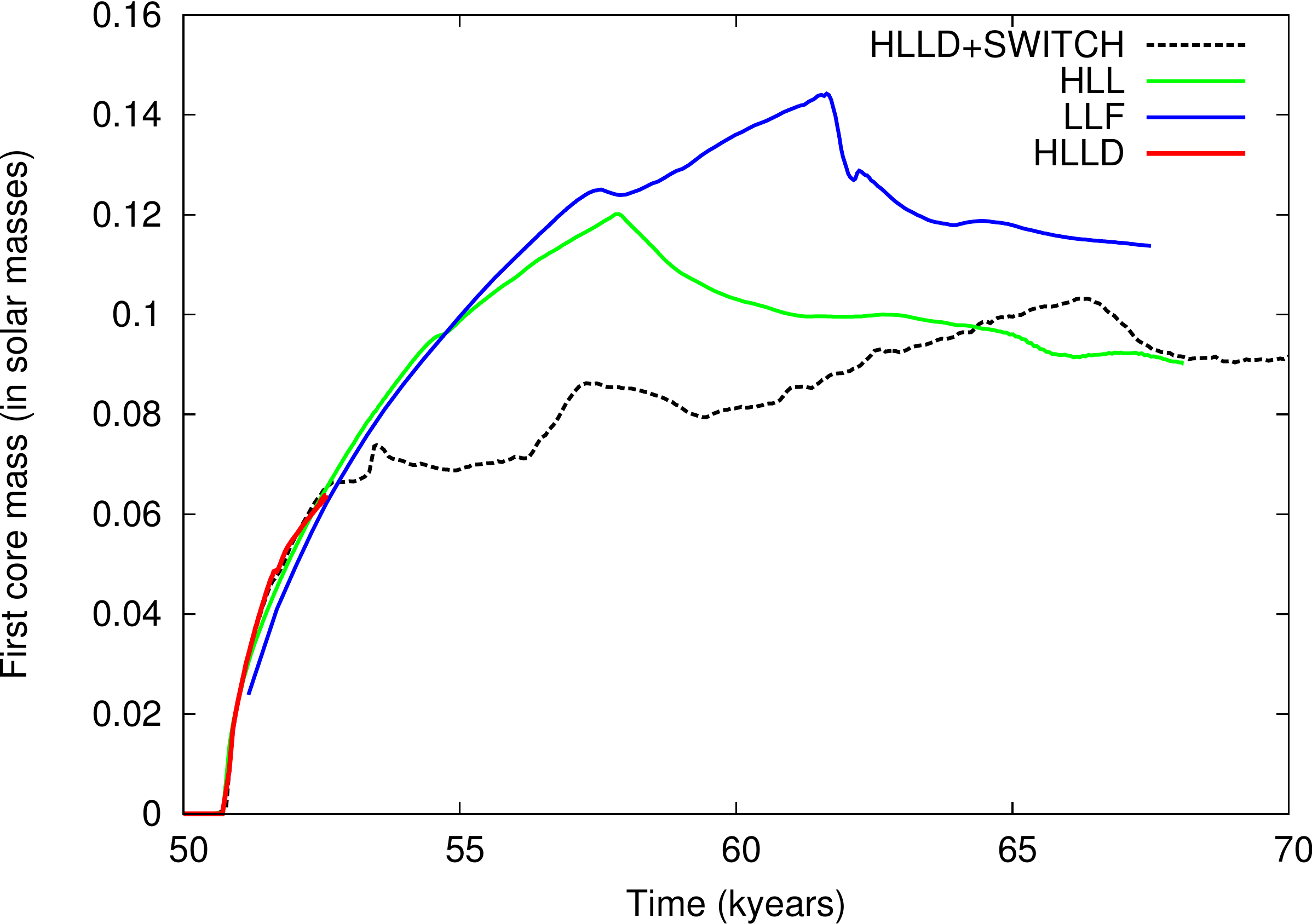}
       \caption{First core mass evolution for various solvers. Standard resolution of 8 cells per Jeans length.}
        \label{fcImhd}
    \end{center}
\end{figure}

In a second calculation, we removed the switch to a more diffusive solver, keeping HLLD in the entire domain. In this case, the discontinuities (mostly magnetic and density related) continue
to sharpen, until
the time integration in the simulation is frozen due to very high Alfv\'en speeds in the outflows or jets (it can reach speeds of the order of $\gtrsim 10^4~\text{km~s}^{-1}$). Whether this is a correct behaviour is open to debate \citep[see also][]{2014ApJ...793..130L}.
We reach convergence of results quicker than in the HLLD+switch case, since we basically prevent numerical diffusion, provided we have enough resolution at a given scale and density. On the other hand, there are physical diffusive processes at play in star formation, either from microphysics (e.g. ambipolar diffusion) or from turbulent reconnection. Therefore, a description that avoids any diffusion lacks physical mechanisms. 
Finally, we performed simulations using
only HLL that takes into account only $\text{three}$ waves instead of $\text{five}$ with HLLD. This case fits in between the two previous ones; convergence is reached with 16 cells per Jeans length, but we do witness diffusion in the outflow and the vicinity or interior of the Larson core. In this case,
Alfv\'en speeds remain below
$10^3~\text{km~s}^{-1}$. For simulations of the second collapse and formation of the second core, for which time integration is critical, our former tests suggest that using HLL with high enough resolution, at least in ideal MHD, is better than using HLLD (with or without the switch) because it enables both numerical convergence and the probing of longer timescales. 
For the sake of comparison, we performed one test case using the Lax-Friedrich solver alone. The results are similar to the HLL case.

Additionally, \refig{fcImhd} shows the first core mass evolution for each simulation, highlighting significant differences in an outcome as well defined as the first core mass.
Until the time when the least diffusive solver (HLLD) does not allow continuing the simulation anymore because the time step
is too small, the results are similar. However, after this point, depending on the choice of solver, results can significantly differ both in the qualitative picture (as seen in \refig{RiemannSolvers}) or for a precise outcome (\refig{fcImhd}). Therefore, depending on the timescale of interest and the precise point of study, a careful choice of the numerical method has to be made.

\section{Disk velocity profile using ideal MHD}\label{app:imhddiskprofile}

In this appendix, we caution about the definition of a Keplerian disk and the estimate of the central core mass obtained when fitting a priori the disk radial velocity distribution 
with a Keplerian profile ($\propto r^{-1/2}$), in particular in structures whose growth is magnetically driven, as is the case for pseudo-disks in ideal MHD. The disk criteria we used, based on \cite{joos}, define a rotationally dominated structure. We then proceeded by studying the radial dependence of the toroidal velocity, and compared it to the classical Keplerian velocity profiles. The velocity profile, the plasma $\beta,$ and the aspect ratio allowed us to estimate how similar the disks we found are to a Keplerian disk. As emphasised below, we find that these disks can depart significantly from the classical Keplerian picture. The relatively high mass of the surroundings of the first core
and the magnetic fields cause the usual Keplerian model to fail. This issue is of prime importance when trying to characterise the properties of the disk structures observed around Class-0 objects \citep[see e.g.][]{2013ApJ...771...48T,2015ApJ...805..125T}.

\begin{figure}
    \begin{center}
       \includegraphics[width=0.33\textwidth, angle=270]{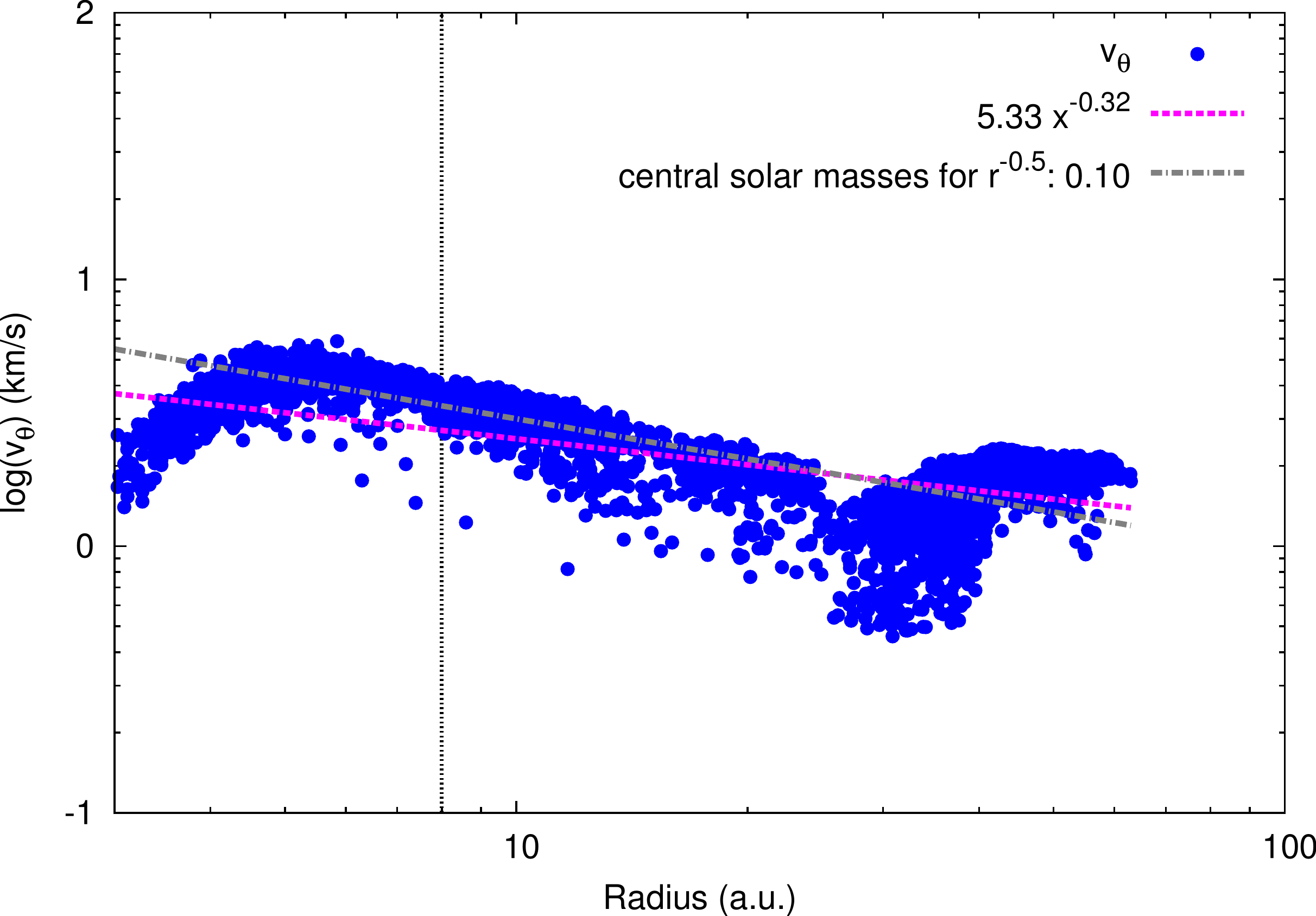}
       \caption{Same as \refig{fig_19_2} for iMHD.}
        \label{estimateMHD}
    \end{center}
\end{figure}

An example is shown in \refig{estimateMHD}. Whereas a Keplerian profile (grey dashed line) can be roughly fitted to the toroidal velocity field (blue dots), 
the best-fitting estimate (dashed red line) gives a significantly different value, $\propto r^{-0.32}$, yielding 
a central mass different from the one inferred from the model of a Keplerian velocity profile around a central point mass. 
This result stresses the fact that fitting a scattered velocity profile, which can depart significantly from a Keplerian one, by such a value
can be very uncertain, casting doubts on estimates of central masses obtained with Keplerian formulae.

\end{document}